\documentclass[usegraphicx]{mn2e}

\usepackage[fleqn]{amsmath}
\numberwithin{equation}{section}
\def\d{{\mathrm{d}}}
\newcommand{\D}{\displaystyle}

\title[General solution Jeans equations]{General solution of the Jeans
  equations for triaxial galaxies with separable potentials} 
\author[Van de Ven et al.]
  {G.~van~de~Ven,$^1$\thanks{E-mail: glenn@strw.leidenuniv.nl}\
  C.~Hunter,$^2$ E.K.~Verolme,$^1$ P.T.~de~Zeeuw$^1$\\
  $^1$Sterrewacht Leiden, Postbus 9513, 2300 RA Leiden, The Netherlands \\
  $^2$Department of Mathematics, Florida State University,
  Tallahassee, FL 32306-4510} 
\date{Accepted 0000 Month 00. Received 0000 Month 00;in original 0000 Month 00}
\pagerange{\pageref{firstpage}--\pageref{lastpage}}
\pubyear{0000}

\begin{document}

\label{firstpage}

\maketitle

% ===== ABSTRACT ===== %

\begin{abstract}
The Jeans equations relate the second-order velocity moments to the
density and potential of a stellar system. 
For general three-dimensional stellar systems, there are three
equations and six independent moments. 
By assuming that the potential is triaxial and of separable St\"ackel
form, the mixed moments vanish in confocal ellipsoidal coordinates. 
Consequently, the three Jeans equations and three remaining
non-vanishing moments form a closed system of three highly-symmetric
coupled first-order partial differential equations in three
variables. 
These equations were first derived by Lynden--Bell, over 40 years ago,
but have resisted solution by standard methods. 
We present the general solution here.

We consider the two-dimensional limiting cases first. 
We solve their Jeans equations by a new method which superposes
singular solutions. 
The singular solutions, which are new, are standard Riemann--Green
functions.
The resulting solutions of the Jeans equations give the second moments
throughout the system in terms of prescribed boundary values of certain
second moments. 
The two-dimensional solutions are applied to non-axisymmetric discs,
oblate and prolate spheroids, and also to the scale-free triaxial
limit. 
There are restrictions on the boundary conditions which we discuss in
detail. 
We then extend the method of singular solutions to the triaxial case,
and obtain a full solution, again in terms of prescribed boundary
values of second moments. 
There are restrictions on these boundary values as well, but the
boundary conditions can all be specified in a single plane. 
The general solution can be expressed in terms of complete
(hyper)elliptic integrals which can be evaluated in a straightforward
way, and provides the full set of second moments which can support a
triaxial density distribution in a separable triaxial potential.
\end{abstract}


% ===== section 1 ===== %

\section{Introduction}
\label{sec:introduction}

Much has been learned about the mass distribution and internal
dynamics of galaxies by modeling their observed kinematics with
solutions of the Jeans equations (e.g., Binney \& Tremaine
1987\nocite{1987gady.book.....B}).  These are obtained
by taking velocity moments of the collisionless Boltzmann equation for
the phase-space distribution function $f$, and connect the second
moments (or the velocity dispersions, if the mean streaming motion is
known) directly to the density and the gravitational potential of the
galaxy, without the need to know $f$. In nearly all cases there are
fewer Jeans equations than velocity moments, so that additional
assumptions have to be made about the degree of anisotropy.
Furthermore, the resulting second moments may not correspond to a
physical distribution function $f\geq 0$. These significant drawbacks
have not prevented wide application of the Jeans approach to
the kinematics of galaxies, even though the results need to be
interpreted with care. 
Fortunately, efficient analytic and numerical methods have been
developed in the past decade to calculate the full range of
distribution functions $f$ that correspond to spherical or 
axisymmetric galaxies, and to fit them directly to kinematic
measurements (e.g., Gerhard 1993\nocite{1993MNRAS.265..213G}; Qian et
al.\ 1995\nocite{1995MNRAS.274..602Q}; Rix et al.\
1997\nocite{1997ApJ...488..702R}; van der Marel et al.\
1998\nocite{1998ApJ...493..613V}). This has provided, for example,
accurate intrinsic shapes, mass-to-light ratios, and central black
hole masses (e.g., Verolme et al.\ 2002\nocite{2002MNRAS..Verolme};
Gebhardt et al.\ 2003\nocite{2002Gebhardt_etal}).

Many galaxy components are not spherical or axisymmetric, but have
triaxial shapes (Binney 1976\nocite{1976MNRAS.177...19B}, 
1978\nocite{1978MNRAS.183..501B}). These include early-type bulges,
bars, and giant elliptical galaxies. In this geometry, there are three
Jeans equations, but little use has been made of them, as they contain
six independent second moments, three of which have to be
chosen ad-hoc (see, e.g., Evans, Carollo \& de Zeeuw
2000\nocite{2000MNRAS.318.1131E}). At the same time,
not much is known about the range of physical solutions, as very few
distribution functions have been computed, and even fewer have been
compared with kinematic data 
(but see Zhao 1996\nocite{1996MNRAS.283..149Z}).

An exception is provided by the special set of triaxial mass models
that have a gravitational potential of St\"ackel form. 
In these systems, the Hamilton--Jacobi equation separates in
orthogonal curvilinear coordinates 
(St\"ackel 1891\nocite{histStackel1891}), 
so that all orbits have three exact integrals of motion, which are
quadratic in the velocities. 
The associated mass distributions can have arbitrary central axis
ratios and a large range of density profiles, but they all have cores
rather than central density cusps, which implies that they do not
provide perfect fits to galaxies 
(de Zeeuw, Peletier \& Franx 1986\nocite{1986MNRAS.221.1001D}). 
Even so, they capture much of the rich internal dynamics of large
elliptical galaxies 
(de Zeeuw 1985a\nocite{1985MNRAS.216..273D}, hereafter Z85; 
Statler 1987\nocite{1987ApJ...321..113S}, 
1991\nocite{1991AJ....102..882S};
Arnold, de Zeeuw \& Hunter 1994\nocite{1994MNRAS.271..924A}).  
Numerical and analytic distribution functions have been constructed
for these models
(e.g., Bishop 1986\nocite{1986ApJ...305...14B};
Statler 1987\nocite{1987ApJ...321..113S};
Dejonghe \& de Zeeuw 1988\nocite{1988ApJ...333...90D};
Hunter \& de Zeeuw 1992\nocite{1992ApJ...389...79H}, hereafter HZ92; 
Mathieu \& Dejonghe 1999\nocite{1999MNRAS.303..455M}), 
and their projected properties have been used to provide constraints
on the intrinsic shapes of individual galaxies (e.g., 
Statler 1994a\nocite{1994ApJ...425..458S}, 
b\nocite{1994ApJ...425..500S}; 
Statler \& Fry 1994\nocite{1994ApJ...425..481S}; 
Statler, DeJonghe \& Smecker-Hane 1999\nocite{1999AJ....117..126S}; 
Bak \& Statler 2000\nocite{2000AJ....120..110B};
Statler 2001\nocite{2001AJ....121..244S}).

The Jeans equations for triaxial St\"ackel systems have
received little attention. This is remarkable, as 
Eddington (1915\nocite{1915MNRAS..76...37E})
already knew that the velocity ellipsoid in these models is everywhere
aligned with the confocal ellipsoidal coordinate system in which the
motion separates. This means that there are only three, and not six,
non-vanishing second-order velocity moments in these coordinates, so
that the Jeans equations form a closed system. However, Eddington, and
later Chandrasekhar (1939\nocite{1939ApJ....90....1C},
1940\nocite{1940ApJ....92..441C}), did not study the velocity moments,
but instead assumed a form for the distribution function, and then
determined which potentials are consistent with it. Lynden--Bell
(1960\nocite{thesisLynden-Bell}) was the first to derive the explicit
form of the Jeans equations for the triaxial St\"ackel models. He
showed that they constitute a highly symmetric set of three
first-order partial differential equations (PDEs) for three unknowns,
each of which is a function of the three confocal ellipsoidal
coordinates, but he did not derive solutions. When it was realized
that the orbital structure in the triaxial St\"ackel models is very
similar to that in the early numerical models for triaxial galaxies
with cores (Schwarzschild 1979\nocite{1979ApJ...232..236S}; Z85),
interest in the second moments increased, and the Jeans equations were
solved for a number of special cases. These include the axisymmetric
limits and elliptic discs (Dejonghe \& de Zeeuw
1988\nocite{1988ApJ...333...90D}; Evans \& Lynden--Bell
1989\nocite{1989MNRAS.236..801E}, hereafter EL89), triaxial galaxies
with only thin tube orbits (HZ92), and, most recently, 
the scale-free limit 
(Evans et al.\ 2000\nocite{2000MNRAS.318.1131E}). 
In all these cases the equations
simplify to a two-dimensional problem, which can be solved with
standard techniques after recasting two first-order equations into a
single second-order equation in one dependent variable. However, these
techniques do not carry over to a single third-order equation in one dependent
variable, which is the best that one could expect to have in the general case.
As a result, the general case has remained unsolved.
 
In this paper, we first present an alternative solution method for the
two-dimensional limiting cases which does not use the standard
approach, but instead uses superpositions of singular solutions. We
show that this approach can be extended to three dimensions, and
provides the general solution for the triaxial case in 
closed form, which we give explicitly. We will apply our solutions in
a follow-up paper, and will use them together with the mean streaming
motions (Statler 1994a\nocite{1994ApJ...425..458S}) to study the
properties of the observed velocity and dispersion fields of triaxial
galaxies.

In \S\ref{sec:jeanseqns4separablemodels}, we define our notation and
derive the Jeans equations for the triaxial St\"ackel models in
confocal ellipsoidal coordinates, together with the continuity
conditions. We summarise the limiting cases, and show that the Jeans
equations for all the cases with two degrees of freedom correspond to
the same two-dimensional problem. We solve this problem in
\S\ref{sec:2Dcases}, first by employing a standard approach with a
Riemann--Green function, and then via the singular solution
superposition method. We also discuss the choice of boundary
conditions in detail. We relate our solution to that derived by EL89
in Appendix \ref{sec:solving4diff}, and explain why it is different. 
In \S\ref{sec:generalcase}, we
extend the singular solution approach to the three-dimensional
problem, and derive the general solution of the Jeans equations for
the triaxial case. 
It contains complete (hyper)elliptic integrals, which we express as
single quadratures that can be numerically evaluated in a 
straightforward way. 
We summarise our conclusions in \S\ref{sec:discconc}.

% ===== section 2 ===== %

\section{The Jeans equations for separable models}
\label{sec:jeanseqns4separablemodels}

We first summarise the essential properties of the triaxial St\"ackel
models in confocal ellipsoidal coordinates. Further details can be
found in Z85.  We show that for these models the mixed second-order 
velocity moments vanish, so that the Jeans equations form a closed
system. We derive the Jeans equations and find the corresponding
continuity conditions for the general case of a triaxial galaxy. We
then give an overview of the limiting cases and show that solving the
Jeans equations for the various cases with two degrees of freedom
reduces to an equivalent two-dimensional problem.

\subsection{Triaxial St\"ackel models}
\label{sec:stackelandconfocalellcoord}

We define confocal ellipsoidal coordinates ($\lambda,\mu,\nu$) as the
three roots for $\tau$ of 
\begin{equation}
  \label{eq:defellipsiodalcoord}
  \frac{x^2}{\tau+\alpha} + \frac{y^2}{\tau+\beta} +
  \frac{z^2}{\tau+\gamma} = 1,  
\end{equation}
with ($x,y,z$) the usual Cartesian coordinates, and with constants
$\alpha,\beta$ and $\gamma$ such that $-\gamma
\leq \nu\leq -\beta \leq \mu \leq -\alpha \leq \lambda$.  
For each point ($x,y,z$), there is a unique set ($\lambda,\mu,\nu$),
but a given combination ($\lambda,\mu,\nu$) generally corresponds
to eight different points ($\pm x, \pm y, \pm z$).
We assume all three-dimensional St\"ackel models in this paper to be
likewise eightfold symmetric. 

\begin{figure}
  \begin{center} 
  \includegraphics[draft=false,scale=0.8,trim=0cm 0.5cm
    0cm 0cm]{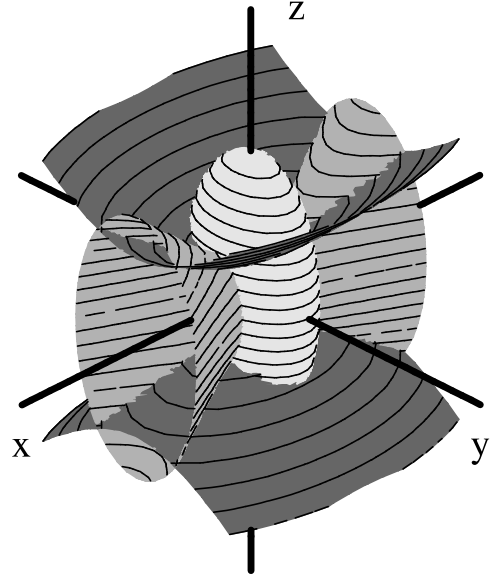} 
  \end{center}
    \caption[]{\slshape Confocal ellipsoidal coordinates.
    Surfaces of constant $\lambda$ are ellipsoids, surfaces of constant
    $\mu$ are hyperboloids of one sheet and surfaces of constant $\nu$
    are hyperboloids of two sheets.}
    \label{fig:ellipsoidalcoordinatesurfaces}
\end{figure}

Surfaces of constant $\lambda$ are ellipsoids, and surfaces of
constant $\mu$ and $\nu$ are hyperboloids of one and two sheets,
respectively (Fig.~\ref{fig:ellipsoidalcoordinatesurfaces}).  
The confocal ellipsoidal coordinates are approximately Cartesian near
the origin and become a conical coordinate system at large radii with
a system of spheres together with elliptic and hyperbolic 
cones (Fig.~\ref{fig:conicalcoordinatecurves}). 
At each point, the three coordinate surfaces are perpendicular to each
other.
Therefore, the line element is of the form
$ds^2=P^2d\lambda^2+Q^2d\mu^2+R^2d\nu^2$, with the metric coefficients
\begin{eqnarray}
  \label{eq:metriccoeffellipsoidal}
  P^2 \hspace{-7pt} & = & \hspace{-7pt} \frac{ (\lambda-\mu)
  (\lambda-\nu) }{ 4 (\lambda+\alpha) (\lambda+\beta) (\lambda+\gamma)
  }, \nonumber \\ 
  Q^2 \hspace{-7pt} & = & \hspace{-7pt} \frac{ (\mu-\nu) (\mu-\lambda)
  }{ 4 (\mu+\alpha) (\mu+\beta) (\mu+\gamma) }, \\  
  R^2 \hspace{-7pt} & = & \hspace{-7pt} \frac{ (\nu-\lambda) (\nu-\mu)
  }{ 4 (\nu+\alpha)(\nu+\beta) (\nu+\gamma) }. \nonumber  
\end{eqnarray}
We restrict attention to models with a gravitational potential
$V_S(\lambda,\mu,\nu)$ of St\"ackel form
(Weinacht 1924\nocite{histWeinacht1924})
\begin{equation}
  \label{eq:formstackelpotential}
  V_S = -\frac{F(\lambda)}{(\lambda\!-\!\mu)(\lambda\!-\!\nu)}
  -\frac{F(\mu)}{(\mu\!-\!\nu)(\mu\!-\!\lambda)}
  -\frac{F(\nu)}{(\nu\!-\!\lambda)(\nu\!-\!\mu)}, 
\end{equation}
where $F(\tau)$ is an arbitrary smooth function. 

Adding any linear function of $\tau$ to $F(\tau)$ changes $V_S$ by at
most a constant, and hence has no effect on the dynamics. Following
Z85, we use this freedom to write
\begin{equation}
\label{eq:defgtau}
F(\tau)=(\tau+\alpha)(\tau+\gamma)G(\tau), 
\end{equation}
where $G(\tau)$ is smooth. It equals the potential along the
intermediate axis. This choice will simplify the analysis of the large
radii behaviour of the various limiting cases.\footnote{Other,
equivalent, choices include
$F(\tau)=-(\tau+\alpha)(\tau+\gamma)G(\tau)$ by HZ92, and
$F(\tau)=(\tau+\alpha)(\tau+\beta)U(\tau)$ by 
de Zeeuw et al.\ (1986\nocite{1986MNRAS.221.1001D}), 
with $U(\tau)$ the potential along the short axis.}

The density $\rho_S$ that corresponds to $V_S$ can be found from
Poisson's equation or by application of Kuzmin's
(1973\nocite{1973Kuzmin}) formula (see de Zeeuw
1985b\nocite{1985MNRAS.216..599D}). This formula shows that, once we
have chosen the central axis ratios and the density along the short
axis, the mass model is fixed everywhere by the requirement of
separability. For centrally concentrated mass models, $V_S$ has the
$x$-axis as long axis and the $z$-axis as short axis. In most cases
this is also true for the associated density 
(de Zeeuw et al.\ 1986\nocite{1986MNRAS.221.1001D}).

%%%FIG
%%%FIG

\subsection{Velocity moments}
\label{sec:velocitymoments}

A stellar system is completely described by its distribution function
(DF), which in general is a time-dependent function $f$ of the six
phase-space coordinates ($\mathbf{x},\mathbf{v}$). Assuming the system
to be in equilibrium ($df/dt=0$) and in steady-state ($\partial
f/\partial t =0$), the DF is independent of time $t$ and satisfies the
(stationary) collisionless Boltzmann equation (CBE). Integration of
the DF over all velocities yields the zeroth-order velocity moment,
which is the density $\rho$ of the stellar system. The first- and
second-order velocity moments are defined as
\begin{eqnarray}
  \label{eq:defvelmoments}
  \langle v_i \rangle (\mathbf{x}) \!\!\!&=&\!\!\!  
  \frac{1}{\rho} \iiint v_i f(\mathbf{x},\mathbf{v}) \; \d^3v, 
                                            \nonumber \\*[-5pt]\\*[-5pt]
  \langle v_i v_j \rangle (\mathbf{x})  \!\!\!&=&\!\!\! 
  \frac{1}{\rho} \iiint v_i v_j f(\mathbf{x},\mathbf{v}) \; \d^3v, \nonumber 
\end{eqnarray}
where $i, j=1, 2, 3$. The streaming motions $\langle v_i \rangle$
together with the symmetric second-order velocity moments $\langle
v_iv_j\rangle$ provide the velocity dispersions $\sigma_{ij}^2 =
\langle v_i v_j \rangle - \langle v_i \rangle \langle v_j \rangle$.

The continuity equation that results from integrating the CBE over
all velocities, relates the streaming motion to the density $\rho$ of
the system. Integrating the CBE over all velocities 
after multiplication by each of the three velocity components,
provides the Jeans equations, which relate the second-order velocity
moments to $\rho$ and $V$, the potential of the system.  
Therefore, if the density and potential are known, we in general have one
continuity equation with three unknown first-order 
velocity moments and three Jeans equations with six unknown
second-order velocity moments.

The potential (\ref{eq:formstackelpotential}) is the most general form
for which the Hamilton--Jacobi equation separates (St\"ackel
1890\nocite{histStackel1890}; Lynden--Bell
1962b\nocite{1962MNRAS.124...95L}; Goldstein
1980\nocite{1980cmbookGoldstein}). All orbits have three exact isolating
integrals
of motion, which are quadratic in the velocities (e.g., Z85). It
follows that there are no irregular orbits, so that Jeans'
(1915\nocite{1915MNRAS..76...70J}) theorem is strictly valid
(Lynden--Bell 1962a\nocite{1962MNRAS.124....1L}; Binney
1982\nocite{1982MNRAS.201...15B}) and the DF is a function of the three
integrals. The orbital motion is a combination of three independent
one-dimensional motions --- either an oscillation or a rotation --- in
each of the three ellipsoidal coordinates.  Different combinations of
rotations and oscillations result in four families of orbits in
triaxial St\"ackel models (Kuzmin 1973\nocite{1973Kuzmin}; Z85): inner
(I) and outer (O) long-axis tubes, short (S) axis tubes and box
orbits.  Stars on box orbits carry out an oscillation in all three
coordinates, so that they provide no net contribution to the mean
streaming. Stars on I- and O-tubes carry out a rotation in $\nu$ and
those on S-tubes a rotation in $\mu$, and oscillations in the other
two coordinates. The fractions of clockwise and counterclockwise stars
on these orbits may be unequal. This means that each of the tube
families can have at most one nonzero first-order velocity moment,
related to $\rho$ by the continuity equation. 
Statler (1994a)\nocite{1994ApJ...425..458S} 
used this property to construct velocity fields for triaxial St\"ackel
models.  
It is not difficult to show by similar arguments (e.g., HZ92) that all
mixed second-order velocity moments also vanish
\begin{equation}
  \label{eq:mixsecondordervelocitymomentsvanish}
  \langle v_\lambda v_\mu \rangle = \langle v_\mu v_\nu \rangle =
  \langle v_\nu v_\lambda \rangle = 0. 
\end{equation}
Eddington (1915\nocite{1915MNRAS..76...37E}) already knew that in a
potential of the form \eqref{eq:formstackelpotential}, the axes of the
velocity ellipsoid at any given point are perpendicular to the
coordinate surfaces, so that the mixed second-order velocity moments are
zero. We are left with three second-order velocity moments, $\langle
v_\lambda^2 \rangle$, $\langle v_\mu^2 \rangle$ and $\langle v_\nu^2
\rangle$, related by three Jeans equations.

\subsection{The Jeans equations}
\label{sec:jeanseqns}

The Jeans equations for triaxial St\"ackel models in confocal
ellipsoidal coordinates were first derived by Lynden--Bell
(1960\nocite{thesisLynden-Bell}). 
We give an alternative derivation here, using the Hamilton equations. 

We first write the DF as a function of ($\lambda,\mu,\nu$) and the
conjugate momenta
\begin{equation}
  \label{eq:conjugatemomenta}
  p_\lambda = P^2\frac{d\lambda}{dt}, \quad
  p_\mu = Q^2\frac{d\mu}{dt}, \quad
  p_\nu = R^2\frac{d\nu}{dt},
\end{equation}
with the metric coefficients $P$, $Q$ and $R$ given in
\eqref{eq:metriccoeffellipsoidal}. 
In these phase-space coordinates the steady-state CBE reads 
\begin{equation}
  \label{eq:continuityeqnellipsoidmomenta}
  \frac{d\tau}{dt}\frac{\partial f}{\partial \tau} + 
  \frac{dp_\tau}{dt}\frac{\partial f}{\partial p_\tau} = 0, 
\end{equation}
where we have used the summation convention with respect to
$\tau=\lambda,\mu,\nu$. The Hamilton equations are
\begin{equation}
  \label{eq:hamiltoneqns}
  \frac{d\tau}{dt} = \frac{\partial H}{\partial p_\tau},  \quad
  \frac{dp_\tau}{dt} = \frac{\partial H}{\partial \tau},  
\end{equation}
with the Hamiltonian defined as 
\begin{equation}
  \label{eq:definitionhamiltonian}
  H = \frac{p_\lambda^2}{2P^2} + \frac{p_\mu^2}{2Q^2} +
  \frac{p_\nu^2}{2R^2} + V(\lambda,\mu,\nu). 
\end{equation}
The first Hamilton equation in \eqref{eq:hamiltoneqns} defines the
momenta \eqref{eq:conjugatemomenta} and gives no new information. The
second gives
\begin{equation}
  \label{eq:resulthamiltonseqnssecondtype}
  \frac{dp_\lambda}{dt} = 
  \frac{p_\lambda^2}{P^3}\frac{\partial P}{\partial \lambda} +
  \frac{p_\mu^2}{Q^3}\frac{\partial Q}{\partial \lambda} +
  \frac{p_\nu^2}{R^3}\frac{\partial R}{\partial \lambda} -
  \frac{\partial V}{\partial \lambda},
\end{equation}
and similar for $p_{\mu}$ and $p_{\nu}$ by replacing the derivatives
with respect to $\lambda$ by derivatives to $\mu$ and $\nu$,
respectively. 

We assume the potential to be of the form $V_S$ defined in
\eqref{eq:formstackelpotential}, and we substitute
\eqref{eq:conjugatemomenta} and
\eqref{eq:resulthamiltonseqnssecondtype} in the CBE
\eqref{eq:continuityeqnellipsoidmomenta}. 
We multiply this equation by $p_\lambda$ and integrate over all
momenta. The mixed second moments vanish 
\eqref{eq:mixsecondordervelocitymomentsvanish}, so that we are left
with
\begin{multline}
  \label{eq:resultintegralcalcjeans}
  \frac{3\langle fp_\lambda^2 \rangle}{P^3} 
  \frac{\partial P}{\partial \lambda} 
  + \frac{\langle fp_\mu^2 \rangle}{Q^3}
  \frac{\partial Q}{\partial \lambda} 
  + \frac{\langle fp_\nu^2 \rangle}{R^3}
  \frac{\partial R}{\partial \lambda} \\
  - \frac{1}{P^2} \frac{\partial  }{\partial \lambda} 
  \langle fp_\lambda^2 \rangle              
  - \langle f \rangle \frac{\partial V_S}{\partial \lambda} = 0,  
\end{multline}
where we have defined the moments 
\begin{eqnarray}
  \label{eq:momentamoments}
  \langle f \rangle \hspace{-7pt} &\equiv& \hspace{-7pt} \int f \d^3 p
  = PQR \, \rho, \nonumber 
  \\*[-5pt] \\*[-5pt]
  \langle fp_\lambda^2 \rangle \hspace{-7pt} &\equiv& \hspace{-7pt}
  \int p_\lambda^2 f \d^3p = P^3QR \, T_{\lambda\lambda}, \nonumber 
\end{eqnarray}
with the diagonal components of the stress tensor
\begin{equation}
  \label{eq:definitionTtautau} 
  T_{\tau\tau} (\lambda,\mu,\nu) \equiv
  \rho\langle v_\tau^2 \rangle, \qquad \tau=\lambda,\mu,\nu.
\end{equation}
The moments $\langle fp_\mu^2 \rangle$ and $\langle fp_\nu^2 \rangle$
follow from $\langle fp_\lambda^2 \rangle$ by cyclic permutation
$\lambda \to \mu \to \nu \to \lambda$, for which
$P\!\to\!Q\!\to\!R\!\to\!P$. We substitute the definitions
\eqref{eq:momentamoments} in eq.~\eqref{eq:resultintegralcalcjeans}
and carry out the partial differentiation in the fourth term. The
first term in \eqref{eq:resultintegralcalcjeans} then cancels, and,
after rearranging the remaining terms and dividing by $PQR$, we obtain
\begin{equation}
  \label{eq:rearrangedjeans}
  \frac{\partial T_{\lambda\lambda}}{\partial \lambda} +
  \frac{T_{\lambda\lambda}\!-\!T_{\mu\mu}}{Q}\frac{\partial
  Q}{\partial \lambda} +
  \frac{T_{\lambda\lambda}\!-\!T_{\nu\nu}}{R}\frac{\partial
  R}{\partial \lambda}= -\rho\frac{\partial V_S}{\partial \lambda}. 
\end{equation}
Substituting the metric coefficients \eqref{eq:metriccoeffellipsoidal}
and carrying out the partial differentiations results in the
Jeans equations  
\begin{subequations}
  \label{eq:jeanstriaxial}
  \begin{equation}
    \label{eq:jeanstriaxial_lambda}
    \frac{\partial T_{\lambda\lambda}}{\partial \lambda} +
    \frac{T_{\lambda\lambda}-T_{\mu\mu}}{2(\lambda-\mu)} +
    \frac{T_{\lambda\lambda}-T_{\nu\nu}}{2(\lambda-\nu)} =
    -\rho\frac{\partial V_S}{\partial \lambda}, 
  \end{equation}
  \begin{equation}
    \label{eq:jeanstriaxial_mu}
    \frac{\partial T_{\mu\mu}}{\partial \mu} +
    \frac{T_{\mu\mu}-T_{\nu\nu}}{2(\mu-\nu)} +
    \frac{T_{\mu\mu}-T_{\lambda\lambda}}{2(\mu-\lambda)}  =
    -\rho\frac{\partial V_S}{\partial \mu}, 
  \end{equation}
  \begin{equation}
    \label{eq:jeanstriaxial_nu}
    \frac{\partial T_{\nu\nu}}{\partial \nu} +
    \frac{T_{\nu\nu}-T_{\lambda\lambda}}{2(\nu-\lambda)} +
    \frac{T_{\nu\nu}-T_{\mu\mu}}{2(\nu-\mu)}  =  
    -\rho\frac{\partial V_S}{\partial \nu}, 
  \end{equation}
\end{subequations}
where the equations for $\mu$ and $\nu$ follow from the one for
$\lambda$ by cyclic permutation. These equations are identical to
those derived by Lynden--Bell (1960\nocite{thesisLynden-Bell}).

In self-consistent models, the density $\rho$ must equal $\rho_S$,
with $\rho_S$ related to the potential $V_S$
\eqref{eq:formstackelpotential} by Poisson's equation.  The Jeans
equations, however, do not require self-consistency. Hence, we make no
assumptions on the form of the density other than that it is
triaxial, i.e., a function of $(\lambda,\mu,\nu)$, and that it tends
to zero at infinity. The resulting solutions for the stresses
$T_{\tau\tau}$ do not all correspond to physical distribution
functions $f\ge 0$. The requirement that the $T_{\tau\tau}$ are
non-negative removes many (but not all) of the unphysical solutions.

\subsection{Continuity conditions}
\label{sec:continuityconditions}

%%%FIG
\begin{figure}
  \begin{center}
    \includegraphics[draft=false,scale=0.55,trim=3cm 4.5cm 3cm
    4.0cm]{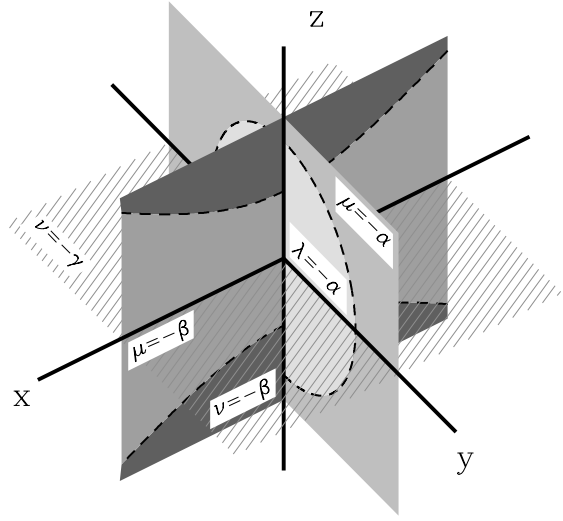}  
  \end{center}
  \caption[]{\slshape Special surfaces inside ($\lambda=-\alpha$) and
    outside ($\mu=-\alpha$) the focal ellipse in the plane $x=0$, and
    inside ($\mu=-\beta$) and outside ($\nu=-\beta$) the two branches
    of the focal hyperbola in the plane $y=0$ and the plane $z=0$
    ($\nu=-\gamma$).} 
   \label{fig:focalplanes}
 \end{figure}
%%%FIG

We saw in \S\ref{sec:velocitymoments} that the velocity ellipsoid is
everywhere aligned with the confocal ellipsoidal coordinates. When
$\lambda\to-\alpha$, the ellipsoidal coordinate surface degenerates
into the area inside the focal ellipse
(Fig.~\ref{fig:focalplanes}). The area outside the focal ellipse is
labeled by $\mu=-\alpha$. Hence, $T_{\lambda\lambda}$ is perpendicular
to the surface \textit{inside} and $T_{\mu\mu}$ is perpendicular to
the surface \textit{outside} the focal ellipse.  On the focal ellipse,
i.e. when $\lambda=\mu=-\alpha$, both stress components therefore have
to be equal. Similarly, $T_{\mu\mu}$ and $T_{\nu\nu}$ are
perpendicular to the area inside ($\mu=-\beta$) and outside
($\nu=-\beta$) the two branches of the focal hyperbola, respectively,
and have to be equal on the focal hyperbola itself
($\mu=\nu=-\beta$). This results in the following two continuity
conditions
\begin{subequations}
  \label{eq:triaxialcontcond}
  \begin{equation}
    \label{eq:triaxialcontcond1}
    T_{\lambda\lambda}(-\alpha,-\alpha,\nu) = T_{\mu\mu}(-\alpha,-\alpha,\nu),
  \end{equation}
  \begin{equation}
    \label{eq:triaxialcontcond2}
    T_{\mu\mu}(\lambda,-\beta,-\beta) = T_{\nu\nu}(\lambda,-\beta,-\beta).
  \end{equation}
\end{subequations}
These conditions not only follow from geometrical arguments, but are
also precisely the conditions necessary to avoid singularities in the
Jeans equations \eqref{eq:jeanstriaxial} when $\lambda=\mu=-\alpha$
and $\mu=\nu=-\beta$.
For the sake of physical understanding, we will also obtain the
corresponding continuity conditions by geometrical arguments for the
limiting cases that follow.

\subsection{Limiting cases}
\label{sec:limitingcases}

When two or all three of the constants $\alpha$, $\beta$ or $\gamma$
are equal, the triaxial St\"ackel models reduce to limiting cases with
more symmetry and thus with fewer degrees of freedom. 
We show in \S\ref{sec:2Dcasessimilar} that solving the Jeans equations
for all the models with two degrees of freedom reduces to the same
two-dimensional problem.  
EL89 first solved this generalised problem and applied it to the disc,
oblate and prolate case. 
Evans et al.\ (2000\nocite{2000MNRAS.318.1131E}) showed
that the large radii case with scale-free DF reduces to the problem
solved by EL89.  
We solve the same problem in a different way in \S\ref{sec:2Dcases},
and obtain a simpler expression than EL89.
In order to make application of the resulting solution
straightforward, and to define a unified notation, we first give an
overview of the limiting cases. \looseness=-2

\subsubsection{Oblate spheroidal coordinates: prolate potentials}
\label{sec:oblatespheroidalcoord}

When $\gamma=\beta$, the coordinate surfaces for constant $\lambda$ and
$\mu$ reduce to oblate spheroids and hyperboloids of revolution around
the $x$-axis. Since the range of $\nu$ is zero, it cannot be used as a
coordinate. The hyperboloids of two sheets are now planes containing
the $x$-axis. We label these planes by an azimuthal angle $\chi$,
defined as $\tan \chi = z/y$. In these oblate spheroidal coordinates
($\lambda, \mu, \chi$) the potential $V_S$ has the form (cf.\
Lynden--Bell 1962b\nocite{1962MNRAS.124...95L})
\begin{equation}
\label{eq:V_Sobspheroidcoords}
  V_S = -\frac{f(\lambda)-f(\mu)}{\lambda\!-\!\mu}
        -\frac{g(\chi)}{(\lambda+\beta)(\mu+\beta)}, 
\end{equation}
where the function $g(\chi)$ is arbitrary, and
$f(\tau)=(\tau+\alpha)G(\tau)$, with $G(\tau)$ as in
eq.~(\ref{eq:defgtau}). The denominator of the second term is
proportional to $y^2+z^2$, so that these potentials are singular along
the entire $x$-axis unless $g(\chi)\equiv0$. In this case, the
potential is prolate axisymmetric, and the associated density $\rho_S$
is generally prolate as well 
(de Zeeuw et al.\  1986\nocite{1986MNRAS.221.1001D}).

The Jeans equations (\ref{eq:jeanstriaxial}) reduce to
\begin{eqnarray}
  \label{eq:jeansprolate}
  \frac{\partial T_{\lambda\lambda}}{\partial \lambda} +
  \frac{T_{\lambda\lambda}-T_{\mu\mu}}{2(\lambda-\mu)} +
  \frac{T_{\lambda\lambda}-T_{\chi\chi}}{2(\lambda+\beta)}
  \hspace{-7pt} & = & \hspace{-7pt}
  -\rho\frac{\partial V_S}{\partial \lambda}, \nonumber \\
  \frac{\partial T_{\mu\mu}}{\partial \mu} +
  \frac{T_{\mu\mu}-T_{\lambda\lambda}}{2(\mu-\lambda)} +
  \frac{T_{\mu\mu}-T_{\chi\chi}}{2(\mu+\beta)} \hspace{-7pt} & = &
  \hspace{-7pt} -\rho\frac{\partial V_S}{\partial \mu}, \\
  \frac{\partial T_{\chi\chi}}{\partial \chi} 
  \hspace{-7pt} & = & \hspace{-7pt}
  - \rho \frac{\partial V_S}{\partial \chi}. \nonumber 
\end{eqnarray}
The continuity condition \eqref{eq:triaxialcontcond1} still holds,
except that the focal ellipse has become a focal circle.  For
$\mu=-\beta$, the one-sheeted hyperboloid degenerates into the
$x$-axis, so that $T_{\mu\mu}$ is perpendicular to the $x$-axis and
coincides with $T_{\chi\chi}$. This gives the following two continuity
conditions
\begin{eqnarray}
  \label{eq:prolatecontcond}
  T_{\lambda\lambda}(-\alpha,-\alpha,\chi)
  \hspace{-7pt} & = & \hspace{-7pt}
  T_{\mu\mu}(-\alpha,-\alpha,\chi), \nonumber \\*[-5pt]  \\*[-5pt]
  T_{\mu\mu}(\lambda,-\beta,\chi)
  \hspace{-7pt} & = & \hspace{-7pt}
  T_{\chi\chi}(\lambda,-\beta,\chi). \nonumber
\end{eqnarray}
By integrating along characteristics, Hunter et al.\
(1990\nocite{1990ApJ...363..367H}) obtained the solution of
(\ref{eq:jeansprolate}) for the special prolate models in which only
the thin I- and O-tube orbits are populated, so that $T_{\mu\mu}
\equiv 0$ and $T_{\lambda\lambda} \equiv 0$, respectively (cf.\
\S\ref{sec:thintubeorbits}).

\subsubsection{Prolate spheroidal coordinates: oblate potentials}
\label{sec:prolatespheroidalcoord}

When $\beta=\alpha$, we cannot use $\mu$ as a coordinate and replace it
by the azimuthal angle $\phi$, defined as $\tan\phi=y/x$.  Surfaces of
constant $\lambda$ and $\nu$ are confocal prolate spheroids and
two-sheeted hyperboloids of revolution around the $z$-axis. The
prolate spheroidal coordinates ($\lambda,\phi,\nu$) follow from the
oblate spheroidal coordinates ($\lambda,\mu,\chi$) by taking
$\mu\!\to\!\nu$, $\chi\!\to\!\phi$ and
$\beta\!\to\!\alpha\!\to\!\gamma$. 
The potential $V_S(\lambda,\phi,\nu)$ is (cf.\ Lynden--Bell
1962b\nocite{1962MNRAS.124...95L})
\begin{equation}
\label{eq:V_Sprospheroidcoords}
  V_S = -\frac{f(\lambda)-f(\nu)}{\lambda\!-\!\nu}
        -\frac{g(\phi)}{(\lambda+\alpha)(\nu+\alpha)}. 
\end{equation}
In this case, the denominator of the second term is proportional to
$R^2=x^2+y^2$, so that the potential is singular along the entire
$z$-axis, unless $g(\phi)$ vanishes. When $g(\phi)\equiv0$, the
potential is oblate, and the same is generally true for the associated
density $\rho_S$.

The Jeans equations (\ref{eq:jeanstriaxial}) reduce to
\begin{eqnarray}
  \label{eq:jeansoblate}
  \frac{\partial T_{\lambda\lambda}}{\partial \lambda} +
  \frac{T_{\lambda\lambda}-T_{\phi\phi}}{2(\lambda+\alpha)} +
  \frac{T_{\lambda\lambda}-T_{\nu\nu}}{2(\lambda-\nu)} 
  \hspace{-7pt} & = & \hspace{-7pt}
  -\rho\frac{\partial V_S}{\partial \lambda}, \nonumber \\
  \frac{\partial T_{\phi\phi}}{\partial \phi} 
  \hspace{-7pt} & = & \hspace{-7pt}
  -\rho \frac{\partial V_S}{\partial \phi}. \\
  \frac{\partial T_{\nu\nu}}{\partial \nu} +
  \frac{T_{\nu\nu}-T_{\lambda\lambda}}{2(\nu-\lambda)} +
  \frac{T_{\nu\nu}-T_{\phi\phi}}{2(\nu+\alpha)}
  \hspace{-7pt} & = & \hspace{-7pt}
  -\rho\frac{\partial V_S}{\partial \nu}. \nonumber
\end{eqnarray}
For $\lambda=-\alpha$, the prolate spheroidal coordinate surfaces
reduce to the part of the $z$-axis between the foci. The part beyond
the foci is reached if $\nu=-\alpha$. 
Hence, in this case, $T_{\lambda\lambda}$ is perpendicular to part of
the $z$-axis between, and $T_{\nu\nu}$ is perpendicular to the part of
the $z$-axis beyond the foci. They coincide at the foci
($\lambda=\nu=-\alpha$), resulting in one continuity condition. Two
more follow from the fact that $T_{\phi\phi}$ is perpendicular to the
(complete) $z$-axis, and thus coincides with 
$T_{\lambda\lambda}$ and $T_{\nu\nu}$ on the part between and beyond
the foci, respectively: 
\begin{eqnarray} 
  \label{eq:oblatecontcond}
  T_{\lambda\lambda}(-\alpha,\phi,-\alpha) \hspace{-7pt} & = & 
  \hspace{-7pt} T_{\nu\nu}(-\alpha,\phi,-\alpha), \nonumber \\
  T_{\lambda\lambda}(-\alpha,\phi,\nu) \hspace{-7pt} & = &
  \hspace{-7pt} T_{\phi\phi}(-\alpha,\phi,\nu), \\  
  T_{\nu\nu}(\lambda,\phi,-\alpha) \hspace{-7pt} & = &
  \hspace{-7pt} T_{\phi\phi}(\lambda,\phi,-\alpha).\nonumber 
\end{eqnarray}
For oblate models with thin S-tube orbits ($T_{\lambda\lambda} \equiv
0$, see \S\ref{sec:thintubeorbits}), the analytical solution of
(\ref{eq:jeansoblate}) was derived by Bishop
(1987\nocite{1987ApJ...322..618B}) and by de Zeeuw \& Hunter
(1990\nocite{1990ApJ...356..365D}). Robijn \& de Zeeuw
(1996\nocite{1996MNRAS.279..673R}) obtained the second-order velocity
moments for models in which the thin tube orbits were thickened
iteratively. Dejonghe \& de Zeeuw (1988\nocite{1988ApJ...333...90D},
Appendix D) found a general solution by integrating along
characteristics. Evans (1990\nocite{evans1990numericalmethod}) gave an
algorithm for solving (\ref{eq:jeansoblate}) numerically, and
Arnold (1995\nocite{1995MNRAS.276..293A}) computed a solution using
characteristics without assuming a separable potential.

\subsubsection{Confocal elliptic coordinates: non-circular discs}
\label{sec:confocalellipticcoord}

In the principal plane $z=0$, the ellipsoidal coordinates reduce to
confocal elliptic coordinates ($\lambda,\mu$), with coordinate curves
that are ellipses ($\lambda$) and hyperbolae ($\mu$), that share their
foci on the symmetry $y$-axis. The potential of the perfect elliptic
disc, with its surface density distribution stratified on concentric
ellipses in the plane $z=0$ ($\nu=-\gamma$), is of St\"ackel form both
in and outside this plane. By a superposition of perfect elliptic
discs, one can construct other surface densities and corresponding
disc potentials that are of St\"ackel form in the plane $z=0$, but not
necessarily outside it (Evans \& de Zeeuw
1992\nocite{1992MNRAS.257..152E}).  The expression for the potential in
the disc is of the form (\ref{eq:V_Sobspheroidcoords}) with
$g(\chi)\equiv0$:
\begin{equation}
\label{eq:discpot}
  V_S = -\frac{f(\lambda)-f(\mu)}{\lambda\!-\!\mu}, 
\end{equation}
where again $f(\tau)=(\tau+\alpha)G(\tau)$, so that $G(\tau)$ equals
the potential along the $y$-axis.

Omitting all terms with $\nu$ in \eqref{eq:jeanstriaxial}, we
obtain the Jeans equations for non-circular St\"ackel discs
\begin{eqnarray}
\label{eq:jeanselliptic}
  \frac{\partial T_{\lambda\lambda}}{\partial \lambda} +
  \frac{T_{\lambda\lambda}-T_{\mu\mu}}{2(\lambda-\mu)}
  \hspace{-7pt} & = & \hspace{-7pt}
  -\rho\frac{\partial V_S}{\partial \lambda}, \nonumber \\*[-5pt]  \\*[-5pt] 
  \frac{\partial T_{\mu\mu}}{\partial \mu} +
  \frac{T_{\mu\mu}-T_{\lambda\lambda}}{2(\mu-\lambda)}
  \hspace{-7pt} & = & \hspace{-7pt}
  -\rho\frac{\partial V_S}{\partial \mu}, \nonumber
\end{eqnarray}
where now $\rho$ denotes a surface density. The parts of the $y$-axis
between and beyond the foci are labeled by $\lambda=-\alpha$ 
and $\mu=-\alpha$, resulting in the continuity condition
\begin{equation}
  \label{eq:continuityconditionelliptic}
  T_{\lambda\lambda}(-\alpha,-\alpha) = T_{\mu\mu}(-\alpha,-\alpha).
\end{equation}

%%%FIG
\begin{figure}
  \begin{center}
    \includegraphics[draft=false,scale=0.8,trim=0cm 0.5cm 0cm
    0cm]{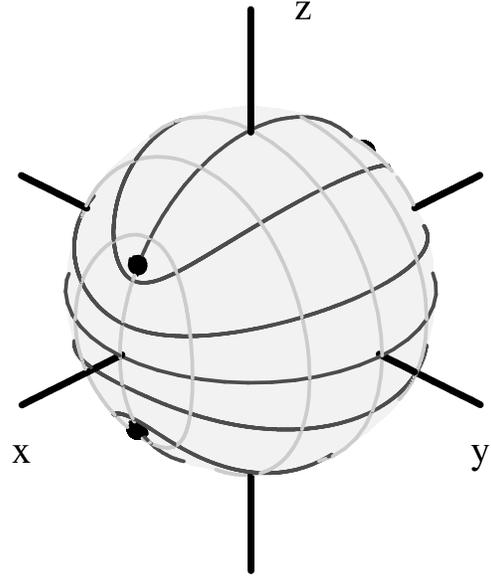}  
  \end{center}
  \caption[]{\slshape Behaviour of the confocal ellipsoidal
    coordinates in the limit of large radii $r$. 
    The surfaces of constant $\lambda$ become spheres. 
    The hyperboloids of constant $\mu$ and $\nu$ approach their
    asymptotic surfaces, and intersect the sphere on the light and
    dark curves, respectively. These form an orthogonal curvilinear
    coordinate system $(\mu, \nu)$ on the sphere.
    The black dots indicate the transition points ($\mu=\nu=-\beta$)
    between both sets of curves.}
  \label{fig:conicalcoordinatecurves}
\end{figure}
%%%FIG

\subsubsection{Conical coordinates: scale-free triaxial limit}
\label{sec:conicalcoord}

At large radii, the confocal ellipsoidal coordinates
($\lambda,\mu,\nu$) reduce to conical coordinates ($r,\mu,\nu$), with
$r$ the usual distance to the origin, i.e., $r^2=x^2+y^2+z^2$ and
$\mu$ and $\nu$ angular coordinates on the sphere
(Fig.~\ref{fig:conicalcoordinatecurves}). The potential
$V_S(r,\mu,\nu)$ is scale-free, and of the form
\begin{equation}
\label{eq:V_Sconicalcoords}
  V_S = -{\tilde F}(r) + \frac{F(\mu)-F(\nu)}{r^2(\mu\!-\!\nu)},
\end{equation}
where ${\tilde F}(r)$ is arbitrary, and $F(\tau)=
(\tau+\alpha)(\tau+\gamma)G(\tau)$, as in eq.~(\ref{eq:defgtau}).

The Jeans equations in conical coordinates follow from the general
triaxial case \eqref{eq:jeanstriaxial} by going to large 
radii. Taking $\lambda \to r^2 \gg -\alpha \geq \mu,\nu$, the stress
components approach each other and we have
\begin{equation}
  \label{eq:ratiosinlargedistancelimit}
   \frac{T_{\lambda\lambda}-T_{\mu\mu}}{2(\lambda-\mu)}, \; 
   \frac{T_{\lambda\lambda}-T_{\nu\nu}}{2(\lambda-\nu)} 
   \sim \frac{1}{r} \to 0, \quad
   \frac{\partial}{\partial \lambda} \to \frac{1}{2r}
   \frac{\partial}{\partial \lambda}.
\end{equation}
Hence, after multiplying \eqref{eq:jeanstriaxial_lambda} by $2r$, the
Jeans equations for scale-free St\"ackel models are
\begin{eqnarray}
  \label{eq:jeansconical}
  \frac{\partial T_{rr}}{\partial r} +
  \frac{2T_{rr}-T_{\mu\mu}-T_{\nu\nu}}{r} \hspace{-7pt} & = &
  \hspace{-7pt} -\rho\frac{\partial V_S}{\partial r}, \nonumber \\ 
  \frac{\partial T_{\mu\mu}}{\partial \mu} +
  \frac{T_{\mu\mu}-T_{\nu\nu}}{2(\mu-\nu)} \hspace{-7pt} & = &
  \hspace{-7pt} -\rho\frac{\partial V_S}{\partial \mu}, \\
  \frac{\partial T_{\nu\nu}}{\partial \nu} +
  \frac{T_{\nu\nu}-T_{\mu\mu}}{2(\nu-\mu)} \hspace{-7pt} & = &
  \hspace{-7pt} -\rho\frac{\partial V_S}{\partial \nu}. \nonumber  
\end{eqnarray}
The general Jeans equations in conical coordinates, as derived by
Evans et al.\ (2000\nocite{2000MNRAS.318.1131E}), 
reduce to \eqref{eq:jeansconical} for vanishing mixed second
moments. At the transition points between the curves of constant $\mu$
and $\nu$ ($\mu=\nu=-\beta$), the tensor components $T_{\mu\mu}$ 
and $T_{\nu\nu}$ coincide, resulting in the continuity condition
\begin{equation}
  \label{eq:conicalcontcond}
  T_{\lambda\lambda}(r,-\beta,-\beta) = T_{\phi\phi}(r,-\beta,-\beta).
\end{equation}

\subsubsection{One-dimensional limits}
\label{sec:onedimensionallimits}

There are several additional limiting cases with more symmetry for
which the form of $V_S$ (Lynden--Bell 1962b\nocite{1962MNRAS.124...95L})
and the associated Jeans equations follow in a straightforward way
from the expressions that were given above. 
We only mention spheres and circular discs.

When $\alpha\!=\!\beta\!=\!\gamma$, the variables $\mu$ and $\nu$ loose their
meaning and the ellipsoidal coordinates reduce to spherical
coordinates $(r, \theta, \phi)$. A steady-state spherical model
without a preferred axis is invariant under a rotation over the angles
$\theta$ and $\phi$, so that we are left with only one Jeans equation
in $r$, and $T_{\theta\theta}=T_{\phi\phi}$.  This equation can
readily be obtained from the CBE in spherical coordinates (e.g.,
Binney \& Tremaine 1987\nocite{1987gady.book.....B}). 
It also follows as a limit from the Jeans equations 
\eqref{eq:jeanstriaxial} for triaxial St\"ackel models or from any 
of the above two-dimensional limiting cases. Consider for example 
the Jeans equations in conical coordinates \eqref{eq:jeansconical}, and
take $\mu\to\theta$ and $\nu\to\phi$. The stress components $T_{rr}$
and $T_{\mu\mu}=T_{\nu\nu}=T_{\phi\phi}=T_{\theta\theta}$ depend only
$r$, so that we are left with \looseness=-1
\begin{equation}
  \label{eq:jeansspherical}
  \frac{dT_{rr}}{dr} + \frac{2(T_{rr}-T_{\theta\theta})}{r} 
  = -\rho\frac{dV_S}{dr}, 
\end{equation}
which is the well-known result for non-rotating spherical systems
(Binney \& Tremaine 1987\nocite{1987gady.book.....B}). 

In a similar way, the one Jeans equation for the circular disc-case
follows from, e.g., the first equation of (\ref{eq:jeanselliptic}) by
taking $\mu=-\alpha$ and replacing $T_{\mu\mu}$ by $T_{\phi\phi}$,
where $\phi$ is the azimuthal angle defined in
\S\ref{sec:prolatespheroidalcoord}.  
With $\lambda+\alpha=R^2$ this gives \looseness=-1
\begin{equation}
  \label{eq:jeanscirculardiscs}
  \frac{dT_{RR}}{dR} + \frac{T_{RR}-T_{\phi\phi}}{R} = -\rho\frac{dV_S}{dR},
\end{equation}
which may be compared with Binney \& Tremaine
(1987)\nocite{1987gady.book.....B}, their eq.~(4.29).

\subsubsection{Thin tube orbits}
\label{sec:thintubeorbits}

Each of the three tube orbit families in a triaxial St\"ackel model 
consists of a rotation in one of the ellipsoidal coordinates
and oscillations in the other two (\S\ref{sec:velocitymoments}). 
The I-tubes, for example, rotate in $\nu$ and oscillate in $\lambda$
and $\mu$, with turning points $\mu_1$, $\mu_2$ and $\lambda_0$, so
that a typical orbit fills the volume \looseness=-2
\begin{equation}
  \label{eq:volumeItube}
  -\gamma \le \nu \le -\beta, \quad
  \mu_1 \le \mu \le \mu_2, \quad
  -\alpha \le \lambda \le \lambda_0.
\end{equation}
When we restrict ourselves to infinitesimally thin I-tubes, i.e.,
$\mu_1=\mu_2$, there is no motion in the $\mu$-coordinate.
Therefore, the second-order velocity moment in this coordinate is
zero, and thus also the corresponding stress component
$T_{\mu\mu}^\mathrm{I} \equiv 0$. 
As a result, eq.~\eqref{eq:jeanstriaxial_mu} reduces to an
algebraic relation between $T_{\lambda\lambda}^\mathrm{I}$ and
$T_{\nu\nu}^\mathrm{I}$. 
This relation can be used to eliminate $T_{\nu\nu}^\mathrm{I}$ and
$T_{\lambda\lambda}^\mathrm{I}$ from the remaining Jeans equations 
\eqref{eq:jeanstriaxial_lambda} and \eqref{eq:jeanstriaxial_nu}
respectively. 

HZ92 solved the resulting two first-order PDEs (their Appendix B) and
showed that the same result is obtained by direct evaluation of the
second-order velocity moments, using the thin I-tube DF. 
They derived similar solutions for thin O- and S-tubes, for which
there is no motion in the $\lambda$-coordinate, so that
$T_{\lambda\lambda}^\mathrm{O} \equiv 0$ and
$T_{\lambda\lambda}^\mathrm{S} \equiv 0$, respectively. 

In St\"ackel discs we have -- besides the flat box orbits -- only one
family of (flat) tube orbits. 
For infinitesimally thin tube orbits $T_{\lambda\lambda} \equiv 0$, 
so that the Jeans equations \eqref{eq:jeanselliptic} reduce to two
different relations between $T_{\mu\mu}$ and the density and
potential. 
In \S \ref{sec:thinlooporbits}, we show how this places restrictions on
the form of the density and we give the solution for $T_{\mu\mu}$.
We also show that the general solution of \eqref{eq:jeanselliptic},
which we obtain in \S \ref{sec:2Dcases}, contains the thin tube result.
The same is true for the triaxial case: the general solution of
\eqref{eq:jeanstriaxial}, which we derive in \S \ref{sec:generalcase},
contains the three thin tube orbit solutions as special cases (\S
\ref{sec:thintubeorbitssolution}).

\subsection{All two-dimensional cases are similar}
\label{sec:2Dcasessimilar}

EL89 showed that the Jeans equations in oblate and prolate spheroidal
coordinates, \eqref{eq:jeansprolate} and \eqref{eq:jeansoblate}, can
be transformed to a system that is equivalent to the two Jeans
equations \eqref{eq:jeanselliptic} in confocal elliptic
coordinates. 
Evans et al.\ (2000\nocite{2000MNRAS.318.1131E}) 
arrived at the same two-dimensional form for
St\"ackel models with a scale-free DF. We introduce a transformation
which differs slightly from that of EL89, but has the advantage that
it removes the singular denominators in the Jeans equations.

The Jeans equations \eqref{eq:jeansprolate} for prolate potentials
can be simplified by introducing as dependent variables
\begin{equation}
  \label{eq:transformationsLandM}
  \mathcal{T}_{\tau\tau}(\lambda,\mu) =
  (\lambda\!+\!\beta)^\frac{1}{2} (\mu\!+\!\beta)^\frac{1}{2}
  ( T_{\tau\tau} \!-\! T_{\chi\chi} ), \quad
  \tau=\lambda,\mu, 
\end{equation}
so that the first two equations in \eqref{eq:jeansprolate} transform to
\begin{eqnarray}
\label{eq:prolateJeanseqnsrewritten}
  \frac{\partial \mathcal{T}_{\lambda\lambda}}{\partial \lambda}  
  \!+\! \frac{ \mathcal{T}_{\lambda\lambda} \!-\!
    \mathcal{T}_{\mu\mu} }{ 2(\lambda\!-\!\mu) } 
  \hspace{-10pt} & = & \hspace{-10pt} 
  - (\lambda\!+\!\beta)^\frac{1}{2} (\mu\!+\!\beta)^\frac{1}{2}
  \!\! \biggl[
  \rho\frac{\partial V_S}{\partial \lambda} \!+\! 
  \frac{\partial T_{\chi\chi}}{\partial \lambda} 
  \! \biggr], \nonumber \\*[-5pt]  \\*[-5pt]
  \frac{\partial \mathcal{T}_{\mu\mu}}{\partial \mu} 
  \!+\! \frac{ \mathcal{T}_{\mu\mu} \!-\!
  \mathcal{T}_{\lambda\lambda} }{ 2(\mu\!-\!\lambda) } 
  \hspace{-10pt} & = & \hspace{-10pt} 
  - (\mu\!+\!\beta)^\frac{1}{2} (\lambda\!+\!\beta)^\frac{1}{2} 
  \!\! \biggl[ 
  \rho\frac{\partial V_S}{\partial \mu} \!+\! \frac{\partial
  T_{\chi\chi}}{\partial \mu}  
  \! \biggr]. \nonumber
\end{eqnarray}
The third Jeans equation \eqref{eq:jeansprolate} can be integrated in
a straightforward fashion to give the $\chi$-dependence of
$T_{\chi\chi}$. It is trivially satisfied for prolate models with
$g(\chi)\equiv0$. Hence if, following EL89, we regard
$T_{\chi\chi}(\lambda,\mu)$ as a function which can be prescribed,
then equations (\ref{eq:prolateJeanseqnsrewritten}) have known right hand
sides, and are therefore of the same form as those of the disc case
\eqref{eq:jeanselliptic}.  The singular denominator $(\mu+\beta)$ of
\eqref{eq:jeansprolate} has disappeared, and there is a boundary
condition
\begin{equation}
  \label{eq:prolatebc}
  \mathcal{T}_{\mu\mu}(\lambda,-\beta)=0,
\end{equation}
due to the second continuity condition of \eqref{eq:prolatecontcond}
and the definition \eqref{eq:transformationsLandM}.

A similar reduction applies for oblate potentials. The middle equation
of \eqref{eq:jeansoblate} can be integrated to give the
$\phi$-dependence of $T_{\phi\phi}$, and is trivially satisfied for
oblate models. The remaining two equations \eqref{eq:jeansoblate}
transform to
\begin{eqnarray}
\label{eq:oblateJeanseqnsrewritten}
  \frac{\partial \mathcal{T}_{\lambda\lambda}}{\partial \lambda}  
  \!+\! \frac{ \mathcal{T}_{\lambda\lambda} \!-\!
    \mathcal{T}_{\nu\nu} }{ 2(\lambda\!-\!\nu) } 
  \hspace{-10pt} & = & \hspace{-10pt} 
  - (\lambda\!+\!\alpha)^\frac{1}{2} (-\alpha\!-\!\nu)^\frac{1}{2}
  \!\! \biggl[
  \rho\frac{\partial V_S}{\partial \lambda} \!+\! 
  \frac{\partial T_{\phi\phi}}{\partial \lambda} 
  \!\! \biggr], \nonumber   \\*[-5pt]  \\*[-5pt]
  \frac{\partial \mathcal{T}_{\nu\nu}}{\partial \nu} 
  \!+\! \frac{ \mathcal{T}_{\nu\nu} \!-\!
  \mathcal{T}_{\lambda\lambda} }{ 2(\nu\!-\!\lambda) } 
  \hspace{-10pt} & = & \hspace{-10pt} 
  - (-\alpha\!-\!\nu)^\frac{1}{2} (\lambda\!+\!\alpha)^\frac{1}{2} 
  \!\! \biggl[ 
  \rho\frac{\partial V_S}{\partial \nu} \!+\! \frac{\partial
  T_{\phi\phi}}{\partial \nu}  
  \!\! \biggr], \nonumber
\end{eqnarray}
in terms of the dependent variables
\begin{equation}
  \label{eq:obtransformationsLandM}
  \mathcal{T}_{\tau\tau}(\lambda,\nu) =
  (\lambda\!+\!\alpha)^\frac{1}{2} (-\alpha\!-\!\nu)^\frac{1}{2}
  ( T_{\tau\tau} \!-\! T_{\phi\phi} ), \quad
  \tau=\lambda,\nu.
\end{equation}
We now have two boundary conditions
\begin{equation}
\mathcal{T}_{\lambda\lambda}(-\alpha,\nu)=0, \quad
\mathcal{T}_{\nu\nu}(\lambda,-\alpha)=0,
\end{equation}
as a consequence of the last two continuity conditions of
\eqref{eq:oblatecontcond} and the definitions
\eqref{eq:obtransformationsLandM}. 

In the case of a scale-free DF, the stress components in the Jeans
equations in conical coordinates \eqref{eq:jeansconical} have the form
$T_{\tau\tau} = r^{-\zeta}\mathcal{T}_{\tau\tau}(\mu,\nu)$, with
$\zeta>0$ and $\tau=r,\mu,\nu$. After substitution and multiplication
by $r^{\zeta+1}$, the first equation of \eqref{eq:jeansconical}
reduces to
\begin{equation}
  \label{eq:conicalJeansscalefreeDF_r}
  (2-\zeta)\mathcal{T}_{rr}+\mathcal{T}_{\mu\mu}+\mathcal{T}_{\nu\nu}
  = r^{\zeta+1}\rho\frac{\partial V_S}{\partial r}. 
\end{equation}
When $\zeta=2$, $\mathcal{T}_{rr}$ drops out, so that the relation
between $\mathcal{T}_{\mu\mu}$ and $\mathcal{T}_{\nu\nu}$ is known and
the remaining two Jeans equations can be readily solved
(Evans et al.\ 2000\nocite{2000MNRAS.318.1131E}). 
In all other cases, $\mathcal{T}_{rr}$ can be obtained from
\eqref{eq:conicalJeansscalefreeDF_r} once we have solved the last two
equations of \eqref{eq:jeansconical} for $\mathcal{T}_{\mu\mu}$ and
$\mathcal{T}_{\nu\nu}$. 
This pair of equations is identical to the system of Jeans equations
\eqref{eq:jeanselliptic} for the case of disc potentials.
The latter is the simplest form of the equivalent two-dimensional
problem for all St\"ackel models with two degrees of freedom. 
We solve it in the next section.

Once we have derived the solution of \eqref{eq:jeanselliptic}, we may
obtain the solution for prolate St\"ackel potentials by replacing all
terms $-\rho\partial V_s/\partial \tau$ $(\tau=\lambda,\mu)$ by the
right-hand side of \eqref{eq:prolateJeanseqnsrewritten} and
substituting the transformations \eqref{eq:transformationsLandM} for
$T_{\lambda\lambda}$ and $T_{\mu\mu}$. Similarly, our unified notation
makes the application of the solution of \eqref{eq:jeanselliptic} to
the oblate case and to models with a scale-free DF straightforward
(\S\ref{sec:applicationdiscsol}).

% ===== section 3 ===== %

\section{The two-dimensional case}
\label{sec:2Dcases}

We first apply Riemann's method to solve the Jeans equations
\eqref{eq:jeanselliptic} in confocal elliptic coordinates for
St\"ackel discs (\S\ref{sec:confocalellipticcoord}).  
This involves finding a Riemann--Green function that describes the
solution for a source point of stress. 
The full solution is then obtained in compact form by representing the
known right-hand side terms as a sum of sources. 
In \S\ref{sec:singularsolution2D}, we introduce an alternative
approach, the singular solution method. 
Unlike Riemann's method, this can be extended to the three-dimensional
case, as we show in \S\ref{sec:generalcase}. 
We analyse the choice of the boundary conditions in detail in
\S\ref{sec:alternativeBCsdisc}.
In \S\ref{sec:applicationdiscsol}, we apply the two-dimensional  
solution to the axisymmetric and scale-free limits, and we also
consider a St\"ackel disc built with thin tube orbits.

\subsection{Riemann's method}
\label{sec:riemannsmethod}

After differentiating the first Jeans equation of
\eqref{eq:jeanselliptic} with respect to $\mu$ and eliminating terms
in $T_{\mu\mu}$ by applying the second equation, we obtain a
second-order partial differential equation (PDE) for
$T_{\lambda\lambda}$ of the form
\begin{equation}
  \label{eq:ellipticJeansas2ndorderPDE}
  \frac{\partial^2 T_{\lambda\lambda}}{\partial \lambda \partial \mu}
  - \frac{3}{2(\lambda\!-\!\mu)} \frac{\partial
  T_{\lambda\lambda}}{\partial \lambda} + \frac{1}{2(\lambda\!-\!\mu)}
  \frac{\partial T_{\lambda\lambda}}{\partial \mu} =
  U_{\lambda\lambda}(\lambda,\mu).
\end{equation}
Here $U_{\lambda\lambda}$ is a known function given by
\begin{equation}
  \label{eq:expressionH1}
  U_{\lambda\lambda} =
  -\frac{1}{(\lambda\!-\!\mu)^\frac{3}{2}}\frac{\partial}{\partial \mu}
  \biggl[
  (\lambda\!-\!\mu)^\frac{3}{2}\rho\frac{\partial V_S}{\partial \lambda}
  \biggr]
  -\frac{\rho}{2(\lambda\!-\!\mu)} \frac{\partial V_S}{\partial \mu}. 
\end{equation}
We obtain a similar second-order PDE for $T_{\mu\mu}$ by interchanging
$\lambda \leftrightarrow \mu$. Both PDEs can be solved by Riemann's
method. To solve them simultaneously, we define the linear
second-order differential operator
\begin{equation}
  \label{eq:defintionoperatorL}
  \mathcal{L} = \frac{\partial^2}{\partial \lambda \partial \mu}
  - \frac{c_1}{\lambda\!-\!\mu} \frac{\partial}{\partial \lambda} +
  \frac{c_2}{\lambda\!-\!\mu} \frac{\partial}{\partial \mu},
\end{equation}
with $c_1$ and $c_2$ constants to be specified. Hence, the more
general second-order PDE
\begin{equation}
  \label{eq:general2ndorderPDE}
  \mathcal{L}\,T = U,
\end{equation}
with $T$ and $U$ functions of $\lambda$ and $\mu$ alone, reduces to
those for the two stress components by taking
\begin{eqnarray}
\label{eq:valuesc_1c_2andU}
  T=T_{\lambda\lambda}\!\!\!\!\! &:& c_1={\textstyle\frac32}, 
                               \quad c_2={\textstyle\frac12}, \quad
  U=U_{\lambda\lambda}, \nonumber  \\*[-5pt]\\*[-5pt]
  T=T_{\mu\mu} \!\!\!\!\!&:& c_1={\textstyle\frac12}, 
                       \quad c_2={\textstyle\frac32}, \quad
  U=U_{\mu\mu}. \nonumber
\end{eqnarray}
In what follows, we introduce a Riemann--Green function $\mathcal{G}$
and incorporate the left-hand side of \eqref{eq:general2ndorderPDE}
into a divergence. Green's theorem then allows us to rewrite the
surface integral as a line integral over its closed boundary, which
can be evaluated if $\mathcal{G}$ is chosen suitably. We determine the
Riemann--Green function $\mathcal{G}$ which satisfies the required
conditions, and then construct the solution.

\subsubsection{Application of Riemann's method}
\label{sec:applyingriemannsmethod}

\begin{figure}
  \begin{center}
    \includegraphics[draft=false,scale=0.55,trim=3cm 3.5cm 0cm
    9cm]{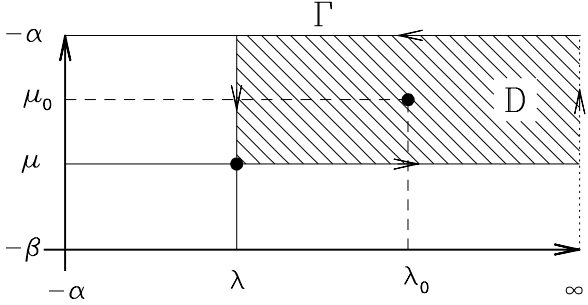} 
  \end{center}
  \caption[]{\slshape The $(\lambda_0, \mu_0)$-plane. The total stress at
    a field point $(\lambda,\mu)$, consists of the weighted
    contributions from source points at $(\lambda_0,\mu_0)$ in the
    domain $D$, with boundary $\Gamma$. }
  \label{fig:pointmasscontribution}
\end{figure}

We form a divergence by defining a linear operator
$\mathcal{L}^\star$, called the \textit{adjoint} of $\mathcal{L}$
(e.g., Copson 1975\nocite{C75}), as
\begin{equation}
    \label{eq:adjointoperatorL*}
    \mathcal{L}^\star =
    \frac{\partial^2 }{\partial \lambda \partial \mu}
    + \frac{\partial}{\partial \lambda} \biggl( \!
    \frac{c_1}{\lambda\!-\!\mu} \! \biggr)
    - \frac{\partial}{\partial \mu} \biggl( \!
    \frac{c_2}{\lambda\!-\!\mu} \! \biggr).
\end{equation}
The combination $\mathcal{G}\mathcal{L}T-T\mathcal{L}^\star
\mathcal{G}$ is a divergence for any twice differentiable function
$\mathcal{G}$ because
\begin{equation}
  \label{eq:divergencewithHandK}
  \mathcal{G}\mathcal{L}T-T\mathcal{L}^\star \mathcal{G} =
  \partial L/\partial \lambda + \partial M/\partial \mu, 
\end{equation}
where
\begin{eqnarray}
    \label{eq:divergencefuncH}
    L(\lambda,\mu) \hspace{-7pt} &=& \hspace{-7pt}
    \frac{\mathcal{G}}{2}\frac{\partial T}{\partial
    \mu} - \frac{T}{2}\frac{\partial \mathcal{G}}{\partial \mu} -
    \frac{c_1\,\mathcal{G}\,T}{\lambda\!-\!\mu}, 
                                               \nonumber \\*[-5pt]\\*[-5pt]
    M(\lambda,\mu) \hspace{-7pt} &=& \hspace{-7pt} 
    \frac{\mathcal{G}}{2}\frac{\partial T}{\partial \lambda} 
    - \frac{T}{2}\frac{\partial \mathcal{G}}{\partial \lambda} 
    + \frac{c_2\,\mathcal{G}\,T}{\lambda\!-\!\mu}. \nonumber
\end{eqnarray}
We now apply the PDE \eqref{eq:general2ndorderPDE} and the definition
\eqref{eq:adjointoperatorL*} in zero-subscripted variables $\lambda_0$
and $\mu_0$. We integrate the divergence
\eqref{eq:divergencewithHandK} over the domain $D=\{
(\lambda_0,\mu_0)$: $\lambda\le\lambda_0\le\infty,
\mu\le\mu_0\le-\alpha \}$, with closed boundary $\Gamma$
(Fig.~\ref{fig:pointmasscontribution}).  It follows by Green's theorem
that
\begin{multline}
  \label{eq:fromdomaintolineintegral}
  \iint\limits_D \hspace{-3pt} \d\lambda_0 \d\mu_0 
  \Bigl( \mathcal{G}\mathcal{L}_0T - T\mathcal{L}_0^\star \mathcal{G} \Bigr) 
  = \\*[-10pt] 
  \oint\limits_\Gamma \! \d\mu_0 \, L(\lambda_0,\mu_0) - 
  \oint\limits_\Gamma \! \d\lambda_0 \, M(\lambda_0,\mu_0), 
\end{multline}
where $\Gamma$ is circumnavigated counter-clockwise.  Here $\mathcal{L}_0$
and $\mathcal{L}_0^\star$ denote the operators
\eqref{eq:defintionoperatorL} and \eqref{eq:adjointoperatorL*} in
zero-subscripted variables.  We shall seek a Riemann--Green function
$\mathcal{G}(\lambda_0,\mu_0)$ which solves the PDE
\begin{equation}
  \label{eq:1stconditionv}
   \mathcal{L}_0^\star \mathcal{G} =0,
\end{equation}
in the interior of $D$.  Then the left-hand side of
\eqref{eq:fromdomaintolineintegral} becomes $\iint_D \d\lambda_0
\d\mu_0 \mathcal{G}(\lambda_0,\mu_0)\, U(\lambda_0,\mu_0)$.  The
right-hand side of \eqref{eq:fromdomaintolineintegral} has a
contribution from each of the four sides of the rectangular boundary
$\Gamma$. We suppose that $M(\lambda_0,\mu_0)$ and $L(\lambda_0,\mu_0)$
decay sufficiently rapidly as $\lambda_0 \to \infty$ so that the
contribution from the boundary at $\lambda_0=\infty$ vanishes and the
infinite integration over $\lambda_0$ converges. Partial integration
of the remaining terms then gives for the boundary integral 
\begin{multline}
  \label{eq:righthandsiderewritten} 
  \int\limits_\lambda^\infty \hspace{-4pt} \d\lambda_0 \Bigl[\!
  \Bigl( \frac{\partial \mathcal{G}}{\partial \lambda_0} \!-\!
  \frac{c_2\,\mathcal{G}}{\lambda_0\!-\!\mu_0} \Bigr) T
  \hspace{-6pt} \underset{\mu_0=\mu}{\Bigr]} 
  \!+\!\!
  \int\limits_\mu^{-\alpha} \hspace{-4pt} \d\mu_0 \Bigl[ \Bigl( 
  \frac{\partial \mathcal{G}}{\partial \mu_0} \!+\! 
  \frac{c_1\,\mathcal{G}}{\lambda_0\!-\!\mu_0} 
  \Bigr) T \hspace{-6pt} \underset{\lambda_0=\lambda}{\Bigr]} \\
  \!+\! \int\limits_\lambda^\infty \hspace{-4pt} \d\lambda_0 \Bigl[
  \Bigl( \frac{\partial T}{\partial \lambda_0} \!+\!
  \frac{c_2\,T}{\lambda_0\!-\!\mu_0} \Bigr) \mathcal{G}
  \hspace{-9pt} \underset{\mu_0=-\alpha}{\Bigr]} \hspace{-10pt}
  + \, \mathcal{G}(\lambda,\mu)T(\lambda,\mu).
\end{multline}
We now impose on $\mathcal{G}$ the additional conditions
\begin{equation}
  \label{eq:2ndconditionv}
  \mathcal{G}(\lambda,\mu)=1,
\end{equation}
and
\begin{eqnarray}
  \label{eq:3rdconditionv}
  \frac{\partial \mathcal{G}}{\partial \lambda_0} 
  - \frac{c_2\,\mathcal{G}}{\lambda_0\!-\!\mu_0} = 0
  \qquad &\mathrm{on}& \quad \mu_0=\mu, \nonumber \\*[-5pt] \\*[-5pt]   
  \frac{\partial \mathcal{G}}{\partial \mu_0} 
  + \frac{c_1\,\mathcal{G}}{\lambda_0\!-\!\mu_0} = 0
  \qquad &\mathrm{on}& \quad \lambda_0=\lambda. \nonumber
\end{eqnarray}
Then eq.~\eqref{eq:fromdomaintolineintegral} gives the explicit
solution
\begin{multline}
  \label{eq:solutionstressT}
  T(\lambda,\mu) = 
  \int\limits_\lambda^\infty \hspace{-4pt} \d\lambda_0 
  \hspace{-4pt} \int\limits_\mu^{-\alpha} \hspace{-5pt} \d\mu_0 \, 
  \mathcal{G}(\lambda_0,\mu_0)\, U(\lambda_0,\mu_0)
  \\*[-10pt] -
  \int\limits_\lambda^\infty \hspace{-4pt} \d\lambda_0
  \Bigl[ \Bigl( \frac{\partial T}{\partial \lambda_0} \!+\! 
  \frac{ c_2\,T}{\lambda_0\!-\!\mu_0} \Bigr) \mathcal{G} 
  \hspace{-10pt} \underset{\mu_0=-\alpha}{\Bigr]} \hspace{10pt},
\end{multline}
for the stress component, once we have found the Riemann--Green
function $\mathcal{G}$.

%%%FIG
%%%FIG

\subsubsection{The Riemann--Green function}
\label{sec:findingtheriemanngreenfunction}

Our prescription for the Riemann--Green function
$\mathcal{G}(\lambda_0,\mu_0)$ is that it satisfies the PDE
\eqref{eq:1stconditionv} as a function of $\lambda_0$ and $\mu_0$, and
that it satisfies the boundary conditions \eqref{eq:2ndconditionv} and
\eqref{eq:3rdconditionv} at the specific values $\lambda_0=\lambda$
and $\mu_0=\mu$.  Consequently $\mathcal{G}$ depends on two sets of
coordinates. Henceforth, we denote it as
$\mathcal{G}(\lambda,\mu;\lambda_0,\mu_0)$.  

An explicit expression for the Riemann--Green function which solves 
\eqref{eq:1stconditionv} is (Copson 1975\nocite{C75})
\begin{equation}
  \label{eq:ansatszforv}
  \mathcal{G}(\lambda,\mu;\lambda_0,\mu_0) =
  \frac{ (\lambda_0\!-\!\mu_0)^{c_2} (\lambda\!-\!\mu_0)^{c_1-c_2} }{
  (\lambda\!-\!\mu)^{c_1} } F(w),  
\end{equation}
where the parameter $w$ is defined as 
\begin{equation}
  \label{eq:definitionofw}
  w = \frac{(\lambda_0\!-\!\lambda) (\mu_0\!-\!\mu) }{
  (\lambda_0\!-\!\mu_0) (\lambda\!-\!\mu) },
\end{equation}
and $F(w)$ is to be determined. Since $w=0$ when $\lambda_0=\lambda$
or $\mu_0=\mu$, it follows from \eqref{eq:2ndconditionv} that the
function $F$ has to satisfy $F(0)=1$. It is straightforward to verify
that $\mathcal{G}$ satisfies the conditions \eqref{eq:3rdconditionv},
and that eq.~\eqref{eq:1stconditionv} reduces to the following
ordinary differential equation for $F(w)$
\begin{equation}
  \label{eq:odeforF}
  w(1\!-\!w)F''+[1\!-\!(2+c_1\!-\!c_2)w]F'-c_1(1\!-\!c_2)F=0. 
\end{equation}
This is a hypergeometric equation (e.g., Abramowitz \& Stegun
1965\nocite{AS65}), and its unique solution satisfying
$F(0)=1$ is
\begin{equation}
  \label{eq:solutionodeforF}
  F(w) = {}_2F_1(c_1,1\!-\!c_2;1;w).
\end{equation}
The Riemann--Green function \eqref{eq:ansatszforv} represents the
influence at a field point at $(\lambda,\mu)$ due to a source point at
$(\lambda_0,\mu_0)$. Hence it satisfies the PDE
\begin{equation}
  \label{eq:pdeforG}
   \mathcal{L}\,\mathcal{G}(\lambda,\mu; \lambda_0,\mu_0) =
   \delta(\lambda_0\!-\!\lambda) \delta(\mu_0\!-\!\mu).  
\end{equation}
The first right-hand side term of the solution \eqref{eq:solutionstressT} 
is a sum over the sources in $D$ which are due to the inhomogeneous term
$U$ in the PDE \eqref{eq:general2ndorderPDE}.  That PDE is hyperbolic
with characteristic variables $\lambda$ and $\mu$. By choosing to
apply Green's theorem to the domain $D$, we made it the domain of
dependence (Strauss 1992\nocite{S92}) of the field point $(\lambda,\mu)$
for \eqref{eq:general2ndorderPDE}, and hence we implicitly decided to
integrate that PDE in the direction of decreasing $\lambda$ and
decreasing $\mu$.

The second right-hand side term of the solution
\eqref{eq:solutionstressT} represents the solution to the homogeneous
PDE $\mathcal{L}\, T =0$ due to the boundary values of $T$ on the part
of the boundary $\mu=-\alpha$ which lies within the domain of
dependence. There is only one boundary term because we implicitly
require that $T(\lambda,\mu) \to 0$ as $\lambda \to \infty$. We verify
in \S\ref{sec:checkingforconsistency} that this requirement is indeed
satisfied.

\subsubsection{The disc solution}
\label{sec:thediscsolution}

We obtain the Riemann--Green functions for $T_{\lambda\lambda}$ and
$T_{\mu\mu}$, labeled as $\mathcal{G}_{\lambda\lambda}$ and
$\mathcal{G}_{\mu\mu}$, respectively, from expressions
\eqref{eq:ansatszforv} and \eqref{eq:solutionodeforF} by substitution
of the values for the constants $c_1$ and $c_2$ from
\eqref{eq:valuesc_1c_2andU}.  The hypergeometric function in
$\mathcal{G}_{\lambda\lambda}$ is the complete elliptic integral of
the second kind\footnote{We use the definition $E(w) = \!\!
\int_0^\frac{\pi}{2} \d\theta \, \sqrt{1-w\sin^2\theta}$}, $E(w)$.
The hypergeometric function in $\mathcal{G}_{\mu\mu}$ can also be
expressed in terms of $E(w)$ using eq.~(15.2.15) of 
Abramowitz \& Stegun (1965\nocite{AS65}), so that we can write 
\begin{subequations}
  \label{eq:riemannfncsforTllandTmm}
  \begin{equation}
    \label{eq:vforTlambdalambda}
    \mathcal{G}_{\lambda\lambda}(\lambda,\mu;\lambda_0,\mu_0) = 
    \frac{ (\lambda_0\!-\!\mu_0)^\frac{3}{2} }{
      (\lambda\!-\!\mu)^\frac{1}{2} }  
    \frac{2E(w)}{\pi (\lambda_0\!-\!\mu) },
  \end{equation}
  \begin{equation}
    \label{eq:vforTmumu}
    \mathcal{G}_{\mu\mu}(\lambda,\mu;\lambda_0,\mu_0) =
    \frac{ (\lambda_0\!-\!\mu_0)^\frac{3}{2} }{
      (\lambda\!-\!\mu)^\frac{1}{2} }  
    \frac{2E(w)}{\pi (\lambda\!-\!\mu_0) },
  \end{equation}
\end{subequations}
Substituting these into \eqref{eq:solutionstressT} gives the solution
of the stress components throughout the disc as 
\begin{subequations}
   \label{eq:solutionTllandTmm}
   \begin{multline}
     \label{eq:solutionTlambdalambda}
     T_{\lambda\lambda}(\lambda,\mu) = 
     \frac{2}{\pi(\lambda\!-\!\mu)^\frac{1}{2}} \Biggl\{ \\ 
     \int\limits_\lambda^\infty \hspace{-5pt} \d\lambda_0 
     \hspace{-6pt} \int\limits_\mu^{-\alpha} \hspace{-4pt} \d\mu_0
     \! \frac{E(w)}{(\lambda_0\!\!-\!\mu)}
     \! \biggl\{\!\! 
     \frac{\partial}{\partial \mu_0} \!\biggl[\!
     - (\lambda_0\!-\!\mu_0)^{\frac{3}{2}} \!
     \rho\frac{\partial V_S}{\partial \lambda_0} \!\biggr]\! -
     \frac{(\lambda_0\!\!-\!\mu_0)^{\frac{1}{2}}}{2} \!
     \rho\frac{\partial V_S}{\partial \mu_0} 
     \!\biggr\} \\
      - \!\! \int\limits_\lambda^\infty \hspace{-4pt} \d\lambda_0 
     \biggl[ \! \frac{E(w)}{(\lambda_0\!-\!\mu)} 
      \hspace{-10pt} \underset{\mu_0=-\alpha}{\biggr]} \hspace{-10pt}
      (\lambda_0\!+\!\alpha) \frac{\d}{\d\lambda_0} \!\Bigl[\!
      (\lambda_0\!+\!\alpha)^{\frac{1}{2}}T_{\lambda\lambda}(\lambda_0,-\alpha)
      \!\Bigr]\! 
      \!\Biggr\},
    \end{multline}
    \vspace{-20pt}
    \begin{multline}
      \label{eq:solutionTmumu}
      T_{\mu\mu}(\lambda,\mu) =  
     \frac{2}{\pi(\lambda\!-\!\mu)^\frac{1}{2}} \Biggl\{ \\
     \int\limits_\lambda^\infty \hspace{-5pt} \d\lambda_0 
     \hspace{-6pt} \int\limits_\mu^{-\alpha} \hspace{-4pt} \d\mu_0
     \! \frac{E(w)}{(\lambda\!-\!\mu_0)} 
     \! \biggl\{\!\!
     \frac{\partial}{\partial \lambda_0} \!\biggl[\!
     -(\lambda_0\!-\!\mu_0)^{\frac{3}{2}} \!
     \rho\frac{\partial V_S}{\partial \mu_0} \!\biggr]\! +
     \frac{(\lambda_0\!\!-\!\mu_0)^{\frac{1}{2}}}{2} \!
     \rho\frac{\partial V_S}{\partial \lambda_0} 
     \!\biggr\} \\
     \hspace{-20pt}
     - \!\! \int\limits_\lambda^\infty
     \hspace{-4pt} \d\lambda_0 \,
     \biggl[ \! \frac{E(w)}{(\lambda\!-\!\mu_0)} 
     \hspace{-10pt} \underset{\mu_0=-\alpha}{\biggr]} \hspace{-10pt}
     \frac{\d}{\d\lambda_0} \!\Bigl[\!
     (\lambda_0\!+\!\alpha)^{\frac{3}{2}}T_{\mu\mu}(\lambda_0,-\alpha)
     \!\Big]\! 
     \!\Biggr\}. 
  \end{multline}
\end{subequations}
This solution depends on $\rho$ and $V_S$, which are assumed to be
known, and on $T_{\lambda\lambda}(\lambda,-\alpha)$ and
$T_{\mu\mu}(\lambda,-\alpha)$, i.e., the stress components on the part
of the $y$-axis beyond the foci. 
Because these two stress components satisfy the first Jeans equation
of \eqref{eq:jeanselliptic} at $\mu=-\alpha$, we are only free to choose
one of them, say $T_{\mu\mu}(\lambda,-\alpha)$. 
$T_{\lambda\lambda}(\lambda,-\alpha)$ then follows by integrating this
first Jeans equation with respect to $\lambda$, using the continuity
condition \eqref{eq:continuityconditionelliptic} and requiring that
$T_{\lambda\lambda}(\lambda,-\alpha) \to 0$ as $\lambda \to \infty$.

\subsubsection{Consistency check}
\label{sec:checkingforconsistency}

We now investigate the behaviour of our solutions at large distances
and verify that our working hypothesis concerning the radial fall-off
of the functions $L$ and $M$ in eq.~(\ref{eq:divergencefuncH}) is
correct. The solution \eqref{eq:solutionstressT} consists of two
components: an area integral due to the inhomogeneous right-hand side
term of the PDE \eqref{eq:general2ndorderPDE}, and a single integral
due to the boundary values. We examine them 
in turn to obtain the conditions for the integrals to converge.  Next,
we parameterise the behaviour of the density and potential at large
distances and apply it to the solution \eqref{eq:solutionTllandTmm}
and to the energy equation \eqref{eq:definitionhamiltonian} to check
if the convergence conditions are satisfied for physical
potential-density pairs.

As $\lambda_0\to\infty$, $w$ tends to the finite limit
$(\mu_0-\mu)/(\lambda-\mu)$. Hence $E(w)$ is finite, and so, by
\eqref{eq:riemannfncsforTllandTmm},
$\mathcal{G}_{\lambda\lambda}=\mathcal{O}(\lambda_0^{1/2})$ and
$\mathcal{G}_{\mu\mu}=\mathcal{O}(\lambda_0^{3/2})$.  Suppose now that
$U_{\lambda\lambda}(\lambda_0,\mu_0)=\mathcal{O}(\lambda_0^{-l_1-1})$
and $U_{\mu\mu}(\lambda_0,\mu_0)=\mathcal{O}(\lambda_0^{-m_1-1})$ as
$\lambda_0\to\infty$. The area integrals in the solution
\eqref{eq:solutionstressT} then converge, provided that
$l_1>\frac{1}{2}$ and $m_1>\frac{3}{2}$.  These requirements place
restrictions on the behaviour of the density $\rho$ and potential
$V_S$ which we examine below.  Since
$\mathcal{G}_{\lambda\lambda}(\lambda,\mu;\lambda_0,\mu_0)$ is
$\mathcal{O}(\lambda^{-1/2})$ as $\lambda\to\infty$, the area integral
component of $T_{\lambda\lambda}(\lambda,\mu)$ behaves as
$\mathcal{O}(\lambda^{-1/2}\int_{\lambda}^{\infty}\lambda_0^{-l_1-1/2}
d\lambda_0)$ and so is $\mathcal{O}(\lambda^{-l_1})$. Similarly, with
$\mathcal{G}_{\mu\mu}(\lambda,\mu;\lambda_0,\mu_0) =
\mathcal{O}(\lambda^{-3/2})$ as $\lambda\to\infty$, the first
component of $T_{\mu\mu}(\lambda,\mu)$ is
$\mathcal{O}(\lambda_0^{-m_1})$.

To analyse the second component of the solution
\eqref{eq:solutionstressT}, we suppose that the boundary value
$T_{\lambda\lambda}(\lambda_0,-\alpha)=\mathcal{O}(\lambda_0^{-l_2})$
and $T_{\mu\mu}(\lambda_0,-\alpha)=\mathcal{O}(\lambda_0^{-m_2})$ as
$\lambda_0\to\infty$. 
A similar analysis then shows that the boundary integrals
converge, provided that $l_2>\frac{1}{2}$ and $m_2>\frac{3}{2}$, and
that the second components of $T_{\lambda\lambda}(\lambda,\mu)$ and
$T_{\mu\mu}(\lambda,\mu)$ are $\mathcal{O}(\lambda^{-l_2})$ and
$\mathcal{O}(\lambda^{-m_2})$ as $\lambda\to\infty$, respectively.

We conclude that the convergence of the integrals in the solution
\eqref{eq:solutionstressT} requires that
$T_{\lambda\lambda}(\lambda,\mu)$ and $T_{\mu\mu}(\lambda,\mu)$ decay
at large distance as 
$\mathcal{O}(\lambda^{-l})$ with $l>\frac{1}{2}$ and
$\mathcal{O}(\lambda^{-m})$ with $m>\frac{3}{2}$, respectively. 
The requirements which we have imposed on
$U(\lambda_0,\mu_0)$ and $T(\lambda_0,-\alpha)$ cause the
contributions to $\oint_\Gamma d\mu_0 L(\lambda_0,\mu_0)$ in Green's
formula \eqref{eq:fromdomaintolineintegral} from the segment of the
path at large $\lambda_0$ to be negligible in all cases.

Having obtained the requirements for the Riemann--Green function analysis
to be valid, we now investigate the circumstances in which they apply.
Following Arnold et al.\ (1994\nocite{1994MNRAS.271..924A}), we 
consider densities $\rho$ that decay as $N(\mu)\lambda^{-s/2}$
at large distances. 
We suppose that the function $G(\tau)$ introduced in
eq. \eqref{eq:defgtau} is $\mathcal{O}(\tau^\delta)$ for
$-\frac{1}{2}\le\delta<0$ as $\tau \to \infty$.
The lower limit $\delta=-\frac{1}{2}$ corresponds to a potential
due to a finite total mass, while the upper limit restricts it to
potentials that decay to zero at large distances.

For the disc potential \eqref{eq:discpot}, we then have that
$f(\tau)=\mathcal{O}(\tau^{\delta+1})$ when $\tau \to \infty$. 
Using the definition \eqref{eq:expressionH1}, we obtain
\looseness=-1
\begin{subequations}
  \label{eq:Utautaustaeckel}
  \begin{equation}
    U_{\lambda\lambda}(\lambda,\mu)=
    \frac{f^{\prime}(\mu)-f^{\prime}(\lambda)}
         {2(\lambda-\mu)^2}\rho +
    \frac{V_S+f^{\prime}(\lambda)}{(\lambda-\mu)}
    \frac{\partial \rho}{\partial \mu},
  \end{equation}
  \begin{equation}
    U_{\mu\mu}(\lambda,\mu)=
    \frac{f^{\prime}(\lambda)-f^{\prime}(\mu)}
         {2(\lambda-\mu)^2}\rho -
    \frac{V_S+f^{\prime}(\mu)}{(\lambda-\mu)}
    \frac{\partial \rho}{\partial \lambda},
  \end{equation}
\end{subequations}
where $\rho$ is the surface density of the disc. 
It follows that $U_{\lambda\lambda}(\lambda,\mu)$ is generally the
larger and is $\mathcal{O}(\lambda^{\delta-s/2-1})$ as $\lambda \to
\infty$, whereas $U_{\mu\mu}(\lambda,\mu)$ is
$\mathcal{O}(\lambda^{-2-s/2})$.  Hence, for the components of the
stresses \eqref{eq:solutionTllandTmm} we have
$T_{\lambda\lambda}=\mathcal{O}(\lambda^{\delta-s/2})$ and
$T_{\mu\mu}=\mathcal{O}(\lambda^{-1-s/2})$.  This estimate for
$U_{\lambda\lambda}$ assumes that $\partial\rho/\partial\mu$ is also
$\mathcal{O}(\lambda^{-s/2})$.  It is too high if the density becomes
independent of angle at large distances, as it does for discs with
$s<3$ (Evans \& de Zeeuw 1992\nocite{1992MNRAS.257..152E}).  
Using these estimates with the requirements for integral convergence
that were obtained earlier, we obtain the conditions $s>2\delta+1$ and
$s>1$, respectively, for inhomogeneous terms in
$T_{\lambda\lambda}(\lambda,\mu)$ and $T_{\mu\mu}(\lambda,\mu)$ to be
valid solutions. The second condition implies the first because
$\delta<0$.

With $V_S(\lambda,\mu)=\mathcal{O}(\lambda^\delta)$ at large
$\lambda$, it follows from the energy equation
\eqref{eq:definitionhamiltonian} for bound orbits that the
second-order velocity moments $\langle v_\tau^2 \rangle$ cannot exceed
$\mathcal{O}(\lambda^\delta)$, and hence that stresses
$T_{\tau\tau}=\rho\langle v_\tau^2 \rangle$ cannot exceed
$\mathcal{O}(\lambda^{\delta-s/2})$. This implies for
$T_{\lambda\lambda}(\lambda,\mu)$ that $s>2\delta+1$, and for
$T_{\mu\mu}(\lambda,\mu)$ we have the more stringent requirement that
$s>2\delta+3$. This last requirement is unnecessarily restrictive, but
an alternative form of the solution is needed to do better. Since that
alternative form arises naturally with the singular solution method,
we return to this issue in \S\ref{sec:generalsolution2D}.

Thus, for the Riemann--Green solution to apply, we find the conditions
$s>1$ and $-\frac{1}{2}\leq\delta <0$. These conditions are satisfied
for the perfect elliptic disk $(s=3, \delta=-\frac{1}{2})$, and for
many other separable discs (Evans \& de Zeeuw
1992\nocite{1992MNRAS.257..152E}).

\subsubsection{Relation to the EL89 analysis}
\label{sec:relation2EL89}

EL89 solve for the difference $\Delta \equiv T_{\lambda\lambda} -
T_{\mu\mu}$ using a Green's function method which is essentially
equivalent to the approach used here. EL89 give the Fourier transform
of their Green's function, but do not invert it. We give the
Riemann--Green function for $\Delta$ in Appendix A, and then rederive
it by a Laplace transform analysis. Our Laplace transform analysis can
be recast in terms of Fourier transforms.  When we do this, we obtain
a result which differs from that of EL89.

\subsection{Singular Solution Superposition}
\label{sec:singularsolution2D}

We have solved the disc problem \eqref{eq:jeanselliptic} by combining
the two Jeans equations into a single second-order PDE in one of the
stress components, and then applying Riemann's method to it.  
However, Riemann's method and other standard techniques do not carry
over to a single third-order PDE in one dependent variable, which is
the best that one could expect to have in the general case. 
We therefore introduce an alternative but equivalent method of
solution, also based on the superposition of source points.
In constrast to Riemann's method, this singular solution method is
applicable to the general case of triaxial St\"ackel models.

\subsubsection{Simplified Jeans equations}
\label{sec:simplifiedjeansequations2D}

We define new independent variables
\begin{eqnarray}
\label{eq:deffunctionsStt}
  S_{\lambda\lambda}(\lambda,\mu)  
  \hspace{-7pt} & = & \hspace{-7pt} 
  |\lambda\!-\!\mu|^\frac{1}{2} \,
  T_{\lambda\lambda}(\lambda,\mu), \nonumber \\*[-5pt] \\*[-5pt]
  S_{\mu\mu}(\lambda,\mu)
  \hspace{-7pt} & = & \hspace{-7pt} 
  |\mu\!-\!\lambda|^\frac{1}{2} \,
  T_{\mu\mu}(\lambda,\mu), \nonumber 
\end{eqnarray}
where $|.|$ denotes absolute value, introduced to make the square root
single-valued with respect to cyclic permutation of $\lambda \to \mu
\to \lambda$. The Jeans equations \eqref{eq:jeanselliptic} can then be
written in the form
\begin{subequations}
  \label{eq:jeanselliptic_alternativeS}
  \begin{equation}
    \label{eq:jeansellipticS_lambda}
    \frac{\partial S_{\lambda\lambda}}{\partial \lambda} -
    \frac{S_{\mu\mu}}{2(\lambda\!-\!\mu)} = 
    -|\lambda\!-\!\mu|^\frac{1}{2}  
    \rho\frac{\partial V_S}{\partial \lambda} 
    \equiv g_1(\lambda,\mu), 
  \end{equation}
  \begin{equation}
    \label{eq:jeansellipticS_mu}
    \frac{\partial S_{\mu\mu}}{\partial \mu} -
    \frac{S_{\lambda\lambda}}{2(\mu\!-\!\lambda)} = 
    -|\mu\!-\!\lambda|^\frac{1}{2} 
    \rho\frac{\partial V_S}{\partial \mu} 
    \equiv g_2(\lambda,\mu).  
  \end{equation}
\end{subequations}
For given density and potential, $g_1$ and $g_2$ are known functions
of $\lambda$ and $\mu$.  
Next, we consider a simplified form of
\eqref{eq:jeanselliptic_alternativeS} by taking for $g_1$ and $g_2$,
respectively  
\begin{equation}
  \label{eq:simplifysetting}
  \tilde{g}_1(\lambda,\mu) = 0, 
  \quad 
  \tilde{g}_2(\lambda,\mu) =
  \delta(\lambda_0\!-\!\lambda) \delta(\mu_0\!-\!\mu), 
\end{equation}
with $-\beta\le\mu\le\mu_0\le-\alpha\le\lambda\le\lambda_0$. A similar
set of simplified equations is obtained by interchanging the
expressions for $\tilde{g}_1$ and $\tilde{g}_2$. 
We refer to solutions of these simplified Jeans equations as
\textit{singular solutions}. 

Singular solutions can be interpreted as contributions to the stresses
at a fixed point $(\lambda,\mu)$ due to a source point in
$(\lambda_0,\mu_0)$ (Fig.~\ref{fig:pointmasscontribution}). The full
stress at the field point can be obtained by adding all source point
contributions, each with a weight that depends on the local density and
potential. In what follows, we derive the singular solutions, and then
use this superposition principle to construct the solution for the
St\"ackel discs in \S\ref{sec:generalsolution2D}.

\subsubsection{Homogeneous boundary problem}
\label{sec:homogeneousboundaryproblem2D}

The choice \eqref{eq:simplifysetting} places constraints on the
functional form of $S_{\lambda\lambda}$ and $S_{\mu\mu}$. 
The presence of the delta-functions in $\tilde{g}_2$ 
requires that $S_{\mu\mu}$ contains a term 
$-\delta(\lambda_0\!-\!\lambda)\mathcal{H}( \mu_0\!-\!\mu )$, 
with the step-function
\begin{equation}
  \label{eq:defstepfunc}
  \mathcal{H}(x\!-\!x_0) = 
  \begin{cases}
    0, & \text{$x < x_0$}, \\
    1, & \text{$x \ge x_0$}.
  \end{cases}
\end{equation}
Since $\mathcal{H}'(y)=\delta(y)$, it follows that, by taking the
partial derivative of
$-\delta(\lambda_0\!-\!\lambda)\mathcal{H}(\mu_0\!-\!\mu)$ with
respect to $\mu$, the delta-functions are balanced. There is no
balance when $S_{\lambda\lambda}$ contains
$\delta(\lambda_0\!-\!\lambda)$, and similarly neither stress
components can contain $\delta(\mu_0\!-\!\mu)$. We can, however, add a
function of $\lambda$ and $\mu$ to both components, multiplied by
$\mathcal{H} ( \lambda_0\!-\!\lambda ) \mathcal{H} ( \mu_0\!-\!\mu
)$. In this way, we obtain a singular solution of the form
\begin{eqnarray} 
  \label{eq:functionalformsSllandSmm}
  S_{\lambda\lambda} 
  \hspace{-9pt} & = & \hspace{-9pt} 
  A(\lambda,\mu) \mathcal{H}(\lambda_0\!-\!\lambda)
  \mathcal{H}(\mu_0\!-\!\mu), 
  \nonumber \\*[-5pt] \\*[-5pt] 
  S_{\mu\mu} \hspace{-9pt} & = & \hspace{-9pt} 
  B(\lambda,\mu) \mathcal{H}(\lambda_0\!-\!\lambda)
  \mathcal{H}(\mu_0\!-\!\mu) \!-\! \delta(\lambda_0\!-\!\lambda)
  \mathcal{H}(\mu_0\!-\!\mu),  
  \nonumber  
\end{eqnarray}
in terms of functions $A$ and $B$ that have to be determined.  
Substituting these forms in the simplified Jeans equations and
matching terms yields two homogeneous equations
\begin{equation}
  \label{eq:dischomeqns}
  \frac{\partial A}{\partial \lambda} - \frac{B}{2(\lambda\!-\!\mu)}
  =0, \quad \frac{\partial B}{\partial \mu} -
  \frac{A}{2(\mu\!-\!\lambda)} = 0, 
\end{equation}
and two boundary conditions  
\begin{equation}
  \label{eq:dischombc1and2}
  A(\lambda_0,\mu) = \frac{1}{2(\lambda_0\!-\!\mu)}, 
  \quad B(\lambda,\mu_0) = 0.
\end{equation}
Two alternative boundary conditions which are useful below can be
found as follows. Integrating the first of the
equations (\ref{eq:dischomeqns}) with respect to $\lambda$ on $\mu=\mu_0$,
where $B(\lambda,\mu_0)=0$, gives
\begin{equation}
  \label{eq:dischombc3}
   A(\lambda,\mu_0) = \frac{1}{2(\lambda_0\!-\!\mu_0)}.
\end{equation}
Similarly, integrating the second of equations \eqref{eq:dischomeqns} with
respect to $\mu$ on $\lambda=\lambda_0$ where $A$ is known gives
\begin{equation}
  \label{eq:dischombc4}
   B(\lambda_0,\mu) = \frac{\mu_0\!-\!\mu}{4
   (\lambda_0\!-\!\mu_0) (\lambda_0\!-\!\mu)}.  
\end{equation}
Even though expressions \eqref{eq:dischombc3} and
\eqref{eq:dischombc4} do not add new information, they will be useful
for identifying contour integral formulas in the analysis which
follows.

We have reduced the problem of solving the Jeans equations
\eqref{eq:jeanselliptic} for St\"ackel discs to a two-dimensional
boundary problem. We solve this problem by first deriving a
one-parameter particular solution (\S \ref{sec:particularsolution2D})
and then making a linear combination of particular solutions with
different values of their free parameter, such that the four boundary
expressions are satisfied simultaneously
(\S\ref{sec:derivinghomogenoussolution2D}). This gives the solution of
the homogeneous boundary problem.

\subsubsection{Particular solution}
\label{sec:particularsolution2D}

To find a particular solution of the homogeneous equations
\eqref{eq:dischomeqns} with one free parameter $z$, we take as an
Ansatz
\begin{eqnarray}
  \label{eq:formAandBpart}
  A(\lambda,\mu) \!\!\!&\propto& \!\!\! (\lambda\!-\!\mu)^{a_1}
  (z\!-\!\lambda)^{a_2} (z\!-\!\mu)^{a_3}, \nonumber 
  \\*[-5pt] \\*[-5pt] 
  B(\lambda,\mu) \!\!\!&\propto&\!\!\! (\lambda\!-\!\mu)^{b_1}
  (z\!-\!\lambda)^{b_2} (z\!-\!\mu)^{b_3}, \nonumber 
\end{eqnarray}
with $a_i$ and $b_i$ $(i=1,2,3)$ all constants. Hence,
\begin{eqnarray} 
  \label{eq:ansatzAandBsubst}
  \frac{\partial A}{\partial \lambda}  
  \hspace{-7pt} & = & \hspace{-7pt} A \biggl(
  \frac{a_1}{\lambda\!-\!\mu} - \frac{a_2}{z\!-\!\lambda} \biggr) 
  =\frac{1}{2(\lambda\!-\!\mu)} \biggl( 
  2a_1A\frac{z\!-\!\mu}{z\!-\!\lambda} \,\biggr), 
  \nonumber \\*[-5pt] \\*[-5pt] 
  \frac{\partial B}{\partial \mu}
  \hspace{-7pt} & = & \hspace{-7pt} B \biggl( 
  \frac{b_1}{\mu\!-\!\lambda}- \frac{b_3}{z\!-\!\mu} \biggr) 
  = \frac{1}{2(\mu\!-\!\lambda)} \biggl(
  2b_1B\frac{z\!-\!\lambda}{z\!-\!\mu} \,\biggr),
  \nonumber 
\end{eqnarray} 
where we have set $a_2=-a_1$ and $b_3=-b_1$. Taking $a_1=b_1=\frac12$,
the homogeneous equations are satisfied if
\begin{equation}
  \label{eq:ratioA/B}
  \frac{z\!-\!\lambda}{z\!-\!\mu} = \frac{A}{B} =
  \frac{(z\!-\!\lambda)^{-\frac{1}{2}-b_2}}{(z\!-\!\mu)^{-\frac{1}{2}-a_3}}, 
\end{equation}
so, $a_3=b_2=-\frac32$. We denote the resulting solutions as
\begin{subequations}
  \label{eq:partsolAandB}
  \begin{equation}
    \label{eq:partsolutionA}
    A^P(\lambda,\mu) = \frac{ |\lambda\!-\!\mu|^\frac{1}{2} }{
    (z\!-\!\lambda)^\frac{1}{2} (z\!-\!\mu)^\frac{3}{2} }, 
  \end{equation}
  \hspace{-2cm}
  \begin{equation}
    \label{eq:partsolutionB}
    B^P(\lambda,\mu) = \frac{ |\mu\!-\!\lambda|^\frac{1}{2} }{
    (z\!-\!\mu)^\frac{1}{2} (z\!-\!\lambda)^\frac{3}{2} }. 
  \end{equation}
\end{subequations}
These particular solutions follow from each other by cyclic
permutation $\lambda \to \mu \to \lambda$, as is required from the
symmetry of the homogeneous equations \eqref{eq:dischomeqns}.

\subsubsection{The homogeneous solution}
\label{sec:derivinghomogenoussolution2D}

%%%FIG
\begin{figure}
  \begin{center}
    \includegraphics[draft=false,scale=0.55,trim=2.5cm 3cm 0cm
    11cm]{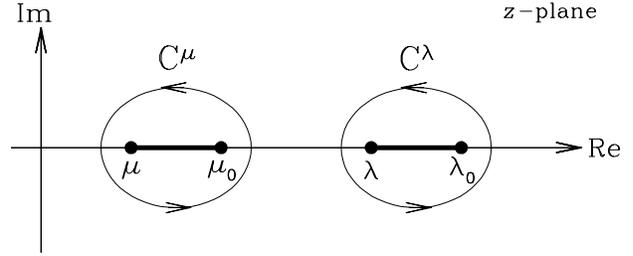}
  \end{center}
  \caption[]{\slshape Contours $C^\mu$ and $C^\lambda$ in the complex
  $z$-plane which appear in the solution (\ref{eq:homsolutions}). The
  two cuts running from $\mu$ to $\mu_0$ and one from $\lambda$ to
  $\lambda_0$ make the integrands single-valued. }
  \label{fig:contourscomplexzplane2D}
\end{figure}
%%%FIG

We now consider a linear combination of the particular solution
\eqref{eq:partsolAandB} by integrating it over the free parameter $z$,
which we assume to be complex. We choose the integration contours in
the complex $z$-plane, such that the four boundary expressions can be
satisfied simultaneously.

We multiply $B^P(\lambda,\mu)$ by~$(z\!-\!\mu_0)^\frac{1}{2}$, and
integrate it over the closed contour $C^\mu$
(Fig.~\ref{fig:contourscomplexzplane2D}). When $\mu=\mu_0$, the
integrand is analytic within $C^\mu$, so that the integral vanishes by
Cauchy's theorem.
Since both the
multiplication factor and the integration are independent of $\lambda$
and $\mu$, it follows from the superposition principle that the
homogeneous equations are still satisfied. In this way, the second of
the boundary expressions \eqref{eq:dischombc1and2} is
satisfied.

Next, we also multiply $B^P(\lambda,\mu)$ by
$(z-\lambda_0)^{-\frac{1}{2}}$, so that the contour $C^\lambda$
(Fig.~\ref{fig:contourscomplexzplane2D}) encloses a double pole when
$\lambda=\lambda_0$. From the Residue theorem
(e.g., Conway 1973\nocite{1973...Conway}), it then follows that
\begin{multline}
  \label{eq:doublepoleBl=l*}
  \oint\limits_{C^\lambda} 
  \frac{(z\!-\!\mu_0)^\frac{1}{2}}{(z\!-\!\lambda_0)^\frac{1}{2}}
  B^P(\lambda_0,\mu) \, \d z 
  = \oint\limits_{C^\lambda} 
  \frac{ (z\!-\!\mu_0)^\frac{1}{2}
  (\lambda_0\!-\!\mu)^\frac{1}{2} }{ (z\!-\!\mu)^\frac{1}{2}  
  (z\!-\!\lambda_0)^2 } \, \d z = \\*[5pt]  
  2\pi i (\lambda_0\!-\!\mu)^\frac{1}{2} 
  \biggl[ \frac{\d}{\d z}
  \biggl( \frac{z\!-\!\mu_0}{z\!-\!\mu} \biggr)^\frac{1}{2}
  \hspace{-7pt} \underset{z=\lambda_0}{\biggr]}
  \hspace{-7pt} 
  = \! \frac{ \pi i (\mu_0\!-\!\mu) }{
  (\lambda_0\!-\!\mu_0)^\frac{1}{2} (\lambda_0\!-\!\mu) },   
\end{multline}
which equals the boundary expression \eqref{eq:dischombc4}, up to the
factor $4\pi i (\lambda_0\!-\!\mu_0)^\frac{1}{2}$.

Taking into account the latter factor, and the ratio
\eqref{eq:ratioA/B} of $A$ and $B$, we postulate as homogeneous
solution \looseness=-1
\begin{subequations}
  \label{eq:homsolutions}
  \begin{equation}
    \label{eq:homsolA}
    A(\lambda,\mu) \! = \!\! \frac{1}{4\pi i}
    \frac{ |\lambda\!-\!\mu|^\frac{1}{2} }{
    |\lambda_0\!-\!\mu_0|^\frac{1}{2} } 
    \hspace{-3pt} \oint\limits_C \hspace{-3pt} 
    \frac{(z\!-\!\mu_0)^\frac{1}{2}\,\d z}{
    (z\!-\!\lambda)^\frac{1}{2} \! (z\!-\!\mu)^\frac{3}{2} \!
    (z\!-\!\lambda_0)^\frac{1}{2}},     
  \end{equation}
  \begin{equation}
    \label{eq:homsolB}
    B(\lambda,\mu) \! = \!\!  \frac{1}{4\pi i}
    \frac{ |\mu\!-\!\lambda|^\frac{1}{2} }{
    |\lambda_0\!-\!\mu_0|^\frac{1}{2} } 
    \hspace{-3pt} \oint\limits_C \hspace{-3pt} 
    \frac{(z\!-\!\mu_0)^\frac{1}{2}\,\d z}{
    (z\!-\!\mu)^\frac{1}{2} \! (z\!-\!\lambda)^\frac{3}{2} \! 
    (z\!-\!\lambda_0)^\frac{1}{2}},  
  \end{equation}
\end{subequations}
with the choice for the contour $C$ still to be specified. 

The integrands in \eqref{eq:homsolutions} consist of multi-valued
functions that all come in pairs 
$(z\!-\!\tau)^{1/2-m}(z\!-\!\tau_0)^{1/2-n}$, 
for integer $m$ and $n$, and for $\tau$ being
either $\lambda$ or $\mu$. Hence, we can make the integrands
single-valued by specifying two cuts in the complex $z$-plane, one
from $\mu$ to $\mu_0$ and one from $\lambda$ to $\lambda_0$. The
integrands are now analytic in the cut plane away from its cuts and
behave as $z^{-2}$ at large distances, so that the integral over a
circular contour with infinite radius is zero\footnote{We evaluate the
  square roots as $(z-\tau)^{\frac{1}{2}} = |z-\tau|\exp
  i\arg(z-\tau)$ with $|\arg(z-\tau)| \le \pi$.}.
Connecting the simple contours $C^\lambda$ and $C^\mu$ with this
circular contour shows that the cumulative contribution from each of
these contours cancels. As a consequence, every time we integrate over
the contour $C^\lambda$, we will obtain the same result by integrating
over $-C^\mu$ instead.
This means we integrate over $C^\mu$ and take the negative of the
result or, equally, integrate over $C^\mu$ in clockwise direction.

%, i.e., integrating over $C^\mu$ and taking the
%negative of the result or, equally, integrating over $C^\mu$ in 
%clockwise direction. 

%%%FIG
\begin{figure}
  \begin{center}
    \includegraphics[draft=false,scale=0.55,trim=2.75cm 4.5cm 0cm
    7.0cm]{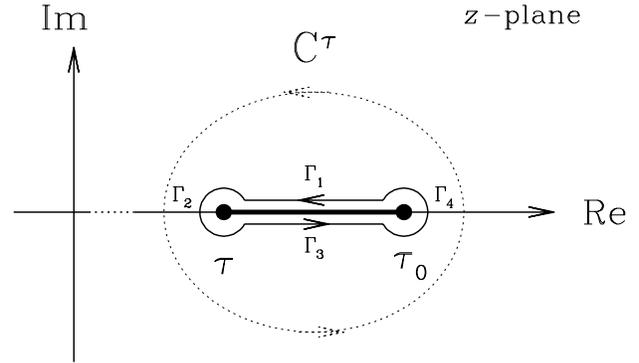}
  \end{center}
  \caption[]{\slshape Integration along the contour $C^\tau$. The
  contour is wrapped around the branch points $\tau$ and $\tau_0$
  ($\tau=\lambda,\mu$), and split into four parts.  $\Gamma_1$ and
  $\Gamma_3$ run parallel to the real axis in opposite
  directions. $\Gamma_2$ and $\Gamma_4$ are two arcs around $\tau$ and
  $\tau_0$, respectively.}
  \label{fig:wrappedcontourszplane2D}
\end{figure}
%%%FIG

For example, we obtained the boundary expression for $B$ in
\eqref{eq:doublepoleBl=l*} by applying the Residue theorem to the
double pole 
enclosed by the contour $C^\lambda$. The evaluation of the integral
becomes less straightforward when we consider the contour $-C^\mu$
instead. Wrapping the contour around the branch points $\mu$ and
$\mu_0$ (Fig.~\ref{fig:wrappedcontourszplane2D}), one may easily
verify that the contribution from the two arcs vanishes if their
radius goes to zero. Taking into account the change in phase when
going around the two branch points, one may show that the
contributions from the two remaining parts of the contour, parallel to
the real axis, are equivalent. Hence, we arrive at the following
(real) integral
\begin{equation}
  \label{eq:evaluationBl=l*ifCmu_1}
  B(\lambda_0,\mu) = \frac{1}{2\pi} 
  \frac{ (\lambda\!-\!\mu_0)^\frac{1}{2} }{
  (\lambda_0\!-\!\mu_0)^\frac{1}{2} }
  \int\limits_\mu^{\mu_0} 
  \frac{\d t}{(\lambda_0\!-\!t)^2} 
  \sqrt{ \frac{\mu_0\!-\!t}{t\!-\!\mu} }.
\end{equation}
The substitution 
\begin{equation}
  \label{eq:substitution4t}
  t = \mu + \frac{ (\mu_0\!-\!\mu) (\lambda_0\!-\!\mu_0)
  \, \sin^2\theta }{ (\mu_0\!-\!\mu) \, s \!-\! (\lambda_0\!-\!\mu) } 
\end{equation}
then indeed gives the correct boundary expression
\eqref{eq:dischombc4}.

When we take $\mu=\mu_0$ in \eqref{eq:homsolB}, we are left with the
integrand $(z-\lambda)^{-3/2} (z-\lambda_0)^{-1/2}$.  This is analytic
within the contour $C^\mu$ and hence it follows from Cauchy's theorem
that there is no contribution. However, if we take the contour
$-C^\lambda$ instead, it is not clear at once that the integral indeed
is zero. To evaluate the complex integral we wrap the contour
$C^\lambda$ around the branch points $\lambda$ and $\lambda_0$
(Fig.~\ref{fig:wrappedcontourszplane2D}). There will be no
contribution from the arc around $\lambda_0$ if its radius goes to
zero. However, since the integrand involves the term $z-\lambda$ with
power $-\frac32$, the contribution from the arc around $\lambda$ is of
the order $\epsilon^{-1/2}$ and hence goes to infinity if its radius
$\epsilon>0$ reduces to zero.  
If we let the two remaining straight parts
of the contour run from $\lambda+\epsilon$ to $\lambda_0$, then their
cumulative contribution becomes proportional to $\tan
\theta(\epsilon)$, with $\theta(\epsilon)$ approaching $\frac{\pi}{2}$
when $\epsilon$ reduces to zero. 
Hence, both the latter contribution and the contribution from
the arc around $\lambda$ approaches infinity. However, careful
investigation of their limiting behaviour shows that they cancel when
$\epsilon$ reaches zero, as is required for the boundary expression
$B(\lambda,\mu_0)=0$.

We have shown that the use of $C^\lambda$ and $-C^\mu$ gives the
same result, but the effort to evaluate the contour integral
varies between the two choices. 
The boundary expressions for $A(\lambda,\mu)$,
\eqref{eq:dischombc1and2} and \eqref{eq:dischombc3} are obtained most
easily if we consider $C^\lambda$ when $\lambda=\lambda_0$ and
$-C^\mu$ when $\mu=\mu_0$. In both cases the integrand in
\eqref{eq:homsolA} has a single pole within the chosen contour, so
that the boundary expressions follow by straightforward application of
the Residue theorem.

We now have proven that the homogeneous solution
\eqref{eq:homsolutions} solves the homogeneous equations
\eqref{eq:dischomeqns}, satisfies the boundary values
\eqref{eq:dischombc1and2}--\eqref{eq:dischombc4} separately and, from
the observation that $C^\lambda$ and $-C^\mu$ produce the same result,
also simultaneously.

\subsubsection{Evaluation of the homogeneous solution}
\label{sec:evaluatinghomogenoussolution2D}

The homogeneous solution \eqref{eq:homsolutions} consists of complex
contour integrals, which we transform to real integrals by wrapping
the contours $C^\lambda$ and $C^\mu$ around the corresponding pair of
branch points (Fig.~\ref{fig:wrappedcontourszplane2D}). To have no
contribution from the arcs around the branch points, we choose the
(combination of) contours such that the terms in the integrand
involving these branch points have powers larger than $-1$. In this
way, we can always evaluate the complex integral as a (real) integral
running from one branch point to the other.

In the homogeneous solution \eqref{eq:homsolA} for $A$ we choose
$C=C^\lambda$ and in \eqref{eq:homsolB} for $B$ we take $C=-C^\mu$.
Taking into account the changes in phase when going around the branch
points, we obtain the following expressions for the homogeneous
solution
\begin{subequations}
  \label{eq:homsolrealintg}
  \begin{equation}
    \label{eq:homsolrealA}
    A(\lambda,\mu) \! = \!\! \frac{1}{2\pi}
    \frac{ |\lambda\!-\!\mu|^\frac{1}{2} }{
    |\lambda_0\!-\!\mu_0|^\frac{1}{2} } 
    \hspace{-3pt} \int\limits_\lambda^{\lambda_0} \hspace{-5pt} 
    \frac{\d t}{t\!-\!\mu}
    \! \sqrt{ \! \frac{t\!-\!\mu_0}{ (t\!-\!\lambda) (t\!-\!\mu)
    (\lambda_0\!-\!t) } },       
  \end{equation}
  \begin{equation}
    \label{eq:homsolrealB}
    B(\lambda,\mu) \! = \!\!  \frac{1}{2\pi}
    \frac{ |\lambda\!-\!\mu|^\frac{1}{2} }{
    |\lambda_0\!-\!\mu_0|^\frac{1}{2} } 
    \hspace{-3pt} \int\limits_\mu^{\mu_0} \hspace{-5pt} 
    \frac{\d t}{\lambda\!-\!t}
    \! \sqrt{ \! \frac{\mu_0\!-\!t}{ (\lambda\!-\!t) (t\!-\!\mu) 
    (\lambda_0\!-\!t) } }.     
  \end{equation}
\end{subequations}
By a parameterisation of the form \eqref{eq:substitution4t}, or by
using an integral table (e.g., Byrd \& Friedman
1971\nocite{BF1971...integralbook}), expressions
\eqref{eq:homsolrealintg} can be written conveniently in terms of the
complete elliptic integral of the second kind, $E$, and its
derivative $E'$ \looseness=-1
\begin{subequations}
  \label{eq:homogeneoussolutions}
  \begin{equation}
    \label{eq:homsolrealAfull}
    A(\lambda,\mu;\lambda_0,\mu_0) = \frac{ E(w) }{ \pi(\lambda_0\!-\!\mu)},
  \end{equation}
  \begin{equation}
    \label{eq:homsolrealBfull}
    B(\lambda,\mu;\lambda_0,\mu_0) =
                   - \frac{ 2w E'(w) }{ \pi (\lambda_0\!-\!\lambda) }.
  \end{equation}
\end{subequations}
with $w$ defined as in \eqref{eq:definitionofw}. The second set of
arguments that were added to $A$ and $B$ make explicit the
position $(\lambda_0,\mu_0)$ of the source point which is causing the
stresses at the field point $(\lambda,\mu)$.

\subsubsection{The disc solution}
\label{sec:generalsolution2D}

The solution of equations \eqref{eq:jeanselliptic_alternativeS} with
right hand sides of the simplified form 
\begin{equation}
  \label{eq:simplifysettingtwo}
  \tilde{g}_1(\lambda,\mu) =  
  \delta(\lambda_0\!-\!\lambda) \delta(\mu_0\!-\!\mu),
  \quad 
  \tilde{g}_2(\lambda,\mu) = 0,
\end{equation}
is obtained from the solution \eqref{eq:functionalformsSllandSmm}
by interchanging $\lambda\leftrightarrow\mu$ and 
$\lambda_0\leftrightarrow\mu_0$. It is
\begin{eqnarray}
  \label{eq:functionalformsSllandSmmswitch}
  S_{\lambda\lambda} 
  \hspace{-9pt} & = & \hspace{-9pt} 
  B(\mu,\lambda;\mu_0,\lambda_0) \mathcal{H}(\lambda_0\!\!-\!\lambda) 
  \mathcal{H}(\mu_0\!\!-\!\mu) \!-\! \delta(\mu_0\!\!-\!\mu)
  \mathcal{H}(\lambda_0\!\!-\!\lambda),   
  \nonumber \\ \\*[-10pt] 
  S_{\mu\mu} 
  \hspace{-9pt} & = & \hspace{-9pt} 
  A(\mu,\lambda;\mu_0,\lambda_0) \mathcal{H}(\lambda_0\!\!-\!\lambda)
  \mathcal{H}(\mu_0\!\!-\!\mu).
  \nonumber
\end{eqnarray}
To find the solution to the full equations
\eqref{eq:jeanselliptic_alternativeS} at $(\lambda,\mu)$, we multiply
the singular solutions \eqref{eq:functionalformsSllandSmm} and
\eqref{eq:functionalformsSllandSmmswitch} by $g_1(\lambda_0,\mu_0)$
and $g_2(\lambda_0,\mu_0)$ respectively and integrate over $D$, the
domain of dependence of $(\lambda,\mu)$. This gives the first two
lines of the two equations \eqref{eq:gensolutions} below. The terms in
the third lines are due to the boundary values of $S_{\mu\mu}$ at
$\mu=-\alpha$. They are found by multiplying the singular solution
\eqref{eq:functionalformsSllandSmm} evaluated for $\mu_0=-\alpha$ by
$-S_{\mu\mu}(\lambda_0,-\alpha)$ and integrating over $\lambda_0$ in
$D$.  It is easily verified that this procedure correctly represents
the boundary values with singular solutions. The final result for the
general solution of the Jeans equations
\eqref{eq:jeanselliptic_alternativeS} for St\"ackel discs, after using
the evaluations \eqref{eq:homogeneoussolutions}, is \looseness=-1
\begin{subequations}
  \label{eq:gensolutions}
  \begin{multline}
    \label{eq:gensolSll}
    S_{\lambda\lambda}(\lambda,\mu) = 
    -\int\limits_\lambda^\infty \hspace{-4pt} \d\lambda_0 
    \; g_1(\lambda_0,\mu) \\
    +\int\limits_\lambda^\infty \hspace{-4pt} \d\lambda_0 
    \hspace{-5pt} \int\limits_\mu^{-\alpha} \hspace{-4pt} \d\mu_0
    \biggl[
    -g_1(\lambda_0,\mu_0)\frac{2wE'(w)}{\pi(\mu_0\!-\!\mu)} 
    \!+\!g_2(\lambda_0,\mu_0)\frac{E(w)}{\pi(\lambda_0\!-\!\mu)}
    \biggr] \\ 
    - \int\limits_\lambda^\infty \hspace{-4pt} \d\lambda_0 
    \, S_{\mu\mu}(\lambda_0,-\alpha) \, 
    \biggl[ \frac{E(w)}{\pi(\lambda_0\!-\!\mu)} \hspace{-10pt}
    \underset{\mu_0=-\alpha}{\biggr]} 
    \hspace{-10pt},
    \hspace{66pt} 
  \end{multline}
  \vspace{-10pt}
  \begin{multline}
    \label{eq:gensolSmm}
    S_{\mu\mu}(\lambda,\mu) = 
    -\int\limits_\mu^{-\alpha} \hspace{-4pt} \d\mu_0
    \; g_2(\lambda,\mu_0) \\ 
    +\int\limits_\lambda^\infty \hspace{-4pt} \d\lambda_0 
    \hspace{-5pt} \int\limits_\mu^{-\alpha} \hspace{-4pt} \d\mu_0
    \biggl[
    -g_1(\lambda_0,\mu_0)\frac{E(w)}{\pi(\lambda\!-\!\mu_0)} 
    \!-\!g_2(\lambda_0,\mu_0)\frac{2wE'(w)}{\pi(\lambda_0\!-\!\lambda)}
    \biggr] \\ 
    + S_{\mu\mu}(\lambda,-\alpha)
    - \int\limits_\lambda^\infty \hspace{-4pt} \d\lambda_0
    \, S_{\mu\mu}(\lambda_0,-\alpha) 
    \biggl[-\frac{2wE'(w)}{\pi(\lambda_0\!-\!\lambda)} 
    \hspace{-10pt} \underset{\mu_0=-\alpha}{\biggr]} 
    \hspace{-10pt}.
  \end{multline} 
\end{subequations}
The terms $(\mu_0\!-\!\mu)^{-1}$ and $(\lambda_0\!-\!\lambda)^{-1}$ do
not cause singularities because they are canceled by components of
$w$. In order to show that equations \eqref{eq:gensolutions} are equivalent
to the solution \eqref{eq:solutionTllandTmm} given by Riemann's
method, integrate the terms in $E'(w)$ by parts, and use the
definitions of $S_{\tau\tau}$, $g_1$ and $g_2$.

\subsubsection{Convergence of the disc solution}
\label{sec:convdiscsol}

We now return to the convergence issues first discussed in
\S\ref{sec:checkingforconsistency}, where we assumed that the density 
$\rho$ decays as $N(\mu)\lambda^{-s/2}$ at large distances 
and the St\"ackel potential as $\mathcal{O}(\lambda^\delta)$.
For the physical reasons given there, the assigned 
boundary stress $T_{\mu\mu}(\lambda,-\alpha)$ cannot exceed
$\mathcal{O}(\lambda^{\delta-s/2})$ at large $\lambda$, giving an  
$S_{\mu\mu}(\lambda,-\alpha)$ of $\mathcal{O}(\lambda^{\delta-s/2+1/2})$. 
It follows that the infinite integrals in $S_{\mu\mu}(\lambda_0,-\alpha)$ 
in the solution \eqref{eq:gensolutions} require only that $s>2\delta+1$ 
for their convergence. This is the less restrictive result to which 
we referred earlier. \looseness=-2

The terms in the boundary stress are seen to contribute terms of the
correct order $\mathcal{O}(\lambda^{\delta-s/2+1/2})$ to
$S_{\lambda\lambda}(\lambda,\mu)$ and $S_{\mu\mu}(\lambda,\mu)$.  
The formulas for the density and potential show that
$g_1(\lambda,\mu)=\mathcal{O}(\lambda^{\delta-s/2-1/2})$ while
$g_2(\lambda,\mu)$ is larger and
$\mathcal{O}(\lambda^{-s/2-1/2})$ as $\lambda \to \infty$.  The
$\lambda_0$ integrations with $g_1$ and $g_2$ in their integrands all
converge provided $s>2\delta+1$. Hence, both
$S_{\lambda\lambda}(\lambda,\mu)$ and $S_{\mu\mu}(\lambda,\mu)$ are
$\mathcal{O}(\lambda^{\delta-s/2+1/2})$, so that the stress components
$T_{\tau\tau}(\lambda,\mu)$ ($\tau=\lambda,\mu$) are
$\mathcal{O}(\lambda^{\delta-s/2})$, 
which is consistent with the physical reasoning of
\S\ref{sec:checkingforconsistency}. 

Hence, all the conditions necessary for \eqref{eq:gensolutions} to be
a valid solution of the Jeans equations
\eqref{eq:jeanselliptic_alternativeS} for a St\"ackel disc are
satisfied provided that $s>2\delta+1$. We have seen in
\S\ref{sec:checkingforconsistency} that $\delta$ must lie in the range
$[-\frac12,0)$. When $\delta\to 0$ the models approach the isothermal
disk, for which also $s=1$ when the density is consistent with the
potential. Only then our requirement $s>2\delta+1$ is violated.

\subsection{Alternative boundary conditions}
\label{sec:alternativeBCsdisc}

We now derive the alternative form of the general disc solution when
the boundary conditions are not specified on $\mu=-\alpha$ but on
$\mu=-\beta$, or on $\lambda=-\alpha$ rather than in the limit
$\lambda\to\infty$. While the former switch is straightforward, the
latter is non-trivial, and leads to non-physical solutions.

\subsubsection{Boundary condition for $\mu$}
\label{sec:switchingoneboundarycond}

The analysis in \S\ref{sec:riemannsmethod} and
\S\ref{sec:singularsolution2D} is that needed when the boundary
conditions are imposed at large $\lambda$ and at $\mu=-\alpha$.  The
Jeans equations \eqref{eq:jeanselliptic} can be solved in a similar
way when one or both of those conditions are imposed instead at the
opposite boundaries $\lambda=-\alpha$ and/or $\mu=-\beta$. The
solution by Riemann's method is accomplished by applying Green's
theorem to a different domain, for example
$D^{\prime}=\{(\lambda_0,\mu_0)$: $\lambda\le\lambda_0\le\infty,
-\beta\le\mu_0\le\mu\}$ when the boundary conditions are at
$\mu=-\beta$ and as $\lambda \to \infty$.  The Riemann--Green
functions have to satisfy the same PDE \eqref{eq:1stconditionv} and
the same boundary conditions \eqref{eq:2ndconditionv} and
\eqref{eq:3rdconditionv}, and so again are given by equations
\eqref{eq:vforTlambdalambda} and \eqref{eq:vforTmumu}. The variable
$w$ is negative in $D^{\prime}$ instead of positive as in $D$, but
this is unimportant. The only significant difference in the solution
of eq.~\eqref{eq:general2ndorderPDE} is that of a sign due to changes
in the limits of the line integrals.  The final result, in place of
eq.~\eqref{eq:solutionstressT}, is
\begin{multline}
  \label{eq:solutionstressTb}
  T(\lambda,\mu) = 
  - \int\limits_\lambda^\infty \hspace{-4pt} \d\lambda_0 
  \hspace{-4pt} \int\limits_{-\beta}^\mu \hspace{-5pt} \d\mu_0
  \mathcal{G}(\lambda_0,\mu_0)\, U(\lambda_0,\mu_0)
  \\*[-10pt]
  - \int\limits_\lambda^\infty \hspace{-4pt} \d\lambda_0
  \Bigl[ \Bigl( \frac{\partial T}{\partial \lambda_0} + 
  \frac{ c_2\,T}{\lambda_0\!-\!\mu_0} \Bigr) \mathcal{G} 
  \hspace{-10pt} \underset{\mu_0=-\beta}{\Bigr]} \hspace{-5pt}.
\end{multline}
To apply the method of singular solutions to solve for the stresses
when the boundary stresses are specified at $\mu=-\beta$ rather than at
$\mu=-\alpha$, we modify the singular solutions 
\eqref{eq:functionalformsSllandSmm} by replacing the step-function 
$\mathcal{H}(\mu_0\!-\!\mu)$ by $-\mathcal{H}(\mu\!-\!\mu_0)$ throughout.
No other change is needed because both functions give $-\delta(\mu-\mu_0)$
on partial differentiation with respect to $\mu$.
The two-dimensional problem for $A$ and $B$ remains the same, and so,
as with Riemann's method, its solution remains the same.
Summing over sources in $D^{\prime}$ now gives
\begin{subequations}
  \label{eq:gensolutionsb}
  \begin{multline}
    \label{eq:gensolSllb}
    S_{\lambda\lambda}(\lambda,\mu) = 
    -\int\limits_\lambda^\infty \hspace{-4pt} \d\lambda_0 
    \; g_1(\lambda_0,\mu) \\
    -\int\limits_\lambda^\infty \hspace{-4pt} \d\lambda_0 
    \hspace{-5pt} \int\limits_{-\beta}^\mu \hspace{-4pt} \d\mu_0 
    \biggl[
    -g_1(\lambda_0,\mu_0)\frac{2wE'(w)}{\pi(\mu_0\!-\!\mu)} 
    +g_2(\lambda_0,\mu_0)\frac{E(w)}{\pi(\lambda_0\!-\!\mu)}
    \biggr] \\ 
    - \int\limits_\lambda^\infty \hspace{-4pt} \d\lambda_0 
    \, S_{\mu\mu}(\lambda_0,-\beta) \, 
    \biggl[ \frac{E(w)}{\pi(\lambda_0\!-\!\mu)} \hspace{-10pt}
    \underset{\mu_0=-\beta}{\biggr]} 
    \hspace{-10pt},
    \hspace{66pt} 
  \end{multline}
  \vspace{-10pt}
  \begin{multline}
    \label{eq:gensolSmmb}
    S_{\mu\mu}(\lambda,\mu) = 
    \int\limits_{-\beta}^\mu \hspace{-4pt} \d\mu_0
    \; g_2(\lambda,\mu_0) \\ 
    -\int\limits_\lambda^\infty \hspace{-4pt} \d\lambda_0 
    \hspace{-5pt} \int\limits_{-\beta}^\mu \hspace{-4pt} \d\mu_0
    \biggl[
    -g_1(\lambda_0,\mu_0)\frac{E(w)}{\pi(\lambda\!-\!\mu_0)} 
    -g_2(\lambda_0,\mu_0)\frac{2wE'(w)}{\pi(\lambda_0\!-\!\lambda)}
    \biggr] \\ 
    + S_{\mu\mu}(\lambda,-\beta)
    - \int\limits_\lambda^\infty \hspace{-4pt} \d\lambda_0
    \, S_{\mu\mu}(\lambda_0,-\beta) 
    \biggl[-\frac{2wE'(w)}{\pi(\lambda_0\!-\!\lambda)} 
    \hspace{-10pt} \underset{\mu_0=-\beta}{\biggr]} 
    \hspace{-10pt}.
    \hspace{2pt} 
  \end{multline} 
\end{subequations}
as an alternative to equations \eqref{eq:gensolutions}.

\subsubsection{Boundary condition for $\lambda$}
\label{sec:boundarycondlargedist}

There is a much more significant difference when one assigns boundary
values at $\lambda=-\alpha$ rather than at $\lambda \to \infty$. 
It is still necessary that stresses decay to zero at large distances.
The stresses induced by arbitrary boundary data at the finite
boundary $\lambda=-\alpha$ do decay to zero as a consequence of
geometric divergence. The issue is that of the rate of this decay.
We find that it is generally less than that required by our analysis in
\S\ref{sec:checkingforconsistency}.

To isolate the effect of boundary data at $\lambda=-\alpha$,
we study solutions of the two-dimensional Jeans equations
\eqref{eq:jeanselliptic} when the inhomogeneous right hand side
terms are set to zero and homogeneous boundary conditions of
zero stress are applied at either $\mu=-\alpha$ or $\mu=-\beta$.
These solutions can be derived either by Riemann's method or
by singular solutions. The solution of the homogeneous PDE
$\mathcal{L} T =0$ is
\begin{equation}
  \label{eq:homsolutionstressTa}
  T(\lambda,\mu) = \hspace{-3pt} -
  \hspace{-3pt} \int\limits_{\mu}^{-\alpha} \hspace{-4pt} \d\mu_0
  \Bigl[ \Bigl( \frac{\partial T}{\partial \mu_0} \!-\!
  \frac{ c_1\,T}{\lambda_0\!-\!\mu_0} \Bigr) 
  \mathcal{G}(\lambda,\mu;\lambda_0,\mu_0)
  \hspace{-10pt} \underset{\lambda_0=-\alpha}{\Bigr]} \hspace{-5pt},
\end{equation}
for the case of zero stress at $\mu=-\alpha$, and
\begin{equation}
  \label{eq:homsolutionstressTb}
  T(\lambda,\mu) = \hspace{-3pt} 
  \hspace{-3pt} \int\limits^{\mu}_{-\beta} \hspace{-4pt} \d\mu_0
  \Bigl[ \Bigl( \frac{\partial T}{\partial \mu_0} \!-\!
  \frac{ c_1\,T}{\lambda_0\!-\!\mu_0} \Bigr) 
                            \mathcal{G}(\lambda,\mu;\lambda_0,\mu_0)
  \hspace{-10pt} \underset{\lambda_0=-\alpha}{\Bigr]} \hspace{-5pt},
\end{equation}
for the case of zero stress at $\mu=-\beta$.

The behaviour of the stresses at large distances is governed by the
behaviour of the Riemann--Green functions $\mathcal{G}$ for distant
field points $(\lambda,\mu)$ and source points at $\lambda_0=-\alpha$.
It follows from equations \eqref{eq:riemannfncsforTllandTmm} that
$T_{\lambda\lambda}(\lambda,\mu)=\mathcal{O}(\lambda^{-1/2})$ and
$T_{\mu\mu}(\lambda,\mu)=\mathcal{O}(\lambda^{-3/2})$.  
As a restult, the radial stresses dominate at large distances and they
decay as only the inverse first power of distance.  
Their rate of decay is less than $\mathcal{O}(\lambda^{\delta-s/2})$
-- obtained in \S\ref{sec:checkingforconsistency} from physical
arguments -- if the requirement $s>2\delta +1$ is satisfied. 
This inequality is the necessary condition which we derived in
\S\ref{sec:generalsolution2D} for \eqref{eq:gensolutions} to be a
valid solution of the disc Jeans equations
\eqref{eq:jeanselliptic_alternativeS}. 
It is violated in the isothermal limit.

There is a physical implication of radial stresses which decay as only
the inverse first power of distance. It implies that net forces of
finite magnitude are needed at an outer boundary to maintain the
system, the finite magnitudes arising from the product of the decaying
radial stresses and the increasing length of the boundary over which
they act. That length grows as the first power of distance.  
Because this situation is perhaps more naturally understood in three
dimensions, we return to it in our discussion of oblate models in
\S\ref{sec:oblatemodels}.  For now, lacking any physical reason for
allowing a stellar system to have such an external constraint, we
conclude that boundary conditions can be applied only at large
$\lambda$ and not at $\lambda=-\alpha$.

\subsubsection{Disc solution for a general finite region}
\label{sec:iscsolforgenfinreg}

We now apply the singular solution method to solve equations
\eqref{eq:jeanselliptic_alternativeS} in some rectangle $\mu_{{\rm 
min}} \leq \mu \leq \mu_{{\rm max}}$, $\lambda_{{\rm min}} \leq
\lambda \leq \lambda_{{\rm max}}$, when the stress $S_{\mu\mu}$ is
given a boundary in $\mu$, and $S_{\lambda\lambda}$ is given on a
boundary in $\lambda$.  This solution includes 
\eqref{eq:gensolutions} and \eqref{eq:gensolutionsb} as special
cases. It will be needed for the large-radii scale-free case of
\S\ref{sec:scalefreeDF}.\looseness=-2

As we saw in \S\ref{sec:switchingoneboundarycond}, singular solutions
can easily be adapted to alternative choices for the domain of
dependence of a field point $(\lambda,\mu)$. Originally this was $D$,
the first of the four quadrants into which $(\lambda_0,\mu_0)$-space
is split by the lines $\lambda_0=\lambda$ and $\mu_0=\mu$
(Fig. \ref{fig:pointmasscontribution}). It has the
singular solution \eqref{eq:functionalformsSllandSmm}. We then
obtained the singular solution for the fourth quadrant $D^{\prime}$
simply by replacing $\mathcal{H}(\mu_0\!-\!\mu)$ by
$-\mathcal{H}(\mu\!-\!\mu_0)$ in \eqref{eq:functionalformsSllandSmm}.
We can similarly find the singular solution for the second quadrant
$\lambda_{{\rm min}} \leq \lambda_0 \leq \lambda$, $\mu \leq \mu_0
\leq \mu_{{\rm max}}$ by replacing
$\mathcal{H}(\lambda_0\!-\!\lambda)$ by
$-\mathcal{H}(\lambda\!-\!\lambda_0)$, and for the third quadrant
$\lambda_{{\rm min}} \leq \lambda_0 \leq \lambda$, $\mu_{{\rm min}}
\leq \mu_0 \leq \mu$ by replacing $\mathcal{H}(\lambda_0\!-\!\lambda)$
by $-\mathcal{H}(\lambda\!-\!\lambda_0)$ {\it and}
$\mathcal{H}(\mu_0\!-\!\mu)$ by $-\mathcal{H}(\mu\!-\!\mu_0)$.  We
find the part of the solution of equations
\eqref{eq:jeanselliptic_alternativeS} due to the right hand side $g$
terms by multiplying the first and second terms of the singular
solutions by $g_1(\lambda_0,\mu_0)$ and $g_2(\lambda_0,\mu_0)$,
respectively, and integrating over the relevant domain.  We use
$\lambda=\lambda_e$ and $\mu=\mu_e$ to denote the boundaries at which
stresses are specified. We find the part of the solution generated by
the boundary values of $S_{\mu\mu}$ by multiplying the singular
solution \eqref{eq:functionalformsSllandSmm}, modified for the domain
and evaluated at $\mu_0=\mu_e$, by $\pm S_{\mu\mu}(\lambda_0,\mu_e)$
and integrating over $\lambda_0$ in the domain. The plus sign is
needed when $\mu_e=\mu_{{\rm min}}$ and the minus when
$\mu_e=\mu_{{\rm max}}$. Similarly, the part of the solution generated
by the boundary values of $S_{\lambda\lambda}$ is obtained by
multiplying the singular solution
\eqref{eq:functionalformsSllandSmmswitch}, modified for the domain and
evaluated at $\lambda_0=\lambda_e$, by $\pm
S_{\lambda\lambda}(\lambda_e,\mu_0)$ and integrating over $\mu_0$ in
the domain. The sign is plus if $\lambda_e=\lambda_{{\rm min}}$ and
minus if $\lambda_e=\lambda_{{\rm max}}$. The final solution is 
\begin{subequations}
  \label{eq:gendiscsolfourdom}
  \begin{multline}
    \label{eq:gendiscsolSll}
    S_{\lambda\lambda}(\lambda,\mu) =
    S_{\lambda\lambda}(\lambda_e,\mu)
    -\! \int\limits_\lambda^{\lambda_e} \hspace{-4pt} \d\lambda_0
    g_1(\lambda_0,\mu) \\
    +\!\!\int\limits_\lambda^{\lambda_e} \hspace{-4pt} \d\lambda_0  
    \!\!\int\limits_\mu^{\mu_e} \hspace{-4pt} \d\mu_0 
    \left[ 
      g_1(\lambda_0,\!\mu_0\!)B(\mu,\!\lambda;\!\mu_0,\!\lambda_0\!)  
      \!+\!
      g_2(\lambda_0,\!\mu_0\!)A(\lambda,\!\mu;\!\lambda_0,\!\mu_0\!) 
    \right] \\
    -\!\! \int\limits_\lambda^{\lambda_e} \hspace{-4pt} \d\lambda_0 
    S_{\mu\mu}(\!\lambda_0,\!\mu_e)
    A(\lambda,\!\mu;\!\lambda_0,\!\mu_e\!)
    -\!\! \int\limits_\mu^{\mu_e} \hspace{-4pt} \d\mu_0
    S_{\lambda\lambda}(\!\lambda_e,\!\mu_0) 
    B(\mu,\!\lambda;\!\mu_0,\!\lambda_e\!),\\*[-10pt]\ 
  \end{multline}
  \vspace{-20pt}
  \begin{multline}
    \label{eq:gendiscsolSmm}
    S_{\mu\mu}(\lambda,\mu)=
    S_{\mu\mu}(\lambda,\mu_e)
    - \int\limits_\mu^{\mu_e} \hspace{-4pt} \d\mu_0 
    g_2(\lambda,\mu_0) \\
    +\!\!\int\limits_\lambda^{\lambda_e} \hspace{-4pt} \d\lambda_0
    \!\! \int\limits_\mu^{\mu_e} \hspace{-4pt} \d\mu_0
    \left[
      g_1(\lambda_0,\!\mu_0\!)A(\mu,\!\lambda;\!\mu_0,\!\lambda_0\!)
      \!+\!
      g_2(\lambda_0,\!\mu_0\!)B(\lambda,\!\mu;\!\lambda_0,\!\mu_0\!) 
    \right] \\
    -\!\! \int\limits_\lambda^{\lambda_e} \hspace{-4pt} \d\lambda_0
    S_{\mu\mu}(\!\lambda_0,\!\mu_e\!) 
    B(\lambda,\!\mu;\!\lambda_0,\!\mu_e\!)
    -\!\! \int\limits_\mu^{\mu_e} \hspace{-4pt} \d\mu_0
    S_{\lambda\lambda}(\!\lambda_e,\!\mu_0\!) \,
    A(\mu,\!\lambda;\!\mu_0,\!\lambda_e\!). \\*[-10pt]\
  \end{multline}
\end{subequations}
This solution is uniquely determined once $g_1$ and $g_2$ are given, 
and the boundary values $S_{\mu\mu}(\lambda_0,\mu_e)$ and
$S_{\lambda\lambda}(\lambda_e,\mu_0)$ are prescribed.  It shows that
the hyperbolic equations \eqref{eq:jeanselliptic_alternativeS} can
equally well be integrated in either direction in the characteristic
variables $\lambda$ and $\mu$.  Solutions \eqref{eq:gensolutions} and
\eqref{eq:gensolutionsb} are obtained by taking $\lambda_e \to
\infty$, $S_{\lambda\lambda}(\lambda_e,\mu_0) \to 0$, setting
$\mu_e=-\alpha$ and $\mu_e=-\beta$ respectively, and evaluating $A$
and $B$ by equations \eqref{eq:homogeneoussolutions}.

\subsection{Applying the disc solution to limiting cases}
\label{sec:applicationdiscsol}
 
We showed in \S\ref{sec:2Dcasessimilar} that the Jeans equations for
prolate and oblate potentials and for three-dimensional St\"ackel
models with a scale-free DF all reduce to a set of two equations
equivalent to those for the St\"ackel disc. Here we apply our solution
for the St\"ackel disc to these special three-dimensional cases, with
particular attention to the behaviour at large radii and the boundary
conditions. This provides further insight in some of the previously
published solutions. We also consider the case of a St\"ackel disc
built with thin tube orbits.

\subsubsection{Prolate potentials}
\label{sec:prolatemodels}

We can apply the disc solution \eqref{eq:gensolutionsb} to solve the
Jeans equations \eqref{eq:prolateJeanseqnsrewritten} by setting
$S_{\lambda\lambda}(\lambda,\mu) = |\lambda-\mu|^\frac{1}{2}
\mathcal{T}_{\lambda\lambda}(\lambda,\mu)$ and 
$S_{\mu\mu}(\lambda,\mu) = |\mu-\lambda|^\frac{1}{2}
\mathcal{T}_{\mu\mu} (\lambda,\mu)$, and taking  
\begin{eqnarray}
 \label{eq:gforprolates}
  g_1(\lambda,\mu) \hspace{-10pt} & = & \hspace{-10pt}
  - |\lambda\!-\!\mu|^\frac{1}{2}(\lambda\!+\!\beta)^\frac{1}{2} 
    (\mu\!+\!\beta)^\frac{1}{2}
  \biggl[
  \rho\frac{\partial V_S}{\partial \lambda} \!+\!
  \frac{\partial T_{\chi\chi}}{\partial \lambda}
  \! \biggr], \nonumber \\*[-5pt]  \\*[-5pt]
  g_2(\lambda,\mu)  \hspace{-10pt} & = & \hspace{-10pt}
  - |\mu\!-\!\lambda|^\frac{1}{2}(\lambda\!+\!\beta)^\frac{1}{2} 
    (\mu\!+\!\beta)^\frac{1}{2}
  \biggl[
  \rho\frac{\partial V_S}{\partial \mu} \!+\! \frac{\partial
  T_{\chi\chi}}{\partial \mu}
  \! \biggr]. \nonumber
\end{eqnarray}
The boundary terms in $S_{\mu\mu}(\lambda,-\beta)$ vanish because of
the boundary condition \eqref{eq:prolatebc}.  As before, we regard the
azimuthal stress $T_{\chi\chi}$ as a variable that can be arbitrarily
assigned, provided that it has the correct behaviour at large
$\lambda$ (\S\ref{sec:checkingforconsistency}). 
The choice of $T_{\chi\chi}$ is also restricted by
the requirement that the resulting solutions for the stresses
$T_{\lambda\lambda}$ and $T_{\mu\mu}$ must be non-negative (see
\S\ref{sec:jeanseqns}).

The analysis needed to show that the solution obtained in this way
is valid requires only minor modifications of that of 
\S\ref{sec:convdiscsol}. We suppose that the prescribed azimuthal 
stresses also decay as $\mathcal{O}(\lambda^{\delta-s/2})$ as
$\lambda \to \infty$. As a result of the extra factor in the 
definitions \eqref{eq:gforprolates}, we now have
$g_1(\lambda,\mu)=\mathcal{O}(\lambda^{\delta-s/2})$
and $g_2(\lambda,\mu)=\mathcal{O}(\lambda^{-s/2})$ as
$\lambda \to \infty$. The $\lambda_0$ integrations
converge provided $s>2\delta+2$, and $S_{\lambda\lambda}$ 
and $S_{\mu\mu}$ are $\mathcal{O}(\lambda^{\delta-s/2+1})$. 
Hence the stresses $T_{\lambda\lambda}$ and $T_{\mu\mu}$, 
which follow from
$T_{\tau\tau} = T_{\chi\chi} + S_{\tau\tau}/\sqrt{ (\lambda\!-\!\mu) 
  (\lambda\!+\!\beta) (\mu\!+\!\beta) }$, are once again 
$\mathcal{O}(\lambda^{\delta-s/2})$. 
The requirement $s>2\delta+2$ is no stronger than the requirement
$s>2\delta+1$ of \S\ref{sec:convdiscsol}; it is simply the
three-dimensional version of that requirement. It also does not break
down until the isothermal limit. That limit is still $\delta \to 0$,
but now $s \to 2$.

\subsubsection{Oblate potentials}
\label{sec:oblatemodels}

The oblate case with Jeans equations
\eqref{eq:oblateJeanseqnsrewritten} differs significantly from the
prolate case. 
Now $S_{\lambda\lambda}(\lambda,\nu) = |\lambda-\nu|^\frac{1}{2} 
\mathcal{T}_{\lambda\lambda}(\lambda,\nu)$ 
vanishes at $\lambda=-\alpha$ and 
$S_{\nu\nu}(\lambda,\nu) = |\nu-\lambda|^\frac{1}{2} 
\mathcal{T}_{\nu\nu}(\lambda,\nu)$
vanishes at $\nu=-\alpha$.  
If one again supposes that the azimuthal stresses $T_{\phi\phi}$ can
be assigned initially, then one encounters the problem discussed in
\S\ref{sec:boundarycondlargedist} of excessively large radial stresses
at large distances.   
To relate that analysis to the present case, we use the solution
\eqref{eq:gensolutions} with $\mu$ replaced by $\nu$, and with zero
boundary value $S_{\nu\nu}(\lambda,-\alpha)$, and for $g_1$ and $g_2$
the right hand side of \eqref{eq:oblateJeanseqnsrewritten} multiplied
by $|\lambda-\nu|^\frac{1}{2}$ and $|\nu-\lambda|^\frac{1}{2}$,
respectively. 

The estimates we obtained for the prolate case are still valid, so the
stresses $T_{\lambda\lambda}$ and $T_{\nu\nu}$ are
$\mathcal{O}(\lambda^{\delta-s/2})$. Difficulties arise when this
solution for $S_{\lambda\lambda}$ does not vanish at
$\lambda=-\alpha$, but instead has some nonzero value $\kappa(\nu)$
there. To obtain a physically acceptable solution, we must add to it a
solution of the homogeneous equations
\eqref{eq:oblateJeanseqnsrewritten} with boundary values
$\mathcal{T}_{\lambda\lambda}(-\alpha,\nu) =
-\kappa(\nu)/\sqrt{-\alpha-\nu}$ and
$\mathcal{T}_{\nu\nu}(\lambda,-\alpha)=0$.  This is precisely the
problem we discussed in \S\ref{sec:boundarycondlargedist} where we showed
that the resulting solution gives $\mathcal{T}_{\lambda\lambda}
(\lambda,\mu)=\mathcal{O}(\lambda^{-1/2})$, and hence
$T_{\lambda\lambda}(\lambda,\mu)=\mathcal{O}(\lambda^{-1})$.  This is
larger than $\mathcal{O}(\lambda^{\delta-s/2})$ when the
three-dimensional requirement $s>2\delta+2$ is met.  We therefore
conclude that the approach in which one first selects the azimuthal
stress $T_{\phi\phi}$ and then calculates the other two stresses will
be unsuccessful unless the choice of $T_{\phi\phi}$ is fortunate, and
leads to $\kappa(\nu) \equiv 0$.  Otherwise, it leads only to models
which either violate the continuity condition $T_{\lambda\lambda}
-T_{\phi\phi}=0$ at $\lambda=-\alpha$, or else have radial stresses
which require external forces at large distances.

The physical implication of radial stresses which decay as only
$\mathcal{O}(\lambda^{-1})$, or the inverse second power of distance,
is that net forces of finite magnitude are needed at an outer boundary
to maintain the system. This finite magnitude arises from the product
of the decaying radial stresses and the increasing surface area of the
boundary over which they act, which grows as the second power of
distance. This situation is analogous to that of an isothermal sphere,
as illustrated in problem 4--9 of Binney \& Tremaine
(1987\nocite{1987gady.book.....B}), for which the contribution 
from an outer surface integral must be taken into account in the
balance between energies required by the virial theorem.

There are, of course, many physical models which satisfy the
continuity condition and whose radial stresses decay in the physically
correct manner at large distances, but some strategy other than that
of assigning $T_{\phi\phi}$ initially is needed to find them. In fact,
only Evans (1990\nocite{evans1990numericalmethod}) used the approach of
assigning $T_{\phi\phi}$ initially. He computed a numerical solution
for a mass model with $s=3$ and $V_S \propto
\mathcal{O}(\lambda^{-1/2}\ln\lambda)$ for large $\lambda$, so that the
stresses there should be $\mathcal{O}(\lambda^{-2}\ln\lambda)$.  He
set $T_{\phi\phi}=-\frac{1}{3}\rho V_S$, which is of this magnitude,
and integrated from $\lambda=-\alpha$ in the direction of increasing
$\lambda$ for a finite range. Evans does not report on the large
$\lambda$ behaviour, and it is possible that his choice of
$T_{\phi\phi}$ gives $\kappa(\nu)=0$, but his Figure 2 especially
shows velocity ellipsoids which become increasingly elongated in the
radial direction, consistent with our prediction that
$T_{\lambda\lambda}$ generally grows as $\mathcal{O}(\lambda^{-1})$
when the boundary value of $T_{\lambda\lambda}$ is assigned at
$\lambda=-\alpha$.

A more common and effective approach to solve the Jeans equations
for oblate models has been to specify the ratio
$T_{\lambda\lambda}/T_{\nu\nu}$, and then to solve for one of those
stresses and $T_{\phi\phi}$ 
(Bacon, Simien \& Monnet 1983\nocite{1983A&A...128..405B}; 
Dejonghe \& de Zeeuw 1988\nocite{1988ApJ...333...90D}; 
Evans \& Lynden--Bell 1991\nocite{1991MNRAS.251..213E}; 
Arnold 1995\nocite{1995MNRAS.276..293A}).
This leads
to a much simpler mathematical problem with just a single first-order
PDE. The characteristics of that PDE have non-negative slopes
$d\lambda/d\nu$, and therefore cut across the coordinate lines of constant
$\lambda$ and $\nu$.  The solution is obtained by integrating in along
the characteristics from large $\lambda$. The continuity conditions
\eqref{eq:oblatecontcond} are taken care of automatically, the region
$-\gamma \le \nu \le -\alpha \le \infty$ is covered, and it is easy to
verify that the stresses so obtained are everywhere positive.

\subsubsection{Large radii limit with scale-free DF}
\label{sec:scalefreeDF}

We found in \S\ref{sec:conicalcoord} that the first of the Jeans
equations in conical coordinates \eqref{eq:jeansconical} reduces to an
algebraic relation for the radial stress $T_{rr}$. The problem that
remains is that of solving the second and third Jeans equations for
$T_{\mu\mu}$ and $T_{\nu\nu}$. Those equations are exactly the same as
those of the disc case after we apply the coordinate permutation
$\lambda \to \mu \to \nu$, and the physical domain is $-\gamma \le \nu
\le -\beta \le \mu \le -\alpha$ with finite ranges of both
variables. Hence, the solution \eqref{eq:gendiscsolfourdom} can be
applied with $T_{\mu\mu}$ assigned at either $\mu_e=-\alpha$ or
$\mu_e=-\beta$, and $T_{\nu\nu}$ at either $\nu_e =-\beta$ or $\nu_e
=-\gamma$.  For $g_1$ and $g_2$ we take the same expressions as for
the disc case, i.e., the right-hand side of
\eqref{eq:jeanselliptic_alternativeS}, but with $\lambda \to \mu \to
\nu$ and multiplied by $r^\zeta$. To obtain $T_{\mu\mu}$ and
$T_{\nu\nu}$ from the $S_{\lambda\lambda}$ and $S_{\mu\mu}$
respectively, we use the transformation
\begin{equation}
  \label{eq:trafoSttconical}
  S_{\tau\tau} = (\mu\!-\!\nu)^\frac{1}{2} \, r^\zeta
  T_{\tau\tau}, \quad \tau=\mu,\nu,
\end{equation}
with $\zeta>0$ the scaling factor. We can choose to specify the stress
components on the two boundaries $\mu=-\beta$ and $\nu=-\beta$. For a
given radius $r$ these boundaries cover the circular cross section
with the $(x,z)$-plane (Fig.~\ref{fig:conicalcoordinatecurves}).
We can consider the $(x,z)$-plane as the starting space for the
solution. It turns out that the latter also applies to the triaxial
solution (\S\ref{sec:alternativeBCstriaxial}) and compares well with
Schwarzschild (1993\nocite{1993ApJ...409..563S}), who used the same
plane to start his numerically calculated orbits from.

\subsubsection{Thin tube orbits}
\label{sec:thinlooporbits}

For infinitesimally thin tube orbits in St\"ackel discs we have that
$S_{\lambda\lambda} \equiv 0$ (\S\ref{sec:thintubeorbits}),
so that equations \eqref{eq:jeanselliptic_alternativeS} reduce to 
\begin{equation}
  \label{eq:thinloopjeanseqns}
  -\frac{S_{\mu\mu}}{2(\lambda\!-\!\mu)} = g_1(\lambda,\mu), 
  \quad
  \frac{\partial S_{\mu\mu}}{\partial \mu} = g_2(\lambda,\mu). 
\end{equation}
A solution is possible only if the right hand side terms satisfy the
subsidiary equation
\begin{equation}
  \label{eq:thinloopeqg1andg2}
  g_2(\lambda,\mu) = -2 \frac{\partial}{\partial \mu} 
  \left[ 
    (\lambda\!-\!\mu) g_1(\lambda,\mu)
  \right].
\end{equation}
We find below that this equation places restrictions on the form of
the (surface) density $\rho$, and we use this relation between $g_1$
and $g_2$ to show that the disc solution \eqref{eq:gensolutions}
yields the right results for the stress components. 

If we write the disc potential \eqref{eq:discpot} as a divided
difference, $V_S=-f[\lambda,\mu]$, we have that 
\begin{equation}
  \label{eq:thinloopexprg1andg2}
  g_1 = (\lambda\!-\!\mu)^\frac{1}{2} \rho f[\lambda,\lambda,\mu], 
  \quad
  g_2 = (\lambda\!-\!\mu)^\frac{1}{2} \rho f[\lambda,\mu,\mu].
\end{equation}
Upon substitution of these expressions in
\eqref{eq:thinloopeqg1andg2} we obtain a PDE in $\mu$, of which the
solution implies the following form for the density 
\begin{equation}
  \label{eq:thinlooprhosol}
  \rho(\lambda,\mu) = \frac{ \tilde{f}(\lambda) }{ (\lambda\!-\!\mu)
  \sqrt{f[\lambda,\lambda,\mu]} },
\end{equation}
where $\tilde{f}(\lambda)$ is an arbitrary function independent of
$\mu$. From \eqref{eq:thinloopjeanseqns} and the definition
\eqref{eq:deffunctionsStt} it then follows that 
$T_{\mu\mu}(\lambda,\mu,\nu) 
= -2\tilde{f}(\lambda) \sqrt{f[\lambda,\lambda,\mu]}$.
The tube density that de Zeeuw, Hunter \& Schwarzschild
(1987\nocite{1987ApJ...317..607D}) derive from the DF for thin tube
orbits in the perfect elliptic disk (their eq.~[4.25]) is indeed of
the form \eqref{eq:thinlooprhosol}.  

To show that the general disc solution \eqref{eq:gensolutions} gives
$S_{\lambda\lambda}(\lambda,\mu)=0$, we substitute
eq.~\eqref{eq:thinloopeqg1andg2} for $g_2(\lambda,\mu)$ in
\eqref{eq:gensolSll}. 
After partial integration and using 
\begin{equation}
  \label{eq:thinloopeqdEdmu0_Ep}
  2(\lambda_0\!-\!\mu_0) \frac{\partial }{\partial \mu_0} 
  \frac{E(w)}{\pi(\lambda_0\!-\!\mu)} = 
  \frac{2wE'(w)}{\pi(\mu_0\!-\!\mu)},
\end{equation}
we find that the area integral reduces to
\begin{equation}
  \label{eq:thinloopleftareaintgSll}
  \int\limits_\lambda^\infty \hspace{-3pt} \d\lambda_0 
  \biggl\{
    g_1(\lambda_0,\mu) - 2(\lambda_0\!+\!\alpha)g_1(\lambda_0,-\alpha)
    \biggl[ \frac{E(w)}{\pi(\lambda_0\!-\!\mu)} \hspace{-10pt}
  \underset{\mu_0=-\alpha}{\biggr]} \hspace{-10pt}
  \biggr\}.
\end{equation}
The first part cancels the first line of \eqref{eq:gensolSll} and
since from \eqref{eq:thinloopjeanseqns} we have that
$-2(\lambda_0\!+\!\alpha)g_1(\lambda_0,-\alpha) =
S_{\mu\mu}(\lambda_0,-\alpha)$, the second part cancels the third
line.  
Hence, we have $S_{\lambda\lambda}(\lambda,\mu)=0$ as required. 
To see that the general disc solution also yields
$S_{\mu\mu}(\lambda,\mu)$ correctly, we apply similar steps to 
\eqref{eq:gensolSmm}, where we use the relation  
\begin{equation}
  \label{eq:thinloopeqdE'dmu0_E}
  -2(\lambda_0\!-\!\mu_0) \frac{\partial }{\partial \mu_0} 
  \frac{2wE'(w)}{\pi(\lambda_0\!-\!\lambda)} = 
  \frac{E(w)}{\pi(\lambda\!-\!\mu_0)}.
\end{equation}
We are finally left with 
\begin{equation}
  \label{eq:thinloopleftofSmmsol}
  S_{\mu\mu}(\lambda,\mu) = S_{\mu\mu}(\lambda,-\alpha) 
  - \int\limits_\mu^{-\alpha} \hspace{-3pt} \d\mu_0 
  g_2(\lambda,\mu_0),
\end{equation}
which is just the second equation of \eqref{eq:thinloopjeanseqns} 
integrated with respect to $\mu$. 

% ===== section 4 ===== %

\section{The general case}
\label{sec:generalcase}

We now solve the system of three Jeans equations
\eqref{eq:jeanstriaxial} for triaxial St\"ackel models by applying the
singular solution superposition method, introduced in
\S\ref{sec:singularsolution2D} for the two-dimensional case.  
Although the calculations are more complex for a triaxial model, the
step-wise solution method is similar to that in two dimensions. 
Specifically, we first simplify the Jeans equations and show that they
reduce to a three-dimensional homogeneous boundary problem. 
We then find a two-parameter particular solution and apply contour
integration to both complex parameters to obtain the general
homogeneous solution. 
The latter yields the three singular solutions of the simplified Jeans
equations, from which, by superposition, we construct the general
solution.

\subsection{Simplified Jeans equations}
\label{sec:simplifiedjeanseqns3D}

We start by introducing the functions
\begin{equation}
  \label{eq:deffuncsStt}
  S_{\tau\tau}(\lambda,\mu,\nu) =
  \sqrt{ (\lambda\!-\!\mu) (\lambda\!-\!\nu) (\mu\!-\!\nu) } \,
  T_{\tau\tau}(\lambda,\mu,\nu),  
\end{equation}
with $\tau=\lambda,\mu,\nu$, to write the Jeans equations for triaxial
St\"ackel models \eqref{eq:jeanstriaxial} in the more convenient form 
\begin{subequations}
  \label{eq:jeanstriaxialS}
  \begin{equation}
    \label{eq:jeanstriaxial_lambdaS}
    \frac{\partial S_{\lambda\lambda}}{\partial \lambda} -
    \frac{S_{\mu\mu}}{2(\lambda\!-\!\mu)} -
    \frac{S_{\nu\nu}}{2(\lambda\!-\!\nu)} = g_1(\lambda,\mu,\nu), 
  \end{equation} 
  \begin{equation}
    \label{eq:jeanstriaxial_muS}
    \frac{\partial S_{\mu\mu}}{\partial \mu} -
    \frac{S_{\nu\nu}}{2(\mu\!-\!\nu)} -
    \frac{S_{\lambda\lambda}}{2(\mu\!-\!\lambda)} =
    g_2(\lambda,\mu,\nu), 
  \end{equation}
  \begin{equation}
    \label{eq:jeanstriaxial_nuS}
    \frac{\partial S_{\nu\nu}}{\partial \nu} -
    \frac{S_{\lambda\lambda}}{2(\nu\!-\!\lambda)} -
    \frac{S_{\mu\mu}}{2(\nu\!-\!\mu)} = g_3(\lambda,\mu,\nu), 
  \end{equation}
\end{subequations}
where the function $g_1$ is defined as
\begin{equation}
  \label{eq:deffunctiong_1}
  g_1(\lambda,\mu,\nu) = 
  - \sqrt{ (\lambda\!-\!\mu) (\lambda\!-\!\nu) (\mu\!-\!\nu) } \, 
  \rho \, \frac{\partial V_S}{\partial \lambda},  
\end{equation}
and $g_2$ and $g_3$ follow by cyclic permutation 
$\lambda\to\mu\to\nu\to\lambda$. \looseness=-1
We keep the three terms $\lambda\!-\!\mu$,
$\lambda\!-\!\nu$ and $\mu\!-\!\nu$ under one square root. With each
cyclic permutation two of the three terms change sign, so that the
combination of the three terms is always positive real. 
Therefore. the square root of the combination is always single-valued,
whereas in the case of three separate square roots we would have a
multi-valued function.

We simplify equations \eqref{eq:jeanstriaxialS} by substituting for
$g_1$, $g_2$ and $g_3$, respectively
\begin{eqnarray}
  \label{eq:rhssimplyfying}
  \tilde{g}_1(\lambda,\mu,\nu) \hspace{-7pt} & = & \hspace{-7pt} 0, 
  \nonumber \\ 
  \tilde{g}_2(\lambda,\mu,\nu) \hspace{-7pt} & = & \hspace{-7pt}
  \, \delta(\lambda_0\!-\!\lambda) \, \delta(\mu_0\!-\!\mu)
  \, \delta(\nu_0\!-\!\nu), \\ 
  \tilde{g}_3(\lambda,\mu,\nu) \hspace{-7pt} & = & \hspace{-7pt} 0, \nonumber   
\end{eqnarray}
with
\begin{equation}
  \label{eq:rangecoords*}
  -\gamma \le \nu \le \nu_0 \le -\beta \le \mu \le \mu_0 \le
   -\alpha \le \lambda \le \lambda_0.  
\end{equation}
We obtain two similar systems of simplified equations by cyclic
permutation of the left-hand side of \eqref{eq:jeanstriaxialS}. 
Once we have obtained the singular solutions of the simplified system
with the right-hand side given by \eqref{eq:rhssimplyfying}, those for
the other two systems follow via cyclic permutation.

\subsection{Homogeneous boundary problem}
\label{sec:homboundaryproblem3D}

The choice \eqref{eq:rhssimplyfying} implies that the functions
$S_{\tau\tau}(\lambda,\mu,\nu)$ \eqref{eq:deffuncsStt} must have the
following forms
\begin{eqnarray}
  \label{eq:functionalformsStt}
    S_{\lambda\lambda} 
    \hspace{-7pt} & = & \hspace{-7pt} 
    A(\lambda,\mu,\nu) 
    \, \mathcal{H}(\lambda_0\!-\!\lambda)
    \mathcal{H}(\mu_0\!-\!\mu) 
    \mathcal{H}(\nu_0\!-\!\nu) 
    \nonumber \\ 
    \hspace{-7pt} &  & \hspace{-7pt} 
    + \; F(\lambda,\mu) 
    \, \delta(\nu_0\!-\!\nu)
    \mathcal{H}(\lambda_0\!-\!\lambda)
    \mathcal{H}(\mu_0\!-\!\mu), 
    \nonumber \\*[3pt]
    S_{\mu\mu} 
    \hspace{-7pt} & = & \hspace{-7pt}
    B(\lambda,\mu,\nu) 
    \, \mathcal{H}(\lambda_0\!-\!\lambda)
    \mathcal{H}(\mu_0\!-\!\mu) 
    \mathcal{H}(\nu_0\!-\!\nu) 
    \nonumber \\ 
    \hspace{-7pt} &  & \hspace{-7pt} 
    + \; G(\lambda,\mu)
    \, \delta(\nu_0\!-\!\nu) 
    \mathcal{H}(\lambda_0\!-\!\lambda)
    \mathcal{H}(\mu_0\!-\!\mu) 
    \nonumber \\*[-7pt] \\*[-7pt]
    \hspace{-7pt} &  & \hspace{-7pt} 
    + \; H(\mu,\nu)
    \, \delta(\lambda_0\!-\!\lambda)
    \mathcal{H}(\mu_0\!-\!\mu) 
    \mathcal{H}(\nu_0\!-\!\nu) 
    \nonumber \\ 
    \hspace{-7pt} &  & \hspace{-7pt} 
    - \; \delta(\lambda_0\!-\!\lambda) 
    \delta(\nu_0\!-\!\nu)
    \mathcal{H}(\mu_0\!-\!\mu),
    \nonumber \\*[3pt]
    S_{\nu\nu} 
    \hspace{-7pt} & = & \hspace{-7pt} 
    C(\lambda,\mu,\nu)
    \, \mathcal{H}(\lambda_0\!-\!\lambda)
    \mathcal{H}(\mu_0\!-\!\mu) 
    \mathcal{H}(\nu_0\!-\!\nu) 
    \nonumber \\ 
    \hspace{-7pt} &  & \hspace{-7pt} 
    + \; I(\mu,\nu)
    \, \delta(\lambda_0\!-\!\lambda) 
    \mathcal{H}(\mu_0\!-\!\mu) 
    \mathcal{H}(\nu_0\!-\!\nu), 
    \nonumber
\end{eqnarray}
with $A$, $B$, $C$ and $F$, $G$, $H$, $I$ yet unknown functions of
three and two coordinates, respectively, and $\mathcal{H}$ the
step-function \eqref{eq:defstepfunc}. After substituting these forms
into the simplified Jeans equations and matching terms we obtain 14
equations. Eight of them comprise the following two homogeneous
systems with two boundary conditions each
\begin{equation}
  \label{eq:2DsysFG}
  \left\{ 
    \begin{array}{lclclcl}
      \D \frac{\partial F}{\partial \lambda} -
      \frac{G}{2(\lambda\!-\!\mu)} 
      & \hspace{-7pt} = \hspace{-7pt} & 0, &  &  
      \D F(\lambda_0,\mu) 
      & \hspace{-7pt} = \hspace{-7pt} &
      \D \frac{1}{2(\lambda_0\!-\!\mu)}, 
      \\*[10pt] 
      \D \frac{\partial G}{\partial \mu} - 
      \frac{F}{2(\mu\!-\!\lambda)} 
      & \hspace{-7pt} = \hspace{-7pt} & 0, &  &  
      \D G(\lambda,\mu_0) 
      & \hspace{-7pt} = \hspace{-7pt} & 0,
    \end{array}
  \right.
\end{equation}
and
\begin{equation}
  \label{eq:2DsysHI}
  \left\{ 
    \begin{array}{lclclcl}
      \D \frac{\partial H}{\partial \mu} - 
      \frac{I}{2(\mu\!-\!\nu)} 
      & \hspace{-7pt} = \hspace{-7pt} & 0, &  & 
      \D H(\mu_0,\nu) 
      & \hspace{-7pt} = \hspace{-7pt} & 0, 
      \\*[10pt] 
      \D \frac{\partial I}{\partial \nu} - 
      \frac{H}{2(\nu\!-\!\mu)} 
      & \hspace{-7pt} = \hspace{-7pt} & 0, &  &  
      \D I(\mu,\nu_0) 
      & \hspace{-7pt} = \hspace{-7pt} & 
      \D \frac{1}{2(\nu_0\!-\!\mu)}.
    \end{array}
  \right.
\end{equation}
We have shown in \S\ref{sec:2Dcases} how to solve these
two-dimensional homogeneous boundary problems in terms of the complete
elliptic integral of the second kind $E$ and its derivative
$E'$. The solutions are
\begin{align}
  \label{eq:solFGHI}
    F(\lambda,\mu) &= \frac{E(w)}{\pi(\lambda_0-\mu)}, 
    & G(\lambda,\mu) &= - \frac{ 2w E'(w) }{
    \pi(\lambda_0-\lambda) },
    \nonumber \\*[-5pt] \\*[-5pt] 
    H(\mu,\nu) &= - \frac{2u E'(u)}{\pi(\nu_0-\nu)},  
    & I(\mu,\nu) &= - \frac{E(u)}{\pi(\mu-\nu_0)}, \nonumber 
\end{align}
where $u$ and similarly $v$, which we will encounter later on,
follow from $w$ \eqref{eq:definitionofw} by cyclic permutation
$\lambda \to \mu \to \nu \to \lambda$ and $\lambda_0 \to \mu_0 \to
\nu_0 \to\lambda_0$, so that
\begin{equation}
  \label{eq:defu*v*w*}
    u =
    \frac{ (\mu_0\!-\!\mu) (\nu_0\!-\!\nu) }{
    (\mu_0\!-\!\nu_0) (\mu\!-\!\nu) }, 
    \quad 
    v =
    \frac{ (\nu_0\!-\!\nu) (\lambda_0\!-\!\lambda) }{
    (\lambda_0\!-\!\nu_0) (\lambda\!-\!\nu) }.
\end{equation}
The remaining six equations form a three-dimensional homogeneous
boundary problem, consisting of three homogeneous Jeans equations
\begin{eqnarray}
  \label{eq:homsystemABC}
    \frac{\partial A}{\partial \lambda} - \frac{B}{2(\lambda\!-\!\mu)} -
    \frac{C}{2(\lambda\!-\!\nu)} &=& 0, \nonumber \\ 
    \frac{\partial B}{\partial \mu} - \frac{C}{2(\mu\!-\!\nu)} -
    \frac{A}{2(\mu\!-\!\lambda)} &=& 0, \\ 
    \frac{\partial C}{\partial \nu} - \frac{A}{2(\nu\!-\!\lambda)} -
    \frac{B}{2(\nu\!-\!\mu)} &=& 0. \nonumber 
\end{eqnarray}
and three boundary conditions, specifically the values of
$A(\lambda_0,\mu,\nu)$, $B(\lambda,\mu_0,\nu)$, and
$C(\lambda,\mu,\nu_0)$. 
As in \S\ref{sec:homogeneousboundaryproblem2D}, it is useful to
supplement these boundary conditions with the values of $A$, $B$, and
$C$ at the other boundary surfaces. 
These are obtained by integrating the pairs of equations
\eqref{eq:homsystemABC} which apply at those surfaces, and using the
boundary conditions. 
This results in the following nine boundary values 
\begin{eqnarray}
  \label{eq:BCshomogeneousproblem}
  A(\lambda_0,\mu,\nu) \hspace{-8pt} & = & \hspace{-8pt} 
  \frac{1}{2\pi} \!\Biggl[\!
  \frac{ E(u) }{ (\lambda_0\!\!-\!\nu) (\mu\!-\!\nu_0) } \!+\!
  \frac{ 2u E'(u)}{ (\lambda_0\!\!-\!\mu)
    (\nu_0\!\!-\!\nu) } 
  \!\Biggr], \nonumber \\ 
  A(\lambda,\mu_0,\nu) \hspace{-8pt} & = & \hspace{-8pt} 
  \frac{1}{2\pi} \!\Biggl[\!
  \frac{ E(v) }{ (\lambda_0\!\!-\!\nu)
  (\mu_0\!\!-\!\nu_0) }
  \!+\! \frac{ 2v E'(v) }{ (\lambda_0\!\!-\!\mu_0)
  (\nu_0\!\!-\!\nu) }
  \!\Biggr], \nonumber \\  
  A(\lambda,\mu,\nu_0) \hspace{-8pt} & = & \hspace{-8pt} 
  \frac{E(w)}{4\pi (\lambda_0\!\!-\!\mu)} \!\Biggl[\!
  \frac{\lambda\!-\!\mu}{ (\lambda\!-\!\nu_0) (\mu\!-\!\nu_0) } \!+\!
  \frac{\lambda_0\!\!-\!\mu_0}{ (\lambda_0\!\!-\!\nu_0)
  (\mu_0\!\!-\!\nu_0) } 
  \!\Biggr], \nonumber \\
  B(\lambda_0,\mu,\nu) \hspace{-8pt} & = & \hspace{-8pt} 
  \frac{ u E'(u) }{ 2\pi (\nu_0\!\!-\!\nu) } \!\Biggl[\!
  \frac{\mu_0\!\!-\!\mu}{ (\lambda_0\!-\!\!\mu_0)
  (\lambda_0\!\!-\!\mu) }
  \!-\! \frac{\nu_0\!\!-\!\nu}{ (\lambda_0\!\!-\!\nu_0)
  (\lambda_0\!\!-\!\nu) }
  \!\Biggr], \nonumber \\*[7pt] 
  B(\lambda,\mu_0,\nu) \hspace{-8pt} & = & \hspace{-8pt} \, 0, \\*[7pt] 
  B(\lambda,\mu,\nu_0) \hspace{-8pt} & = & \hspace{-8pt} 
  \frac{ w E'(w) }{ 2\pi (\lambda_0\!\!-\!\lambda) }
  \!\Biggl[\! 
  \frac{\mu_0\!\!-\!\mu}{ (\mu_0\!\!-\!\nu_0)
  (\mu\!-\!\nu_0) }
  \!-\! \frac{\lambda_0\!\!-\!\lambda}{(\lambda_0\!\!-\!\nu_0)
  (\lambda\!-\!\nu_0) }
  \!\Biggr], \nonumber \\
  C(\lambda_0,\mu,\nu) \hspace{-8pt} & = & \hspace{-8pt} 
  \frac{ E(u) }{ 4\pi (\mu\!-\!\nu_0) } \!\Biggl[\!
  \frac{\mu\!-\!\nu}{ (\lambda_0\!\!-\!\mu)
  (\lambda_0\!\!-\!\nu) } \!+\! 
  \frac{\mu_0\!\!-\!\nu_0}{ (\lambda_0\!\!-\!\mu_0)
  (\lambda_0\!\!-\!\nu_0) }
  \!\Biggr], \nonumber \\ 
  C(\lambda,\mu_0,\nu) \hspace{-8pt} & = & \hspace{-8pt} 
  \frac{1}{2\pi} \!\Biggl[\!
  \frac{ E(v) }{ (\lambda_0\!\!-\!\mu_0)
  (\lambda\!-\!\nu_0) }
  \!-\! \frac{ 2v E'(v) }{ (\mu_0\!\!-\!\nu_0)
  (\lambda_0\!\!-\!\lambda) }
  \!\Biggr], \nonumber \\  
  C(\lambda,\mu,\nu_0) \hspace{-8pt} & = & \hspace{-8pt} 
  \frac{1}{2\pi} \!\Biggl[\!
  \frac{ E(w) }{ (\lambda_0\!\!-\!\mu)
  (\lambda\!-\!\nu_0) } \!-\! \frac{ 2w E'(w) }{
  (\mu\!-\!\nu_0) 
  (\lambda_0\!\!-\!\lambda) }
  \!\Biggr]. \nonumber  
\end{eqnarray}
If we can solve the three homogeneous equations
\eqref{eq:homsystemABC} and satisfy the nine boundary expressions
\eqref{eq:BCshomogeneousproblem} simultaneously, then we obtain the
singular solutions \eqref{eq:functionalformsStt}. 
By superposition, we can then construct the solution of the Jeans
equations for triaxial St\"ackel models.

\subsection{Particular solution}
\label{sec:particularsolution3D}

By analogy with the two-dimensional case, we look for particular
solutions of the homogeneous equations \eqref{eq:homsystemABC} and by
superposition of these particular solutions we try to satisfy the
boundary expressions \eqref{eq:BCshomogeneousproblem} simultaneously,
in order to obtain the homogeneous solution for $A$, $B$ and $C$.

\subsubsection{One-parameter particular solution}
\label{sec:oneparparticularsol}

By substitution one can verify that 
\begin{equation}
  \label{eq:partsol_1par}
  A^P(\lambda,\mu,\nu) =
  \frac{ \sqrt{ (\lambda\!-\!\mu) (\lambda\!-\!\nu)
  (\mu\!-\!\nu) } }
  { (\lambda\!-\!\mu) (\lambda\!-\!\nu)}
  \frac{(z\!-\!\lambda)}{(z\!-\!\mu)(z\!-\!\nu)},   
\end{equation}
with $B^P$ and $C^P$ following from $A^P$ by cyclic permutation,
solves the homogeneous equations \eqref{eq:homsystemABC}. To satisfy
the nine boundary expressions \eqref{eq:BCshomogeneousproblem}, we
could integrate this particular solution over its free parameter $z$,
in the complex plane. From \S\ref{sec:derivinghomogenoussolution2D}, it
follows that, at the boundaries, this results in simple polynomials in
$(\lambda,\mu,\nu)$ and $(\lambda_0,\mu_0,\nu_0)$. This means that the
nine boundary expressions \eqref{eq:BCshomogeneousproblem} cannot be
satisfied, since in addition to these simple polynomials they also
contain $E$ and $E'$. The latter are functions of one variable, so
that at least one extra freedom is necessary. Hence, we look for a
particular solution with \textit{two} free parameters.

\subsubsection{Two-parameter particular solution}
\label{sec:twoparparticularsol}

A particular solution with two free parameters $z_1$ and $z_2$
can be found by splitting the $z$-dependent terms
of the one-parameter solution \eqref{eq:partsol_1par} into two similar
parts and then relabelling them. The result is the following
two-parameter particular solution
\begin{eqnarray}
  \label{eq:2parpartsol}
  A^P \hspace{-7pt} &=& \hspace{-7pt} 
  \frac{\sqrt{ (\lambda\!-\!\mu) (\lambda\!-\!\nu) (\mu\!-\!\nu) } } 
  { (\lambda\!-\!\mu) (\lambda\!-\!\nu) } 
  \prod_{i=1}^2 \frac{ (z_i\!-\!\lambda)^\frac{1}{2} }
  { (z_i\!-\!\mu)^\frac{1}{2} (z_i\!-\!\nu)^\frac{1}{2} }, 
  \nonumber \\
  B^P \hspace{-7pt} &=& \hspace{-7pt}
  \frac{\sqrt{ (\lambda\!-\!\mu) (\lambda\!-\!\nu) (\mu\!-\!\nu) } }  
  { (\mu\!-\!\nu) (\mu\!-\!\lambda) } 
  \prod_{i=1}^2 \frac{ (z_i\!-\!\mu)^\frac{1}{2} }
  { (z_i\!-\!\nu)^\frac{1}{2} \! (z_i\!-\!\lambda)^\frac{1}{2} },
  \\
  C^P \hspace{-7pt} &=& \hspace{-7pt} 
  \frac{\sqrt{ (\lambda\!-\!\mu) (\lambda\!-\!\nu) (\mu\!-\!\nu) } }
  { (\nu\!-\!\lambda) (\nu\!-\!\mu) }
  \prod_{i=1}^2  \frac{ (z_i\!-\!\nu)^\frac{1}{2} }
  { (z_i\!-\!\lambda)^\frac{1}{2} (z_i\!-\!\mu)^\frac{1}{2} }. 
  \nonumber 
\end{eqnarray}
These functions are cyclic in $(\lambda,\mu,\nu)$, as is required from
the symmetry of the homogeneous equations \eqref{eq:homsystemABC}.
The presence of the square roots, such as occurred earlier in the
solution \eqref{eq:formAandBpart} for the disc case, allows us
to fit boundary values that contain elliptic integrals.

To show that this particular solution solves the homogeneous Jeans
equations, we calculate the derivative of $A^P(\lambda,\mu,\nu)$ with
respect to $\lambda$:
\begin{equation}
  \label{eq:chechpartsol1}
  \frac{\partial A^P}{\partial \lambda} = \frac{A^P}{2} \Biggl(
  \frac{1}{\lambda\!-\!z_1} + \frac{1}{\lambda\!-\!z_2} -
  \frac{1}{\lambda\!-\!\mu} - \frac{1}{\lambda\!-\!\nu} \Biggr). 
\end{equation}
This can be written as
\begin{eqnarray}
  \label{eq:chechpartsol2}
  \frac{\partial A^P}{\partial \lambda} \hspace{-7pt} &=& \hspace{-7pt} 
  \frac{1}{ 2 (\lambda\!-\!\mu) } \Biggl[ - \frac{ (z_1\!-\!\mu)
    (z_2\!-\!\mu) (\lambda\!-\!\nu) }{ (z_1\!-\!\lambda)
    (z_2\!-\!\lambda) (\mu\!-\!\nu) } A^P \Biggr] \nonumber 
  \\*[-5pt] \\*[-5pt]
  \hspace{-7pt} && \hspace{-7pt} + \; \frac{1}{ 2 (\lambda\!-\!\nu) }
  \Biggl[ \frac{ (z_1\!-\!\nu) (z_2\!-\!\nu) (\lambda\!-\!\mu) }{
  (z_1\!-\!\lambda) (z_2\!-\!\lambda) (\mu\!-\!\nu) } A^P
  \Biggr]. \nonumber  
\end{eqnarray}
From the two-parameter particular solution we have 
\begin{eqnarray}
  \label{eq:ratiospartsolB/AandC/A}
  \frac{B^P}{A^P} \hspace{-7pt} &=& \hspace{-7pt} -\frac{ (z_1-\mu)
  (z_2-\mu) (\lambda-\nu) }{ (z_1-\lambda) (z_2-\lambda) (\mu-\nu) },
  \nonumber \\*[-5pt] \\*[-5pt] 
  \frac{C^P}{A^P} \hspace{-7pt} &=& \hspace{-7pt} \frac{ (z_1-\nu)
  (z_2-\nu) (\lambda-\mu) }{ (z_1-\lambda) (z_2-\lambda) (\mu-\nu)
  },\nonumber  
\end{eqnarray}
so that, after substitution of these ratios, the first homogeneous
equation of \eqref{eq:homsystemABC}, is indeed satisfied. The
remaining two homogeneous equations can be checked in the same way.

\subsection{The homogeneous solution}
\label{sec:derivinghomogenoussolution3D}

In order to satisfy the four boundary expressions of the
two-dimensional case, we multiplied the one-parameter particular
solution by terms depending on $\lambda_0$, $\mu_0$ and the free
complex parameter $z$, followed by contour integration over the
latter.  Similarly, in the triaxial case we multiply the two-parameter
particular solution \eqref{eq:partsolAandB} by terms depending on
$\lambda_0$, $\mu_0$, $\nu_0$ and the two free parameters $z_1$ and
$z_2$, in such a way that by contour integration over the latter two
complex parameters the nine boundary expressions
\eqref{eq:BCshomogeneousproblem} can be satisfied. Since these terms
and the integration are independent of $\lambda$, $\mu$ and $\nu$, it
follows from the superposition principle that the homogeneous
equations \eqref{eq:homsystemABC} remain satisfied.

The contour integrations over $z_1$ and $z_2$ are mutually
independent, since we can separate the two-parameter particular
solution \eqref{eq:2parpartsol} with respect to these two
parameters. This allows us to choose a pair of contours, one contour
in the $z_1$-plane and the other contour in the $z_2$-plane, and
integrate over them separately. We consider the same simple contours
as in the disk case (Fig.~\ref{fig:contourscomplexzplane2D}) around
the pairs of branch points $(\lambda,\lambda_0)$ and $(\mu,\mu_0)$,
and a similar contour around $(\nu,\nu_0)$. We denote these contours
by $C_i^\lambda$, $C_i^\mu$ and $C_i^\nu$ respectively, with $i=1,2$
indicating in which of the two complex planes we apply the contour
integration.

\subsubsection{Boundary expressions for $B$}
\label{sec:boundaryexpr4B}

It follows from \eqref{eq:BCshomogeneousproblem} that $B=0$ at the
boundary $\mu=\mu_0$. From Cauchy's theorem, $B$ would indeed vanish
if, in this case, in either the $z_1$-plane or $z_2$-plane the
integrand for $B$ is analytic within the chosen integration contour.
The boundary expression for $B$ at $\nu=\nu_0$ follows from the one at
$\lambda=\lambda_0$ by taking $\nu \leftrightarrow \lambda$ and $\nu_0
\leftrightarrow \lambda_0$. In addition to this symmetry, also the
form of both boundary expressions puts constraints on the solution for
$B$. The boundary expressions can be separated in two parts, one
involving the complete elliptic integral $E'$ and the other
consisting of a two-component polynomial in $\tau$ and $\tau_0$
($\tau=\lambda,\mu,\nu$). Each of the two parts follows from a contour
integration in one of the two complex planes. For either of the
complex parameters, $z_1$ or $z_2$, the integrands will consist of a
combination of the six terms $z_i-\tau$ and $z_i-\tau_0$ with powers
that are half-odd integers, i.e., the integrals are of
\textit{hyper}elliptic form. If two of the six terms cancel on one of
the boundaries, we will be left with an elliptic integral. We expect
the polynomial to result from applying the Residue theorem to a double
pole, as this would involve a first derivative and hence give two
components. This leads to the following Ansatz
\begin{multline}
  \label{eq:formforB}
  B(\lambda,\mu,\nu) \propto   
  \frac{\sqrt{ (\lambda\!-\!\mu) (\lambda\!-\!\nu) (\mu\!-\!\nu) }}{
    (\mu\!-\!\nu) (\mu\!-\!\lambda) } \times \\
  \hspace{20pt} \oint\limits_{C_1} \frac{ (z_1\!-\!\mu)^\frac{1}{2}
    (z_1\!-\!\lambda_0)^\frac{1}{2} \, \d z_1 }{
    (z_1\!-\!\nu)^\frac{1}{2} (z_1\!-\!\lambda)^\frac{1}{2}
    (z_1\!-\!\mu_0)^\frac{1}{2} (z_1\!-\!\nu_0)^\frac{3}{2} }
  \times \\ 
  \oint\limits_{C_2} \frac{ (z_2\!-\!\mu)^\frac{1}{2}
  (z_2\!-\!\nu_0)^\frac{1}{2} \, \d z_2 }{ (z_2\!-\!\nu)^\frac{1}{2}
  (z_2\!-\!\lambda)^\frac{1}{2} (z_2\!-\!\mu_0)^\frac{1}{2} 
  (z_2\!-\!\lambda_0)^\frac{3}{2} }.  
\end{multline}
Upon substitution of $\mu=\mu_0$, the terms involving $\mu_0$ cancel
in both integrals, so that the integrands are analytic in both
contours $C_1^\mu$ and $C_2^\mu$. Hence, by choosing either of these
contours as integration contour, the boundary expression
$B(\lambda,\mu_0,\nu)=0$ is satisfied.

When $\lambda=\lambda_0$, the terms with $\lambda_0$ in the first
integral in \eqref{eq:formforB} cancel, while in the second integral we
have $(z_2-\lambda_0)^{-2}$. The first integral is analytic within
$C_1^\lambda$, so that there is no contribution from this
contour. However, the integral over $C_1^\mu$ is elliptic and can be
evaluated in terms of $E'$ (cf.\
\S\ref{sec:evaluatinghomogenoussolution2D}). We apply the Residue
theorem to the second integral, for which there is a double pole
inside the contour $C_2^\lambda$. Considering $C_1^\mu$ and
$C_2^\lambda$ as a pair of contours, the expression for $B$ at
$\lambda=\lambda_0$ becomes
\begin{multline}
  \label{eq:expBiflmabda=lambda*}
  B(\lambda,\mu,\nu) \propto  
  - 16 \pi^2 \frac{ \sqrt{ (\lambda_0\!-\!\mu_0)
  (\lambda_0\!-\!\nu_0) (\mu_0\!-\!\nu_0) } }{
  (\mu_0\!-\!\nu_0) (\mu_0\!-\!\lambda_0) }  
  \times \\  
  \frac{u E'(u)}{ 2\pi (\nu_0\!\!-\!\nu) }
  \!\Biggl[\! 
  \frac{\mu_0\!\!-\!\mu}{ (\lambda_0\!\!-\!\mu_0)
    (\lambda_0\!\!-\!\mu) } \!-\!
  \frac{\nu_0\!\!-\!\nu}{
    (\lambda_0\!\!-\!\nu_0) (\lambda_0\!\!-\!\nu) } 
  \!\Biggr],
\end{multline}
which is the required boundary expression up to a scaling factor.  As
before, we keep the terms $\lambda_0\!-\!\mu_0$, $\lambda_0\!-\!\nu_0$
and $\mu_0\!-\!\nu_0$ under one square root, so that it is
single-valued with respect to cyclic permutation in these coordinates.

The boundary expression for $B$ at $\nu=\nu_0$ is symmetric with the
one at $\lambda=\lambda_0$, so that a similar approach can be used. 
In this case, for the second integral, there is no contribution from
$C_2^\nu$, whereas it can be expressed in terms of $E'$ if
$C_2=C_2^\mu$.     
The first integrand has a double pole in $C_1^\nu$.
The total contribution from the pair ($C_1^\nu$,$C_2^\mu$) gives the
correct boundary expression, up to a scaling factor that is the same
as in \eqref{eq:expBiflmabda=lambda*}.

Taking into account the latter scaling factor, this shows that the
Ansatz \eqref{eq:formforB} for $B$ produces the correct boundary
expressions and hence we postulate it as the homogeneous solution for
$B$. The expressions for $A$ and $C$ then follow from the ratios
\eqref{eq:ratiospartsolB/AandC/A}. Absorbing the minus sign in
\eqref{eq:expBiflmabda=lambda*} into the pair of contours, i.e.,
either of the two contours we integrate in clockwise direction, we
postulate the following homogeneous solution
\begin{eqnarray}
  \label{eq:homsolform4A}
   A(\lambda,\mu,\nu) \hspace{-9pt} &=& \hspace{-9pt}
  \frac{(\mu_0\!\!-\!\nu_0\!)
  (\mu_0\!\!-\!\lambda_0\!) }{ 16\pi^2 (\lambda\!-\!\mu)
  (\lambda\!-\!\nu) }   
  \sqrt{\! \frac{ (\lambda\!-\!\mu) (\lambda\!-\!\nu)
  (\mu\!-\!\nu) }{ (\lambda_0\!\!-\!\mu_0\!)
  (\lambda_0\!\!-\!\nu_0\!) (\mu_0\!\!-\!\nu_0\!) } }   
  \times \nonumber \\  
  \oint\limits_{C_1} \hspace{-7pt} & & \hspace{-20pt}
  \frac{ (z_1\!-\!\lambda)^\frac{1}{2}
  (z_1\!-\!\lambda_0)^\frac{1}{2} \, \d z_1 }{
  (z_1\!-\!\mu)^\frac{1}{2} 
  (z_1\!-\!\nu)^\frac{1}{2} (z_1\!-\!\mu_0)^\frac{1}{2}
  (z_1\!-\!\nu_0)^\frac{3}{2} } \times \nonumber \\ 
  \oint\limits_{C_2} \hspace{-7pt} & & \hspace{-20pt}
  \frac{ (z_2\!-\!\lambda)^\frac{1}{2}
  (z_2\!-\!\nu_0)^\frac{1}{2} \, \d z_2 }{
  (z_2\!-\!\mu)^\frac{1}{2} 
  (z_2\!-\!\nu)^\frac{1}{2} (z_2\!-\!\mu_0)^\frac{1}{2}
  (z_2\!-\!\lambda_0)^\frac{3}{2} },    
\end{eqnarray}
\begin{eqnarray}
  \label{eq:homsolform4B}
  B(\lambda,\mu,\nu) \hspace{-9pt} &=& \hspace{-9pt}
  \frac{ (\mu_0\!\!-\!\nu_0\!)
  (\mu_0\!\!-\!\lambda_0\!) }{ 16\pi^2 (\mu\!-\!\nu)
  (\mu\!-\!\lambda) } 
  \sqrt{\! \frac{ (\lambda\!-\!\mu) (\lambda\!-\!\nu)
  (\mu\!-\!\nu) }{ (\lambda_0\!\!-\!\mu_0\!)
  (\lambda_0\!\!-\!\nu_0\!) (\mu_0\!\!-\!\nu_0\!) } }   
  \times \nonumber \\  
  \oint\limits_{C_1} \hspace{-7pt} & & \hspace{-20pt}
  \frac{ (z_1\!-\!\mu)^\frac{1}{2} 
  (z_1\!-\!\lambda_0)^\frac{1}{2} \, \d z_1 }{
  (z_1\!-\!\nu)^\frac{1}{2} 
  (z_1\!-\!\lambda)^\frac{1}{2} (z_1\!-\!\mu_0)^\frac{1}{2}
  (z_1\!-\!\nu_0)^\frac{3}{2} } 
  \times \nonumber \\
  \oint\limits_{C_2} \hspace{-7pt} & & \hspace{-20pt}
  \frac{ (z_2\!-\!\mu)^\frac{1}{2}
  (z_2\!-\!\nu_0)^\frac{1}{2} \, \d z_2 }{
  (z_2\!-\!\nu)^\frac{1}{2} 
  (z_2\!-\!\lambda)^\frac{1}{2} (z_2\!-\!\mu_0)^\frac{1}{2}
  (z_2\!-\!\lambda_0)^\frac{3}{2} },
\end{eqnarray}
\begin{eqnarray}
  \label{eq:homsolform4C}
  C(\lambda,\mu,\nu) \hspace{-9pt} &=& \hspace{-9pt}
  \frac{ (\mu_0\!\!-\!\nu_0\!)
  (\mu_0\!\!-\!\lambda_0\!) }{ 16\pi^2 (\nu\!-\!\lambda)
  (\nu\!-\!\mu) }  
  \sqrt{\! \frac{ (\lambda\!-\!\mu) (\lambda\!-\!\nu)
  (\mu\!-\!\nu) }{ (\lambda_0\!\!-\!\mu_0\!)
  (\lambda_0\!\!-\!\nu_0\!) (\mu_0\!\!-\!\nu_0\!) } }   
  \times \nonumber \\  
  \oint\limits_{C_1} \hspace{-7pt} & & \hspace{-20pt}
  \frac{ (z_1\!-\!\nu)^\frac{1}{2}
  (z_1\!-\!\lambda_0)^\frac{1}{2} \, \d z_1 }{
  (z_1\!-\!\lambda)^\frac{1}{2} 
  (z_1\!-\!\mu)^\frac{1}{2} (z_1\!-\!\mu_0)^\frac{1}{2}
  (z_1\!-\!\nu_0)^\frac{3}{2} } 
  \times \nonumber \\
  \oint\limits_{C_2} \hspace{-7pt} & & \hspace{-20pt}
  \frac{( z_2\!-\!\nu)^\frac{1}{2}
  (z_2\!-\!\nu_0)^\frac{1}{2} \, \d z_2 }{ (z_2\!-\!\lambda)^\frac{1}{2}
  (z_2\!-\!\mu)^\frac{1}{2} (z_2\!-\!\mu_0)^\frac{1}{2}
  (z_2\!-\!\lambda_0)^\frac{3}{2} }.
\end{eqnarray}
The integrands consist of multi-valued functions that all come in pairs
of the form $(z-\tau)^{\frac{1}{2}-m} (z-\tau_0)^{\frac{1}{2}-n}$, for
integers $m$ and $n$, with $\tau$ equal to $\lambda$, $\mu$ or $\nu$.
Hence, completely analogous to our procedure in
\S\ref{sec:derivinghomogenoussolution2D}, we can make the integrands
single-valued by specifying, in the complex $z_1$-plane and
$z_2$-plane, three cuts running between the three pairs
$(\lambda,\lambda_0)$, $(\mu,\mu_0)$, $(\nu,\nu_0)$ of branch points,
that are enclosed by the simple contours. The integrands are now
analytic in the cut plane away from its cuts and behave again as
$z_i^{-2}$ at large distances, so that the integral over a circular
contour with radius going to infinity, will be zero. Hence, connecting
the simple contours $C_i^\lambda$, $C_i^\mu$ and $C_i^\nu$ with this
circular contour, shows that their cumulative contribution cancels
\begin{equation}
  \label{eq:canceling3contours}
  C_i^\nu + C_i^\mu + C_i^\lambda = 0, \qquad i=1,2.
\end{equation}
This relation will allow us to make a combination of contours, so that
the nine boundary expressions \eqref{eq:BCshomogeneousproblem} can be
satisfied \textit{simultaneously}
(\S\ref{sec:combinationofcontours}). Before doing so, we first
establish whether, with the homogeneous solution for $A$ and $C$ given
by \eqref{eq:homsolform4A} and \eqref{eq:homsolform4C}, respectively,
we indeed satisfy their corresponding boundary expressions separately.

\subsubsection{Boundary expressions for $A$ and $C$}
\label{sec:boundaryexpr4AandC}

The boundary expressions and the homogeneous solution of $C$,
follow from those of $A$ by taking $\lambda \leftrightarrow \nu$ and
$\lambda_0 \leftrightarrow \nu_0$. Henceforth, once we have
checked the boundary expressions for $A$, those for $C$ can be checked
in a similar way.   

Upon substitution of $\lambda=\lambda_0$ in the expression for 
$A$ \eqref{eq:homsolform4A}, the first integrand is proportional to
$z_1-\lambda'$ and thus is analytic within the contour
$C_1^\lambda$. The contribution to the boundary expression 
therefore needs to come from either $C_1^\mu$ or $C_1^\nu$. The
substitution
\begin{equation}
  \label{eq:substAl=l*}
  z_1-\lambda_0 = \frac{ \lambda_0\!-\!\nu }{ \mu\!-\!\nu }
  (z_1\!-\!\mu) - \frac{ \lambda_0\!-\!\mu }{ \mu\!-\!\nu }
  (z_1\!-\!\nu), 
\end{equation}
splits the first integral into two complete elliptic integrals  
\begin{multline}
  \label{eq:splitAl=l*}
  \frac{ \lambda_0\!-\!\nu }{ \mu\!-\!\nu } \oint\limits_{C_1}
  \frac{ (z_1\!-\!\mu)^\frac{1}{2} \, \d z_1 }{ (z_1\!-\!\nu)^\frac{1}{2}
  (z_1\!-\!\mu_0)^\frac{1}{2} (z_1\!-\!\nu_0)^\frac{3}{2} } \\
  - \frac{ \lambda_0\!-\!\mu }{ \mu\!-\!\nu } \oint\limits_{C_1}
  \frac{ (z_1\!-\!\nu)^\frac{1}{2} \, \d z_1 }{ (z_1\!-\!\mu)^\frac{1}{2}
  (z_1\!-\!\mu_0)^\frac{1}{2} (z_1\!-\!\nu_0)^\frac{3}{2} }. 
\end{multline}
Within the contour $C_1^\mu$, the integrals can be evaluated in terms
of $E'(u)$ and $E(u)$ respectively. When $\lambda = \lambda_0$, the
second integral in \eqref{eq:homsolform4A} has a single pole
contribution from the contour $C_2^\lambda$. Together, $-C_1^\mu
C_2^\lambda$, exactly reproduces the boundary expression
$A(\lambda_0,\mu,\nu)$ in \eqref{eq:BCshomogeneousproblem}.

When $\mu=\mu_0$, both integrands in the expression for $A$ have a
single pole within the contour $C_i^\mu$. However, the combination
$C_1^\mu C_2^\mu$ does not give the correct boundary expression.  We
again split both integrals to obtain the required complete elliptic
integrals. In the first we substitute
\begin{equation}
  \label{eq:subst1Am=m*}
  z_1-\lambda_0 = \frac{ \lambda_0\!-\!\nu_0 }{
  \mu_0\!-\!\nu_0 } (z_1\!-\!\mu_0) - \frac{
  \lambda_0\!-\!\mu_0 }{ \mu_0\!-\!\nu_0 }
  (z_1\!-\!\nu_0). 
\end{equation}
For the contour $C_1^\lambda$, the first integral after the split can
be evaluated in terms of $E'(v)$, but the second integral we
leave unchanged. For the integral in the $z_2$-plane we substitute 
\begin{equation}
  \label{eq:subst2Am=m*}
  z_2-\nu_0 = \frac{ \lambda_0\!-\!\nu_0 }{
  \lambda_0\!-\!\mu_0 } (z_2\!-\!\mu_0) - \frac{
  \mu_0\!-\!\nu_0 }{ \lambda_0\!-\!\mu_0 }
  (z_2\!-\!\lambda_0). 
\end{equation}
We take $C_2^\nu$ as contour, and evaluate the first integral after
the split in terms of $E(v)$. We again leave the second integral
unchanged. Except for the contour choice, it is of the same form as
the integral we left unchanged in the $z_1$-plane.

To obtain the required boundary expression for $A$ at $\mu=\mu_0$, it
turns out that we have to add the contribution of {\em three} pairs of
contours, $C_1^\lambda C_2^\mu$, $C_1^\mu C_2^\nu$ and $C_1^\mu
C_2^\mu$. With the above substitutions \eqref{eq:subst1Am=m*} and
\eqref{eq:subst2Am=m*}, the first two pairs together provide the
required boundary expression, but in addition we have two similar
contour integrals
\begin{equation} 
  \label{eq:leftoverintegralAm=m*}
  \frac{i/8\pi}{ (\lambda_0\! \!-\! \nu_0\!)^\frac{1}{2}
    (\lambda \!-\! \nu)^\frac{1}{2} }
  \hspace{-3pt} \oint\limits_{C^\tau} \hspace{-3pt} 
  \frac{ (z \!-\! \lambda)^\frac{1}{2} \, \d z }{ 
       (z \!-\! \nu)^\frac{1}{2} 
    \! (z \!-\! \lambda_0\!)^\frac{1}{2} 
    \! (z \!-\! \nu_0\!)^\frac{1}{2}  
    \! (z \!-\! \mu_0\!) },
\end{equation}
with contours $C^\lambda$ and $C^\nu$, respectively.  The third pair,
$C_1^\mu C_2^\mu$, involves the product of two single pole
contributions. The resulting polynomial
\begin{equation}
  \label{eq:doublesingpole_Am=m*}
  \frac{i/8\pi}{ (\lambda_0\! \!-\! \nu_0\!)^\frac{1}{2}
    (\lambda \!-\! \nu)^\frac{1}{2} } \,
  \frac{ 2\pi i \, (\lambda \!-\! \mu_0\!)^\frac{1}{2} }{  
    \! (\mu_0\! \!-\! \nu)^\frac{1}{2} 
    \! (\lambda_0\! \!-\! \mu_0\!)^\frac{1}{2} 
    \! (\mu_0\! \!-\! \nu_0\!)^\frac{1}{2} }, 
\end{equation}
can be written in the same form as \eqref{eq:leftoverintegralAm=m*},
with contour $C^\mu$. As a result, we now have the same integral over
all three contours, so that from \eqref{eq:canceling3contours}, the
cumulative result vanishes and we are left with the required boundary
expression.

The expression for $A$ at $\nu=\nu_0$ resembles the one for $B$ at the
same boundary. This is expected since their boundary expressions in
\eqref{eq:BCshomogeneousproblem} are also very similar.  The first
integral now has a contribution from a double pole in the contour
$C_1^\nu$. The second integral has no contribution from the contour
$C_2^\nu$. However, within $C_2^\mu$, the second integral can be
evaluated in terms of $E(w)$. We obtain the correct boundary
expression $A(\lambda,\mu,\nu_0)$ by considering the pair $-C_1^\nu
C_2^\mu$.

\subsubsection{Combination of contours}
\label{sec:combinationofcontours}

In the previous paragraphs we have constructed a homogeneous solution
for $A$, $B$ and $C$, and we have shown that with this solution all
nine boundary expressions can be satisfied. For each boundary
expression separately, we have determined the required pair of
contours and also contours from which there is no contribution. Now we
have to find the right combination of all these contours to fit the
boundary expressions simultaneously.

We first summarise the required and non-contributing pairs of contours
per boundary expression
\begin{eqnarray}
  \label{eq:req&allcontours}
  A(\lambda_0,\mu,\nu) \!\!\!\!\!&:&  - C_1^\mu C_2^\lambda \pm
  C_1^\lambda C_2^\tau, \nonumber \\ 
  A(\lambda,\mu_0,\nu) \!\!\!\!\!&:&  + C_1^\mu C_2^\nu + C_1^\lambda
  C_2^\mu + C_1^\mu C_2^\mu, \nonumber \\ 
  A(\lambda,\mu,\nu_0) \!\!\!\!\!&:& - C_1^\nu C_2^\mu \pm C_1^\tau
  C_2^\nu, \nonumber \\*[10pt] 
  B(\lambda_0,\mu,\nu) \!\!\!\!\!&:& - C_1^\mu C_2^\lambda \pm
  C_1^\lambda C_2^\tau, \nonumber \\ 
  B(\lambda,\mu_0,\nu) \!\!\!\!\!&:&  
                        \pm C_1^\mu C_2^\tau \pm C_1^\tau C_2^\mu, \\
  B(\lambda,\mu,\nu_0) \!\!\!\!\!&:&  - C_1^\nu C_2^\mu \pm C_1^\tau
  C_2^\nu, \nonumber \\*[10pt] 
  C(\lambda_0,\mu,\nu) \!\!\!\!\!&:& - C_1^\mu C_2^\lambda \pm
  C_1^\lambda C_2^\tau, \nonumber \\  
  C(\lambda,\mu_0,\nu) \!\!\!\!\!&:& + C_1^\mu C_2^\nu + C_1^\lambda
  C_2^\mu + C_1^\mu C_2^\mu, \nonumber \\ 
  C(\lambda,\mu,\nu_0) \!\!\!\!\!&:&  - C_1^\nu C_2^\mu \pm C_1^\tau
  C_2^\nu, \nonumber  
\end{eqnarray}
where $\tau$ can be $\lambda$, $\mu$ or $\nu$. At each boundary
separately, $\lambda=\lambda_0$, $\mu=\mu_0$ and $\nu=\nu_0$, the
allowed combination of contours matches between $A$, $B$ and $C$. This
leaves the question how to relate the combination of contours at the
different boundaries relative to each other.

From \eqref{eq:canceling3contours}, we know that in both the complex
$z_1$-plane and $z_2$-plane, the cumulative contribution of the three
simple contours cancels. As a consequence, each of the following three
combinations of integration contours
\begin{equation}
  \label{eq:combcontour}
  C_1^\mu C_2^\mu = 
  - \, C_1^\mu \, (\, C_2^\lambda + C_2^\nu\, ) =
  - \, ( \, C_1^\lambda + C_1^\nu \, ) \, C_2^\mu,
\end{equation}
will give the same result. Similarly, we can add to each combination the
pairs $C_1^\lambda C_2^\mu$ and $C_1^\mu C_2^\nu$, to obtain
\begin{equation}
  \label{eq:contourcombination}
   C_1^\mu C_2^\nu \!+\! C_1^\lambda C_2^\mu \!+\! C_1^\mu C_2^\mu
   \!=\! C_1^\lambda C_2^\mu \!-\! C_1^\mu C_2^\lambda \!=\!
   C_1^\mu C_2^\nu \!-\! C_1^\nu C_2^\mu.
\end{equation}
The first combination of contour pairs matches the allowed range for
$\mu=\mu_0$ in \eqref{eq:req&allcontours} and the second and third
match the boundary expressions $\lambda=\lambda_0$ and
$\nu=\nu_0$. This completes the proof that the expressions
\eqref{eq:homsolform4A}--\eqref{eq:homsolform4C} for $A$, $B$ and $C$
solve the homogeneous equations \eqref{eq:homsystemABC} {\em and}
satisfy the nine boundary expressions \eqref{eq:BCshomogeneousproblem}
simultaneously when the integration contour is any of the three
combinations \eqref{eq:contourcombination}.  
We shall see below that the first of these combinations is preferred
in numerical evaluations.

\subsection{Evaluation of the homogeneous solutions}
\label{sec:evaluatinghomogenoussolution3D}

We write the complex contour integrals in the homogeneous solutions
$A$, $B$ and $C$ (\ref{eq:homsolform4A}--\ref{eq:homsolform4C}) as
real integrals. 
The resulting complete hyperelliptic integrals are expressed as single
quadratures, which can be evaluated numerically in a straightforward
way. 
We also express the complete elliptic integrals in the two-dimensional
homogeneous solutions $F$, $G$, $H$ and $I$ \eqref{eq:solFGHI} in this
way to facilitate their numerical evaluation.

\subsubsection{From complex to real integrals}
\label{sec:complex2realintegrals}

To transform the complex contour integrals in
\eqref{eq:homsolform4A}--\eqref{eq:homsolform4C} in real integrals we
wrap the contours $C^\lambda$, $C^\mu$ and $C^\nu$ around the
corresponding pair of branch points 
(Fig.~\ref{fig:wrappedcontourszplane2D}).   
The integrands consist of terms $z-\tau$ and $z-\tau_0$, all with
powers larger than $-1$, except $z_1-\nu_0$ and $z_2-\lambda_0$, both
of which occur to the power $-\frac32$.  
This means that for all simple contours $C_i^\tau$
$(\tau=\lambda,\mu,\nu; i=1,2)$, except for $C_1^\nu$ and
$C_2^\lambda$, the contribution from the arcs around the branch points
vanishes. 
In the latter case, we are left with the parts parallel to the real
axis, so that we can rewrite the complex integrals as real integrals
with the branch points as integration limits. 
The only combination of contours of the three given in
\eqref{eq:contourcombination} that does not involve both $C_1^\nu$ and
$C_2^\lambda$, is 
\begin{equation}
  \label{eq:triplecombofcontourpairs}
  S \equiv C_1^\mu C_2^\nu + C_1^\lambda C_2^\mu + C_1^\mu C_2^\mu.
\end{equation}
We have to be careful with the changes in phase when wrapping each of
the simple contours around the branch points. 
One can verify that the phase changes per contour are the same for all
three the homogeneous solutions $A$, $B$ and $C$, and also that the
contribution from the parts parallel to the real axis is equivalent. 
This gives a factor 2 per contour and thus a factor 4 for the
combination of contour pairs in $S$. 
In this way, we can transform the double complex contour integration
into the following combination of real integrals 
\begin{equation} 
  \label{eq:triplerealcombination}
  \iint\limits_S \hspace{-3pt} \d z_1 \d z_2 = 
  4 (
  \int\limits_\lambda^{\lambda_0} \hspace{-3pt} \d t_1 
  \! \int\limits_\mu^{\mu_0} \hspace{-3pt} \d t_2 +
  \! \int\limits_\mu^{\mu_0} \hspace{-3pt} \d t_1 
  \! \int\limits_\nu^{\nu_0} \hspace{-3pt} \d t_2 -
  \! \int\limits_\mu^{\mu_0} \hspace{-3pt} \d t_1 
  \! \int\limits_\mu^{\mu_0} \hspace{-3pt} \d t_2 
  ),
\end{equation}
with $t_i$ the real part of $z_i$. 

We apply this transformation to 
\eqref{eq:homsolform4A}--\eqref{eq:homsolform4C}, and we absorb the
factor of 4 left in the denominators into the integrals, so that we
can write  
\begin{eqnarray}
  \label{eq:ABCfunc}
  A(\!\lambda,\!\mu,\!\nu;\!\lambda_0,\!\mu_0,\!\nu_0\!)
  \hspace{-9pt} & = & \hspace{-10pt}   
  \frac{ (\mu_0\!\!-\!\nu_0\!) (\mu_0\!\!-\!\lambda_0\!) \Lambda }
  { \pi^2 (\lambda\!-\!\mu) (\lambda\!-\!\nu) }   
  \! \left( A_1 A_2 \!+\! A_3 A_4 \!-\! A_2 A_3 \!\right)\!, 
  \nonumber \\
  B(\!\lambda,\!\mu,\!\nu;\!\lambda_0,\!\mu_0,\!\nu_0\!)
  \hspace{-9pt} & = & \hspace{-10pt}   
  \frac{ (\mu_0\!\!-\!\nu_0\!) (\mu_0\!\!-\!\lambda_0\!) \Lambda } 
  { \pi^2 (\mu\!-\!\nu) (\mu\!-\!\lambda) } 
  \! \left( B_1 B_2 \!+\! B_3 B_4 \!-\! B_2 B_3 \!\right)\!, 
  \nonumber \\
  C(\!\lambda,\!\mu,\!\nu;\!\lambda_0,\!\mu_0,\!\nu_0\!)
  \hspace{-9pt} & = & \hspace{-10pt}   
  \frac{ (\mu_0\!\!-\!\nu_0\!) (\mu_0\!\!-\!\lambda_0\!) \Lambda }
  { \pi^2 (\nu\!-\!\lambda) (\nu\!-\!\mu) }  
  \! \left( C_1 C_2 \!+\! C_3 C_4 \!-\! C_2 C_3 \!\right)\!, 
  \nonumber \\*[-5pt] 
\end{eqnarray}
where $A_i$, $B_i$ and $C_i$ ($i=1,2,3,4$) are complete hyperelliptic 
integrals, for which we give expressions below, and 
\begin{equation}
  \label{eq:definitionofDelta}
  \Lambda^2 = \frac{ (\lambda\!-\!\mu) (\lambda\!-\!\nu) 
  (\mu\!-\!\nu) }{ (\lambda_0\!\!-\!\mu_0\!)
  (\lambda_0\!\!-\!\nu_0\!) (\mu_0\!\!-\!\nu_0\!) }.
\end{equation}
The second set of arguments added to $A$, $B$ and $C$ make explicit
the position $(\lambda_0,\mu_0,\nu_0)$ of the source point which is
causing the stresses at the field point $(\lambda,\mu,\nu)$.

\subsubsection{The complete hyperelliptic integrals}
\label{sec:complHYPERellintegrals}

With the transformation described in the previous section the
expression for, e.g., the complete hyperelliptic integral $A_2$ is of
the form 
\begin{equation}
  \label{eq:expressionA1_1}
  A_2 = \frac{1}{2}
  \int\limits_\mu^{\mu_0} \hspace{-5pt} 
  \frac{\d t}{\lambda_0\!-\!t} 
  \sqrt{ \frac{ (\lambda\!-\!t) (t\!-\!\nu_0) }
    { (\mu_0\!-\!t) (t\!-\!\mu) (\lambda_0\!-\!t) (t\!-\!\nu) } }.     
\end{equation}
The integrand has two singularities, one at the lower integration limit
$t=\mu$ and one at the upper integration limit $t=\mu_0$. 
The substitution $t=\mu+(\mu_0-\mu)\cos^2\theta$ removes
both singularities, since $\d t/\sqrt{ (\mu_0\!-\!t) (t\!-\!\mu) } =
2(\mu_0-\mu)\d \theta$. 

All complete hyperelliptic integrals $A_i$, $B_i$ and $C_i$
($i=1,2,3,4$) in \eqref{eq:ABCfunc} are of the form
\eqref{eq:expressionA1_1} and have at most two singularities at either
of the integration limits. 
Hence, we can apply a similar substitution to remove the
singularities. 
This results in the following expressions
\begin{subequations}
  \label{eq:AiBiCihyperell}
\begin{align}
   A_1 &\!=\! (\lambda_0\!-\!\lambda)^2 \hspace{-5pt} \int\limits_0^{\pi/2} 
  \!\! \frac{\sin^2\theta \cos^2\theta \d\theta}{x_3\Delta_x},  
  & A_2 &\!=\! \hspace{-5pt} \int\limits_0^{\pi/2} 
  \!\! \frac{y_1y_4\d\theta}{y_3\Delta_y},
  \nonumber \\*[-5pt]\\*[-5pt]
  A_4 &\!=\! (\nu_0\!-\!\nu) \hspace{-5pt} \int\limits_0^{\pi/2}
  \!\! \frac{z_2\sin^2\theta \d\theta}{z_1\Delta_z}, 
  & A_3 &\!=\! \hspace{-5pt} \int\limits_0^{\pi/2} 
  \!\! \frac{y_3y_4\d\theta}{y_1\Delta_y}, 
  \nonumber \label{eq:Aihyperell}
\end{align}
\begin{align}
  B_1 &\!=\! (\lambda_0\!-\!\lambda) \hspace{-5pt} \int\limits_0^{\pi/2} 
  \!\! \frac{x_2\sin^2\theta \d\theta}{x_3\Delta_x}, 
  & B_2 &\!=\! (\mu_0\!-\!\mu) \hspace{-5pt} \int\limits_0^{\pi/2}
  \!\! \frac{y_1\cos^2\theta \d\theta}{y_3\Delta_y}, 
  \nonumber \\*[-5pt]\\*[-5pt]
  B_4 &\!=\! (\nu_0\!-\!\nu) \hspace{-5pt} \int\limits_0^{\pi/2} 
  \!\! \frac{z_4\sin^2\theta \d\theta}{z_1\Delta_z}, 
  & B_3 &\!=\! (\mu_0\!-\!\mu) \hspace{-5pt} \int\limits_0^{\pi/2} 
  \!\! \frac{y_3\cos^2\theta \d\theta}{y_1\Delta_y}, 
  \nonumber \label{eq:Bihyperell}
\end{align}
\begin{align}
  C_1 &\!=\! (\lambda_0\!-\!\lambda) \hspace{-5pt} \int\limits_0^{\pi/2} 
  \!\! \frac{x_4\sin^2\theta \d\theta}{x_3\Delta_x}, 
  & C_2 &\!=\! \hspace{-5pt} \int\limits_0^{\pi/2} 
  \!\! \frac{y_1y_2\d\theta}{y_3\Delta_y}, 
  \nonumber \\*[-5pt]\\*[-5pt]
  C_4 &\!=\! (\nu_0\!-\!\nu)^2 \hspace{-5pt} \int\limits_0^{\pi/2} 
  \!\! \frac{\sin^2\theta \cos^2\theta \d\theta}{z_1\Delta_z}, 
  & C_3 &\!=\! \hspace{-5pt} \int\limits_0^{\pi/2} 
  \!\! \frac{y_2y_3\d\theta}{y_1\Delta_y}, 
  \nonumber \label{eq:Cihyperell}
\end{align}
\end{subequations}
where we have defined
\begin{equation}
  \label{eq:deltadefs}
  \Delta_x^2 = x_1x_2x_3x_4, \quad
  \Delta_y^2 = y_1y_2y_3y_4, \quad
  \Delta_z^2 = z_1z_2z_3z_4,
\end{equation}
and the factors $x_i$, $y_i$ and $z_i$ $(i=1,2,3,4)$ are given by
\begin{align}
  x_1 &\!=\! (\lambda\!-\!\mu_0) \!+\! 
  (\lambda_0\!-\!\lambda) \cos^2\theta, 
  & x_2 &\!=\! (\lambda\!-\!\mu) \!+\! 
  (\lambda_0\!-\!\lambda) \cos^2\theta,
  \nonumber \\
  x_3  &\!=\! (\lambda\!-\!\nu_0) \!+\! 
  (\lambda_0\!-\!\lambda) \cos^2\theta,
  & x_4 &\!=\! (\lambda\!-\!\nu) \!+\! 
  (\lambda_0\!-\!\lambda) \cos^2\theta,
  \nonumber \\
  \label{eq:xyzfac}
  y_1 &\!=\! (\mu\!-\!\nu_0) \!+\!     
  (\mu_0\!-\!\mu) \cos^2\theta,
  & y_2 &\!=\!   (\mu\!-\!\nu) \!+\!      
  (\mu_0\!-\!\mu) \cos^2\theta,
  \nonumber \\
  y_3 &\!=\! (\mu\!-\!\lambda_0) \!+\! 
  (\mu_0\!-\!\mu) \cos^2\theta,
  & y_4 &\!=\! (\mu\!-\!\lambda) \!+\!   
  (\mu_0\!-\!\mu) \cos^2\theta,
  \nonumber \\
  z_1 &\!=\! (\nu\!-\!\lambda_0) \!+\! 
  (\nu_0\!-\!\nu) \cos^2\theta,
  & z_2 &\!=\! (\nu\!-\!\lambda) \!+\!  
  (\nu_0\!-\!\nu) \cos^2\theta,
  \nonumber \\
  z_3 &\!=\! (\nu\!-\!\mu_0) \!+\!  
  (\nu_0\!-\!\nu) \cos^2\theta,
  & z_4 &\!=\! (\nu\!-\!\mu) \!+\!  
  (\nu_0\!-\!\nu) \cos^2\theta.
\end{align}
For each $i$ these factors follow from each other by cyclic
permutation of $\lambda \to \mu \to \nu \to \lambda$ and at the same
time $\lambda_0 \to \mu_0 \to \nu_0 \to \lambda_0$. 
Half of the factors -- all $x_i$, $y_1$ and $y_2$ -- are always
positive, whereas the other factors are always negative. 
The latter implies that one has to be careful with the signs of the
factors under the square root when evaluating the single quadratures
numerically.

\subsubsection{The complete elliptic integrals}
\label{sec:complellintegrals}

The two-dimensional homogeneous solutions $F$, $G$, $H$ and $I$ are
given in \eqref{eq:solFGHI} in terms of the Legendre complete elliptic
integrals $E(m)$ and $E'(m)=[E(m)-K(m)]/2m$.  
Numerical routines for $E(m)$ and $K(m)$ (e.g., Press et
al. 1999\nocite{Press99..numrecipies}) generally require the argument
to be $0 \le m < 1$. 
In the allowed range of the confocal ellipsoidal coordinates, 
the arguments $u$ \eqref{eq:defu*v*w*} and $w$
\eqref{eq:definitionofw} become larger than unity.
In these cases we can use transformations to express $E(m)$ and $K(m)$
in terms of $E(1/m)$ and $K(1/m)$ (e.g., Byrd \& Friedman
1971\nocite{BF1971...integralbook}). 

We prefer, however, to write the complete elliptic integrals as single
quadratures similar to the above expressions for the hyperelliptic
integrals. 
These quadratures can easily be evaluated numerically and apply to 
the full range of the confocal ellipsoidal coordinates. 
The resulting expressions for the two-dimensional homogeneous
solutions are
\begin{eqnarray}
  \label{eq:FGHIfunc}
  F(\lambda,\mu;\lambda_0,\mu_0) 
  \hspace{-7pt} & = & \hspace{-7pt}   
  \frac{1}{\pi} \sqrt{\!\frac{\lambda\!-\!\mu}{\lambda_0\!-\!\mu_0}} 
  \int\limits_0^{\pi/2} 
  \!\! \frac{x_1d\theta}{x_2\sqrt{x_1x_2}}, 
  \nonumber \\ 
  G(\lambda,\mu;\lambda_0,\mu_0) 
  \hspace{-7pt} & = & \hspace{-7pt} 
  \frac{1}{\pi} \sqrt{\!\frac{\lambda\!-\!\mu}{\lambda_0\!-\!\mu_0}} 
  (\mu_0\!-\!\mu) \!\!
  \int\limits_0^{\pi/2} 
  \!\! \frac{\sin^2\theta d\theta}{y_4\sqrt{y_3y_4}},  
  \nonumber \\
  H(\mu,\nu;\mu_0,\nu_0) 
  \hspace{-7pt} & = & \hspace{-7pt} 
  \frac{1}{\pi} \sqrt{\!\frac{\mu\!-\!\nu}{\mu_0\!-\!\nu_0}} 
  (\mu_0\!-\!\mu) \!\!
  \int\limits_0^{\pi/2} 
  \!\! \frac{\sin^2\theta d\theta}{y_2\sqrt{y_1y_2}},  
  \nonumber \\
  I(\mu,\nu;\mu_0,\nu_0) 
  \hspace{-7pt} & = & \hspace{-7pt} 
  \frac{1}{\pi} \sqrt{\!\frac{\mu\!-\!\nu}{\mu_0\!-\!\nu_0}} 
  \int\limits_0^{\pi/2} 
  \!\! \frac{z_3d\theta}{z_4\sqrt{z_3z_4}}.
\end{eqnarray}
Again we have added two arguments to make the position of the unit
source explicitly.  
We note that the homogeneous solutions
$A(\lambda,\mu;\lambda_0,\mu_0)$ and $B(\lambda,\mu;\lambda_0,\mu_0)$ 
for the disc case \eqref{eq:homogeneoussolutions} are equivalent to
$F$ and $G$ respectively.

\subsection{General triaxial solution}
\label{sec:generaltriaxialsolution}

We now construct the solution of the Jeans equations for triaxial
St\"ackel models \eqref{eq:jeanstriaxialS}, by superposition of
singular solutions, which involve the homogeneous solution derived in
the above. 
We match the solution to the boundary conditions at $\mu=-\alpha$ and
$\nu=-\beta$, and check for convergence of the solution when $\lambda
\to \infty$. 
Next, we consider alternative boundary conditions and present the
triaxial solution for a general finite region.
We also show that the general solution yields the correct result in
the case of thin tube orbits and the triaxial Abel models of 
Dejonghe \& Laurent (1991\nocite{1991MNRAS.252..606D}).
Finally, we describe a numerical test of the triaxial solution to a
polytrope model.

\subsubsection{Superposition of singular solutions}
\label{sec:superpositionofsingsol}

Substitution of the functions $A$, $B$, $C$ \eqref{eq:ABCfunc} and the
functions $F$, $G$, $H$, $I$ \eqref{eq:FGHIfunc} in expression 
\eqref{eq:functionalformsStt}, provides the three singular solutions
of the system of simplified Jeans equations, with the right-hand side
given by \eqref{eq:rhssimplyfying}. 
We denote these by $S_2^{\tau\tau}$ $(\tau=\lambda,\mu,\nu)$. 
The singular solutions of the two similar simplified systems, with the
triplet of delta functions at the right-hand side of the
\textit{first} and \textit{third} equation, $S_1^{\tau\tau}$ and
$S_3^{\tau\tau}$ then follow from $S_2^{\tau\tau}$ by cyclic
permutation. 
This gives
\begin{subequations}
  \label{eq:ninesingularsolutions}
  \begin{eqnarray}
    \label{eq:singsolsyst1}
    S_1^{\lambda\lambda} 
    \hspace{-9pt} &=& \hspace{-8pt} 
    B(\nu,\lambda,\mu;\nu_0,\lambda_0,\mu_0) 
    \!+\! G(\nu,\lambda;\nu_0,\lambda_0)\delta(\mu_0\!-\!\mu)
    \nonumber \\
    \hspace{-9pt} & & \hspace{-8pt} 
    + H(\lambda,\mu;\lambda_0,\mu_0)\delta(\nu_0\!-\!\nu) 
    \!-\! \delta(\mu_0\!-\!\mu)\delta(\nu_0\!-\!\nu), 
    \nonumber \\
    S_1^{\mu\mu} 
    \hspace{-9pt} &=& \hspace{-8pt} 
    C(\nu,\lambda,\mu;\nu_0,\lambda_0,\mu_0) 
    \!+\! I(\lambda,\mu;\lambda_0,\mu_0)\delta(\nu_0\!-\!\nu) 
    \nonumber \\
    S_1^{\nu\nu} 
    \hspace{-9pt} &=& \hspace{-8pt} 
    A(\nu,\lambda,\mu;\nu_0,\lambda_0,\mu_0) 
    \!+\! F(\nu,\lambda;\nu_0,\lambda_0)\delta(\mu_0\!-\!\mu),
  \end{eqnarray}
  \begin{eqnarray}
    \label{eq:singsolsyst2}
    S_2^{\lambda\lambda} 
    \hspace{-9pt} &=& \hspace{-8pt} 
    A(\lambda,\mu,\nu;\lambda_0,\mu_0,\nu_0) 
    \!+\! F(\lambda,\mu;\lambda_0,\mu_0)\delta(\nu_0\!-\!\nu), 
    \nonumber \\
    S_2^{\mu\mu} 
    \hspace{-9pt} &=& \hspace{-8pt} 
    B(\lambda,\mu,\nu;\lambda_0,\mu_0,\nu_0) 
    \!+\! G(\lambda,\mu;\lambda_0,\mu_0)\delta(\nu_0\!-\!\nu)
    \nonumber \\
    \hspace{-9pt} & & \hspace{-8pt} 
    + H(\mu,\nu;\mu_0,\nu_0)\delta(\lambda_0\!-\!\lambda) 
    \!-\! \delta(\nu_0\!-\!\nu)\delta(\lambda_0\!-\!\lambda), 
    \nonumber \\
    S_2^{\nu\nu} 
    \hspace{-9pt} &=& \hspace{-8pt} 
    C(\lambda,\mu,\nu;\lambda_0,\mu_0,\nu_0) 
    \!+\! I(\mu,\nu;\mu_0,\nu_0)\delta(\lambda_0\!-\!\lambda) 
  \end{eqnarray}
  \begin{eqnarray}
    \label{eq:singsolsyst3}
    S_3^{\lambda\lambda} 
    \hspace{-9pt} &=& \hspace{-8pt} 
    C(\mu,\nu,\lambda;\mu_0,\nu_0,\lambda_0) 
    \!+\! I(\nu,\lambda;\nu_0,\lambda_0)\delta(\mu_0\!-\!\mu),
    \nonumber \\
    S_3^{\mu\mu} 
    \hspace{-9pt} &=& \hspace{-8pt} 
    A(\mu,\nu,\lambda;\mu_0,\nu_0,\lambda_0) 
    \!+\! F(\mu,\nu;\mu_0,\nu_0)\delta(\lambda_0\!-\!\lambda) 
    \nonumber \\
    S_3^{\nu\nu} 
    \hspace{-9pt} &=& \hspace{-8pt} 
    B(\mu,\nu,\lambda;\mu_0,\nu_0,\lambda_0) 
    \!+\! G(\mu,\nu;\mu_0,\nu_0)\delta(\lambda_0\!-\!\lambda) 
    \nonumber \\
    \hspace{-9pt} & & \hspace{-8pt} 
    + H(\nu,\lambda;\nu_0,\lambda_0)\delta(\mu_0\!-\!\mu)
    \!-\! \delta(\lambda_0\!-\!\lambda)\delta(\mu_0\!-\!\mu). 
  \end{eqnarray}
\end{subequations}
These singular solutions describe the contribution of a source point
in $(\lambda_0, \mu_0, \nu_0)$ to $(\lambda, \mu, \nu)$. To find the
solution of the full equations \eqref{eq:jeanstriaxialS}, we multiply
the singular solutions \eqref{eq:singsolsyst1},
\eqref{eq:singsolsyst2} and \eqref{eq:singsolsyst3} by
$g_1(\lambda_0,\mu_0,\nu_0)$, $g_2(\lambda_0,\mu_0,\nu_0)$ and
$g_3(\lambda_0,\mu_0,\nu_0)$, respectively, so that the contribution
from the source point naturally depends on the local density and
potential (cf.\ eq.~[\ref{eq:deffunctiong_1}]). 
Then, for each coordinate $\tau=\lambda,\mu,\nu$, we add the three
weighted singular solutions, and integrate over the volume $\Omega$,
defined as  
\begin{equation}
  \label{eq:volumeOmega}
  \Omega \!=\! \left\{ (\lambda_0,\mu_0,\mu_0\!) \!:\! 
    \lambda \!\le\! \lambda_0 \!\!<\! \!\infty, 
    \mu \!\le\! \mu_0 \!\!\le\! \!-\alpha,
    \nu \!\le\! \nu_0 \!\!\le\! \!-\beta
  \right\} \!, 
\end{equation}
which is the three-dimensional extension of the integration domain $D$
in Fig.~\ref{fig:pointmasscontribution}.  
The resulting solution solves the inhomogeneous Jeans equations
\eqref{eq:jeanstriaxialS}, but does not give the correct values at the
boundaries $\mu=-\alpha$ and $\nu=-\beta$. 
They are found by multiplying the singular solutions
\eqref{eq:singsolsyst2} evaluated at $\mu_0=-\alpha$, and, similarly,
the singular solutions \eqref{eq:singsolsyst3} evaluated at 
$\nu_0=-\beta$, by $-S_{\mu\mu}(\lambda_0,-\alpha,\nu_0)$ and
$-S_{\nu\nu}(\lambda_0,\mu_0,-\beta)$, respectively, and integrating
in $\Omega$ over the coordinates that are not fixed. 
One can verify that this procedure represents the boundary values
correctly.  
The final result for the general solution of the Jeans equations
\eqref{eq:jeanstriaxialS} for triaxial St\"ackel models is
\begin{eqnarray}
  \label{eq:gentriaxsolStt}
  S_{\tau\tau} 
  (\lambda,\hspace{-11pt} & \mu & \hspace{-11pt},\!\nu) \!=\!
  \!\! \int\limits_\lambda^{\infty} \hspace{-5pt} \d\lambda_0 
  \!\!\! \int\limits_\mu^{-\alpha} \hspace{-5pt} \d\mu_0  
  \!\!\! \int\limits_\nu^{-\beta} \hspace{-5pt} \d\nu_0
  \!\! \sum_{i=1}^3 g_i(\lambda_0\!,\!\mu_0\!,\!\nu_0)
  S_i^{\tau\tau} \! (\lambda,\!\mu,\!\nu;\!\lambda_0\!,\!\mu_0\!,\!\nu_0) 
  \nonumber \\*[-3pt]
  \hspace{-7pt} & - & \hspace{-7pt}
  \!\!\! \int\limits_\nu^{-\beta} \hspace{-4pt} \d\nu_0    
  \!\!\! \int\limits_\lambda^{\infty} \hspace{-4pt} \d\lambda_0 
  \, S_{\mu\mu}(\lambda_0,\!-\alpha,\!\nu_0) \, 
  S_2^{\tau\tau} \!
  (\lambda,\!\mu,\!\nu;\!\lambda_0,\!-\alpha,\!\nu_0)  
  \nonumber \\*[-3pt]
  \hspace{-7pt} & - & \hspace{-7pt}
  \!\!\! \int\limits_\lambda^{\infty} \hspace{-4pt} \d\lambda_0 
  \!\!\! \int\limits_\mu^{-\alpha} \hspace{-4pt} \d\mu_0    
  \, S_{\nu\nu}(\lambda_0,\!\mu_0,\!-\beta) \, 
  S_3^{\tau\tau} \!
  (\lambda,\!\mu,\!\nu;\!\lambda_0\!,\!\mu_0\!,\!-\beta),  
\end{eqnarray}
where $\tau=(\lambda, \mu, \nu)$.  
This gives the stresses everywhere, once we have specified
$S_{\mu\mu}(\lambda,-\alpha,\nu)$ and
$S_{\nu\nu}(\lambda,\mu,-\beta)$. 
At both boundaries $\mu=-\alpha$ and $\nu=-\beta$, the three stress
components are related by a set of two Jeans equations, i.e.,
\eqref{eq:jeanstriaxialS} evaluated at $\mu=-\alpha$ and $\nu=-\beta$
respectively. 
From \S\ref{sec:2Dcases}, we know that the solution of these
two-dimensional systems both will involve a (boundary) function of one
variable. 
We need this latter freedom to satisfy the continuity conditions
\eqref{eq:triaxialcontcond}. 
This means it is sufficient to specify any of the three stress
components at $\mu=-\alpha$ and $\nu=-\beta$.

\subsubsection{Convergence of the general triaxial solution}
\label{sec:convergencetrisol}

As in \S\S\ref{sec:checkingforconsistency}, \ref{sec:convdiscsol} and
\ref{sec:applicationdiscsol} we suppose
$G(\tau)=\mathcal{O}(\tau^\delta)$ when $\tau\to \infty$, with $\delta$
in the range $[-\frac12,0)$. 
This implies that the potential $V_S$ \eqref{eq:formstackelpotential}
is also $\mathcal{O}(\tau^\delta)$.  
We assume that the density $\rho$, which does not need to be the
density $\rho_S$ which generates $V_S$, is of the form $N(\mu,
\nu)\lambda^{-s/2}$ when $\lambda\to\infty$. 
In the special case where $\rho=\rho_S$, we have $s\le 4$ except
possibly along the $z$-axis. 
When $s=4$ the models remain flattened out to the largest radii, but
when $s< 4$ the function $N(\mu,\nu)\to 1$ in the limit $\lambda\to
\infty$ (de Zeeuw et al. 1986\nocite{1986MNRAS.221.1001D}). 

From the definition \eqref{eq:deffunctiong_1}, we find that 
$g_1(\lambda_0,\mu_0,\nu_0) = \mathcal{O}(\lambda_0^{\delta-s/2})$
as $\lambda_0\to\infty$, while $g_2(\lambda_0,\mu_0,\nu_0)$ and
$g_3(\lambda_0,\mu_0,\nu_0)$ are larger and both
$\mathcal{O}(\lambda_0^{-s/2})$. 
To investigate the behaviour of the singular solutions
\eqref{eq:ninesingularsolutions} at large distance, we have to
carefully analyse the complete hyperelliptic \eqref{eq:AiBiCihyperell}
and elliptic \eqref{eq:FGHIfunc} integrals as $\lambda_0\to\infty$.
This is simplified by writing them as Carlson's $R$-functions (Carlson
1977\nocite{C77..carlsonfunc}). 
We finally find for the singular solutions that
$S_1^{\tau\tau}=\mathcal{O}(1)$ when $\lambda_0\to\infty$, whereas 
$S_2^{\tau\tau}$ and $S_3^{\tau\tau}$ are smaller and 
$\mathcal{O}(\lambda_0^{-1})$, with $\tau=\lambda,\mu,\nu$. 
This shows that for the volume integral in the triaxial solution
\eqref{eq:gentriaxsolStt} to converge, we must have 
$\delta-s/2+1<0$. 
This is equivalent to the requirement $s>2\delta+2$ we obtained in \S
\ref{sec:applicationdiscsol} for the limiting cases of prolate and
oblate potentials and for the large radii limit with scale-free DF.   
From the convergence of the remaining two double integrals in
\eqref{eq:gentriaxsolStt}, we find that the boundary stresses
$S_{\mu\mu}(\lambda,-\alpha,\nu)$ and 
$S_{\nu\nu}(\lambda,\mu,-\beta)$ cannot exceed 
$\mathcal{O}(1)$ when $\lambda\to\infty$. 

The latter is in agreement with the large $\lambda$ behaviour of
$S_{\tau\tau}(\lambda,\mu,\nu)$ that follows from the volume
integral. 
The singular solutions $S_i^{\lambda\lambda}=\mathcal{O}(1)$
($i=1,2,3$) when $\lambda\to\infty$, larger than $S_i^{\mu\mu}$ and
$S_i^{\nu\nu}$, which are all $\mathcal{O}(\lambda^{-1})$.  
Evaluating the volume integral at large distance then gives 
$S_{\tau\tau}(\lambda,\mu,\nu)=\mathcal{O}(\lambda^{\delta-s/2+1})$,
i.e., not exceeding $\mathcal{O}(1)$ if the requirement $s>2\delta+2$
is satisfied. 
We obtain the same behaviour and requirement from the energy equation
\eqref{eq:definitionhamiltonian}. 
 
We conclude that for the general triaxial case, as well as for the
limiting cases with a three-dimensional shape, the stress components 
$T_{\tau\tau}(\lambda,\mu,\nu)$ are 
$\mathcal{O}(\lambda^{\delta-s/2})$ at large distance, with the
requirement that $s>2\delta+2$ for $-\frac12\le\delta<0$. 
We obtained the same result for the stresses in the disc case, except
that then $s>2\delta+1$.  
Both the three-dimensional and two-dimensional requirements are met
for many density distributions $\rho$ and potentials $V_S$ of
interest. 
They do not break down until the isothermal limit $\delta\to0$, with
$s=1$ (disc) and $s=2$ (three-dimensional) is reached.

\subsubsection{Alternative boundary conditions}
\label{sec:alternativeBCstriaxial}

Our solution for the stress components at each point $(\lambda, \mu,
\nu)$ in a triaxial model with a St\"ackel potential consists of the
weighted contribution of all sources outwards of this point.
Accordingly, we have integrated with respect to $\lambda_0$, $\mu_0$
and $\nu_0$, with lower limits the coordinates of the chosen point and
upper limits $\infty$, $-\alpha$ and $-\beta$, respectively. To obtain
the correct expressions at the outer boundaries, the stresses must
vanish when $\lambda \to \infty$ and they have to be specified at
$\mu=-\alpha$ and $\nu=-\beta$.

The integration limits $\lambda$, $\mu$ and $\nu$ are fixed, but for
the other three limits we can, in principle, equally well choose
$-\alpha$, $-\beta$ and $-\gamma$ respectively. The latter choices
also imply the specification of the stress components at these
boundaries instead. Each of the eight possible combinations of these
limits corresponds to one of the octants into which the physical
region $-\gamma \le \nu_0 \le -\beta \le \mu_0 \le -\alpha \le
\lambda_0 < \infty$ is split by the lines through the point $(\lambda,
\mu, \nu)$. 
By arguments similar to those given in \S\ref{sec:alternativeBCsdisc},
one may show that in all octants the expressions \eqref{eq:ABCfunc}
for $A$, $B$, $C$, and \eqref{eq:solFGHI} for $F$, $G$, $H$, $I$ are
equivalent. 
Hence, again the only differences in the singular
solutions are due to possible changes in the sign of the
step-functions, but the changes in the integration limits cancel the
sign differences between the corresponding singular solutions.
However, as in \S\ref{sec:alternativeBCsdisc} for the two-dimensional
case, it is not difficult to show that while switching the boundary
conditions $\mu$ and $\nu$ is indeed straightforward, the switch from
$\lambda\to\infty$ to $\lambda=-\alpha$ again leads to solutions which
generally have the incorrect radial fall-off, and hence are
non-physical.

\subsubsection{Triaxial solution for a general finite region}
\label{sec:trisolfiniteregion}

If we denote non-fixed integration limits by $\lambda_e$, $\mu_e$ and
$\nu_e$ respectively, we can write the triaxial solution for a
general finite region as
\begin{eqnarray}
  \label{eq:generaltriaxsolStt}
  S_{\tau\tau} 
  (\lambda,\hspace{-11pt} & \mu & \hspace{-11pt},\!\nu) \!=\!
  \!\! \int\limits_\lambda^{\lambda_e} \hspace{-4pt} \d\lambda_0 
  \!\!\! \int\limits_\mu^{\mu_e} \hspace{-4pt} \d\mu_0  
  \!\!\! \int\limits_\nu^{\nu_e} \hspace{-4pt} \d\nu_0
  \!\! \sum_{i=1}^3 g_i(\lambda_0\!,\!\mu_0\!,\!\nu_0)
  S_i^{\tau\tau} \! (\lambda,\!\mu,\!\nu;\!\lambda_0\!,\!\mu_0\!,\!\nu_0) 
  \nonumber \\*[-3pt]
  \hspace{-7pt} & -  & \hspace{-7pt}
  \!\!\! \int\limits_\mu^{\mu_e} \hspace{-4pt} \d\mu_0 
  \!\!\! \int\limits_\nu^{\nu_e} \hspace{-4pt} \d\nu_0  
  \, S_{\lambda\lambda}(\lambda_e,\!\mu_0\!,\!\nu_0) \, 
  S_1^{\tau\tau} \!
  (\lambda,\!\mu,\!\nu;\!\lambda_e,\!\mu_0\!,\!\nu_0)  
  \nonumber \\*[-3pt]
  \hspace{-7pt} & - & \hspace{-7pt}
  \!\!\! \int\limits_\nu^{\nu_e} \hspace{-4pt} \d\nu_0    
  \!\!\! \int\limits_\lambda^{\lambda_e} \hspace{-4pt} \d\lambda_0 
  \, S_{\mu\mu}(\lambda_0\!,\!\mu_e,\!\nu_0) \, 
  S_2^{\tau\tau} \!
  (\lambda,\!\mu,\!\nu;\!\lambda_0\!,\!\mu_e,\!\nu_0)  
  \nonumber \\*[-3pt]
  \hspace{-7pt} & - & \hspace{-7pt}
  \!\!\! \int\limits_\lambda^{\lambda_e} \hspace{-4pt} \d\lambda_0 
  \!\!\! \int\limits_\mu^{\mu_e} \hspace{-4pt} \d\mu_0    
  \, S_{\nu\nu}(\lambda_0\!,\!\mu_0\!,\!\nu_e) \, 
  S_3^{\tau\tau} \!
  (\lambda,\!\mu,\!\nu;\!\lambda_0\!,\!\mu_0\!,\!\nu_e),  
\end{eqnarray}
with, as usual, $\tau=\lambda, \mu, \nu$. The weight functions
$g_i$ ($i=1,2,3$) are defined in \eqref{eq:deffunctiong_1} and the
singular solutions $S_i^{\tau\tau}$ are given by
\eqref{eq:ninesingularsolutions}. The non-fixed integration limits are
chosen in the corresponding physical ranges, i.e., 
$\lambda_e\in[-\alpha,\infty]$, $\mu_e\in[-\beta,-\alpha]$ and
$\nu_e\in[-\gamma,-\beta]$, but $\lambda_e \ne -\alpha$ (see \S
\ref{sec:alternativeBCstriaxial}).  
The solution requires the specification of
the stress components on the boundary surfaces $\lambda=\lambda_e$,
$\mu=\mu_e$ and $\nu=\nu_e$. 
On each of these surfaces the three stress components are related by two
of the three Jeans equations \eqref{eq:jeanstriaxialS} and the
continuity conditions \eqref{eq:triaxialcontcond}. 
Hence, once one of the stress components is prescribed on three
boundary surfaces, the solution \eqref{eq:gentriaxsolStt} yields all
three stresses everywhere in the triaxial St\"ackel galaxy. 
The stresses on the remaining three boundary surfaces then follow as the
limits of the latter solution. \looseness=-2

\subsubsection{Physical solutions}
\label{sec:physical solutions}

Statler (1987\nocite{1987ApJ...321..113S}) and HZ92 showed that many
different DFs are consistent with a triaxial density $\rho$ in the
potential $V_S$.  Specifically, the boundary plane $\nu=-\beta$, i.e.,
the area outside the focal hyperbola in the ($x,z$)-plane
(Fig.~\ref{fig:focalplanes}), is only reached by inner (I) and outer
(O) long-axis tube orbits. A split between the contribution of both
orbit families to the density in this plane has to be chosen, upon
which the DF for both the I and O orbits is fixed in case only thin
tubes are populated, but many other possibilities exist when the full
set of I- and O-orbits is included. For each of these DFs, the density
provided by the I- and O-tubes can then in principle be found
throughout configuration space. In the area outside the focal ellipse
in the ($y,z$)-plane ($\mu=-\alpha$), only the O-tubes and S-tubes
contribute to the density. Subtracting the known density of the
O-orbits leaves the density to be provided by the S-tubes in this
plane, from which their DF can be determined. This is again unique
when only thin orbits are used, but is non-unique otherwise. The
density that remains after subtracting the I-, O-, and S-tube
densities from $\rho$ must be provided by the box (B) orbits. Their DF
is now fixed, and can be found by solving a system of linear
equations, starting from the outside ($\lambda \to \infty$).  

The total DF is the sum of the DFs of the four orbit families, and is
hence highly non-unique. All these DFs give rise to a range of
stresses $T_{\lambda\lambda}, T_{\mu\mu}, T_{\nu\nu}$, and our
solution of the Jeans equations must be sufficiently general to
contain them as a subset. This is indeed the case, as we are allowed
to choose the stresses on the special surfaces $\nu=-\beta$ and
$\mu=-\alpha$. However, not all choices will correspond to physical
DFs.  The requirement $T_{\tau\tau}\ge 0$ is necessary but not
sufficient for the associated DF to be non-negative everywhere.

\subsubsection{The general solution for thin tube orbits}
\label{sec:thintubeorbitssolution}

For each of the three tube families in case of infinitesimally thin
orbits one of the three stress components vanishes everywhere (see
\S\ref{sec:thintubeorbits}).  
We are left with two non-zero stress components related to the density
and potential by three reduced Jeans equations
\eqref{eq:jeanstriaxialS}. 
We thus have subsidiary conditions on the three right hand side terms
$g_1$, $g_2$ and $g_3$. 
 
HZ92 solved for the two non-trivial stresses and showed that they can
be found by single quadratures (with integrands involving no worse
than complete elliptic integrals), once the corresponding stress had
been chosen at $\nu=-\beta$ (for I- and O-tubes) or at $\mu=-\alpha$
(for S-tubes).   

By analogy with the reasoning for the thin tube orbits in the disc
case (\S \ref{sec:thinlooporbits}), we can show that for each of the
three tube families in the case of thin orbits the general triaxial
solution   \eqref{eq:generaltriaxsolStt} gives the stress components
correctly. 
Consider, e.g., the thin I-tubes, for which $S_{\mu\mu} \equiv 0$.
Apply the latter to \eqref{eq:generaltriaxsolStt}, substitute for
$g_1$, $g_2$ and $g_3$ the subsidiary conditions that follow from the
reduced Jeans equations \eqref{eq:jeanstriaxialS} and substitute for
the singular solutions the expressions
\eqref{eq:ninesingularsolutions}.  
After several partial integrations, we use that the homogeneous
solutions $A$, $B$ and $C$ solve a homogeneous system similar to
\eqref{eq:homsystemABC}, but now with respect to the source point 
coordinates $(\lambda_0,\mu_0,\nu_0)$ 
\begin{equation}
  \label{eq:homsystemABC_0}
  \frac{ \partial B(\nu,\!\lambda,\!\mu;\!\nu_0,\!\lambda_0,\!\mu_0\!) }
  { \partial \lambda_0 } \!=\!
  \frac{ A(\lambda,\!\mu,\!\nu;\!\lambda_0,\!\mu_0,\!\nu_0\!) }
  { 2(\lambda_0\!-\!\mu_0) } 
  + \frac{ C(\mu,\!\nu,\!\lambda;\!\mu_0,\!\nu_0,\!\lambda_0\!) }
  { 2(\lambda_0\!-\!\nu_0) }\!,
\end{equation}
where other relations follow by cyclic permutation of $\lambda \to
\mu \to \nu \to \lambda$ and $\lambda_0 \to \mu_0 \to \nu_0 \to
\lambda_0$.
And similar for the two-dimensional homogeneous solutions $F$, $G$,
$H$ and $I$ the relations follow from 
\begin{eqnarray}
  \label{eq:2DsysFGHI_0}
  \frac{ \partial G(\mu,\lambda;\mu_0,\lambda_0) }
  { \partial \lambda_0 }   
  \hspace{-7pt} & = & \hspace{-7pt}  
  \frac{ F(\lambda,\mu;\lambda_0,\mu_0) }
  { 2(\lambda_0\!-\!\mu_0) }, 
  \nonumber \\*[-5pt]\\*[-5pt]
  \frac{ \partial H(\mu,\nu;\mu_0,\nu_0) }
  {\partial \mu_0}
  \hspace{-7pt} & = & \hspace{-7pt}  
  \frac{ I(\nu,\mu;\nu_0,\mu_0) }
  { 2(\mu_0\!-\!\nu_0) }. 
  \nonumber
\end{eqnarray}
It indeed turns out that for $S_{\mu\mu}(\lambda,\mu,\nu)$ all terms
cancel on the right hand side of \eqref{eq:generaltriaxsolStt}. 
The terms that are left in the case of $S_{\lambda\lambda}$ and
$S_{\nu\nu}$ are just eq.~\eqref{eq:jeanstriaxial_lambdaS} integrated
with respect to $\lambda$ and eq.~\eqref{eq:jeanstriaxial_nuS}
integrated with respect to $\nu$, respectively, and using that
$S_{\mu\mu} \equiv 0$.
A similar analysis as above shows that also for thin O- and S-tubes
-- for which $S_{\lambda\lambda} \equiv 0$ in both cases -- the
general triaxial solution yields the correct result.

\subsubsection{Triaxial Abel models}
\label{sec:abelmodels}

For a galaxy with a triaxial potential of St\"ackel form, the DF is a
function of the three exact isolating  integrals of motion,
$f(\mathbf{x},\mathbf{v})=f(E,I_2,I_3)$ 
(see also \S \ref{sec:velocitymoments}). 
The expressions for $E$, $I_2$ and $I_3$ in terms of the phase-space
coordinates ($\mathbf{x},\mathbf{v}$) can be found in e.g. Z85.
We can thus write the velocity moments of the DF as a triple integral
over $E$, $I_2$ and $I_3$. 
Assuming that the DF is function of only one variable 
\begin{equation}
  \label{eq:abel_S}
  S \equiv E + wI_2 + uI_3,
\end{equation}
with $w$ and $u$ constants, Dejonghe \& Laurent
(1991)\nocite{1991MNRAS.252..606D} show that the triple integration
simplifies to a one-dimensional Abel integration over $S$. 
Even though a DF of this form can only describe a self-consistent
model in the spherical case (ellipsoidal hypothesis, see, e.g.,  
Eddington 1915\nocite{1915MNRAS..76...37E}), the Jeans equations do
not require self-consistency. 

The special Abel form results in a simple analytical relation between
the three stress components (Dejonghe \& Laurent
1991\nocite{1991MNRAS.252..606D},  their eq.~[5.6])
\begin{equation}
  \label{eq:abel_stressrelations}
  T_{\mu\mu} = T_{\lambda\lambda} a_{\mu\nu} / a_{\lambda\nu}, 
  \qquad
  T_{\nu\nu} = T_{\lambda\lambda} a_{\mu\nu} / a_{\mu\lambda}, 
\end{equation}
with 
\begin{equation}
  \label{eq:abel_a_sigtau}
  a_{\sigma\tau} = (\gamma\!-\!\alpha) +
  (\sigma\!+\!\alpha)(\tau\!+\!\alpha)w -
  (\sigma\!+\!\gamma)(\tau\!+\!\gamma)u, 
\end{equation}
and $\sigma,\tau=\lambda,\mu,\nu$. With these relations we find that
\begin{equation}
  \label{eq:abel_ratiosinjeanseq}
  \frac{T_{\lambda\lambda}\!-\!T_{\mu\mu}}{\lambda\!-\!\mu} = 
  \frac{T_{\lambda\lambda}}{a_{\lambda\nu}} 
  \frac{\partial a_{\lambda\nu}}{\partial \lambda},
  \quad
  \frac{T_{\lambda\lambda}\!-\!T_{\nu\nu}}{\lambda\!-\!\nu} = 
  \frac{T_{\lambda\lambda}}{a_{\lambda\mu}} 
  \frac{\partial a_{\lambda\mu}}{\partial \lambda}.
\end{equation}
The first Jeans equation \eqref{eq:jeanstriaxial_lambda} 
now becomes a first-order partial
differential equation for $T_{\lambda\lambda}$. 
This equation can be solved in a straightforward way and provides an
elegant and simple expression for the radial stress component
\begin{multline}
  \label{eq:abel_Tll}
  T_{\lambda\lambda}(\lambda,\mu,\nu) = 
  \sqrt{ \frac{a_{\lambda_e\mu}a_{\lambda_e\nu}}
    {a_{\lambda\mu}a_{\lambda\nu}} } \;
  T_{\lambda\lambda}(\lambda_e,\mu,\nu)
  \\
  + \int\limits_\lambda^{\lambda_e} \hspace{-3pt} \d\lambda_0  
  \biggl[
  \sqrt{ \frac{a_{\lambda_0\mu}a_{\lambda_0\nu}}
    {a_{\lambda\mu}a_{\lambda\nu}} } \; 
  \rho \frac{\partial V_S}{\partial \lambda_0}
  \biggr].
\end{multline}
The expressions for $T_{\mu\mu}$ and $T_{\nu\nu}$ follow by
application of the ratios \eqref{eq:abel_stressrelations}. 
If we let the boundary value $\lambda_e \to \infty$, the first term on
the right-hand side of \eqref{eq:abel_Tll} vanishes. 

The density $\rho$, which does not need to be the density $\rho_S$
which generates $V_S$, is of the Abel form as given in eq.~(3.11) of 
Dejonghe \& Laurent (1991\nocite{1991MNRAS.252..606D}).  
If we substitute this form in \eqref{eq:abel_Tll}, we obtain, after
changing the order of integration and evaluating the integral with
respect to $\lambda$, again a single Abel integral that is equivalent
to the expression for $T_{\lambda\lambda}$ that follows from
eq.~(3.10) of by Dejonghe \& Laurent
(1991\nocite{1991MNRAS.252..606D}). 
Using the relations \eqref{eq:abel_stressrelations} and the
corresponding subsidiary conditions for $g_1$, $g_2$ and $g_3$, it can
be shown that the general triaxial solution
\eqref{eq:generaltriaxsolStt} gives the stress components correctly.

\subsubsection{Numerical test}
\label{sec:numericaltest}

We have numerically implemented the general triaxial solution
\eqref{eq:generaltriaxsolStt}, and tested it on a polytrope dynamical
model, for which the DF depends only on energy $E$ as $f(E) \propto
E^{n-3/2}$, with $n>\frac{1}{2}$.
Integration of this DF over velocity $v$, with $E=-V-\frac{1}{2}v^2$
for a potential $V\le0$, shows that the density $\rho \propto
(-V)^n$ (e.g., Binney \& Tremaine 1987\nocite{1987gady.book.....B},
p. 223).  
This density is not consistent with the St\"ackel potentials we use but,
as noted in \S\ref{sec:jeanseqns},
the Jeans equations do not require self-consistency.
The first velocity moments and the mixed second moments 
of the DF are all zero.
The remaining three moments all equal $-V/(n+1)$, so that the
isotropic stress of the polytrope model 
$T_{\mathrm{pol}} \propto (-V)^{n+1}$.   

We take the potential $V$ to be of St\"ackel form $V_S$
\eqref{eq:formstackelpotential}, and consider two different choices
for $G(\tau)$ in \eqref{eq:defgtau}. 
The first is the simple form
$G(\tau)=-GM/(\sqrt{\tau}+\sqrt{-\alpha})$ that is related to
H\'enon's isochrone 
(de Zeeuw \& Pfenniger 1988\nocite{1988MNRAS.235..949D}).  
The second is the form for the perfect ellipsoid, for which
$G(\tau)$ is given in Z85 in terms of complete elliptic integrals. 
The partial derivatives of $V_S(\lambda,\mu,\nu)$, that appear in the
weights $g_1$, $g_2$ and $g_3$, can be obtained in terms of $G(\tau)$
and its derivative in a straightforward way by using the expressions
derived by de Zeeuw et al.\ (1986\nocite{1986MNRAS.221.1001D}).  

The calculation of the stresses is done in the following way.
We choose the polytrope index $n$, and fix the triaxial St\"ackel
model by choosing $\alpha$, $\beta$ and $\gamma$.
This gives $T_{\mathrm{pol}}$.
Next, we obtain successively the stresses $T_{\lambda\lambda}$,
$T_{\mu\mu}$ and $T_{\nu\nu}$ from the general triaxial solution
\eqref{eq:generaltriaxsolStt} by numerical integration, where the
relation between $S_{\tau\tau}$ and $T_{\tau\tau}$
is given by \eqref{eq:deffuncsStt}. 
We first fix the upper integration limits $\lambda_e$, $\mu_e$ and
$\nu_e$.  
All integrands contain the singular solutions
\eqref{eq:ninesingularsolutions}, that involve the homogeneous
solutions $A$, $B$, $C$, $F$, $G$, $H$ and $I$, for which we
numerically evaluate the single quadratures (eq.~[\ref{eq:ABCfunc}],
[\ref{eq:AiBiCihyperell}] and [\ref{eq:FGHIfunc}]).  
The weights $g_1$, $g_2$ and $g_3$ \eqref{eq:deffunctiong_1} involve
the polytrope density and St\"ackel potential. 
This leaves the boundary stresses in the integrands, 
for which we use the polytrope stress $T_{\mathrm{pol}}$ that follows from the
choice of the DF, evaluated at the corresponding boundary surfaces. 
We then evaluate the general solution away from these boundaries, and
compare it with the known result. 

We carried out the numerical calculations for different choices of
$n$, $\alpha$, $\beta$ and $\gamma$ and at different field points
$(\lambda,\mu,\nu)$. 
In each case the resulting stresses $T_{\lambda\lambda}$, $T_{\mu\mu}$
and $T_{\nu\nu}$ -- independently calculated -- were equivalent to
high precision and equal to $T_{\mathrm{pol}}$. 
This agreement provides a check on the accuracy of both our formulae
and their numerical implementation, and demonstrates the feasibility
of using our methods for computing triaxial stress distributions. 
That will be the subject of a follow-up paper.
 
% ===== section 5 ===== %

\section{Discussion and conclusions}
\label{sec:discconc}

Eddington (1915\nocite{1915MNRAS..76...37E}) showed that the velocity
ellipsoid in a triaxial galaxy with a separable potential of St\"ackel
form is everywhere aligned with the confocal ellipsoidal coordinate
system in which the equations of motion separate. Lynden--Bell
(1960\nocite{thesisLynden-Bell}) derived the three Jeans equations which
relate the three principal stresses to the potential and the
density. They constitute a highly-symmetric set of first-order partial
differential equations in the three confocal coordinates. Solutions
were found for the various two-dimensional limiting cases, but with
methods that do not carry over to the general case, which, as a
consequence, remained unsolved.

In this paper, we have introduced an alternative solution method,
using superposition of singular solutions. 
We have shown that this approach not only provides an elegant
alternative to the standard Riemann--Green method for the
two-dimensional limits, but also, unlike the standard methods, can be
generalised to solve the three-dimensional system. 
The resulting solutions contain complete (hyper)elliptic integrals
which can be evaluated in a straightforward way. 
In the derivation, we have recovered (and in some cases corrected) all
previously known solutions for the various two-dimensional limiting
cases with more symmetry, as well as the two special solutions known
for the general case, and have also clarified the restrictions on the
boundary values. 
We have numerically tested our solution on a polytrope model. 

The general Jeans solution is not unique, but requires specification
of principal stresses at certain boundary surfaces, given a separable
triaxial potential, and a triaxial density distribution (not
necessarily the one that generates the potential). 
We have shown that these boundary surfaces can be taken to be the
plane containing the long and the short axis of the galaxy, and, more
specifically, the part that is crossed by all three families of tube
orbits and the box orbits. 
This is not unexpected, as HZ92 demonstrated that the phase-space
distribution functions of these triaxial systems are defined by
specifying the population of each of the three tube orbit families in
a principal plane. 
Once the tube orbit populations have been defined in this way, the
population of the box orbits is fixed, as it must reproduce the
density not contributed by the tubes, and there is only one way to do
this. 
While HZ92 chose to define the population of inner and outer long axis
tubes in a part of the $(x,z)$-plane, and the short axis tubes in a
part of the $(y,z)$-plane, it is in fact also possible to specify all
three of them in the appropriate parts of the $(x,z)$-plane, just as
is needed for the stresses.

The set of all Jeans solutions \eqref{eq:generaltriaxsolStt} contains
all the stresses that are associated with the physical distribution 
functions $f \geq 0$, but, as in the case of spherical and 
axisymmetric models, undoubtedly also contains solutions which are
unphysical, e.g., those associated with distribution functions that
are negative in some parts of phase space. 
The many examples of the use of spherical and axisymmetric Jeans
models in the literature suggest nevertheless that the Jeans solutions
can be of significant use.

While triaxial models with a separable potential do not provide an
adequate description of the nuclei of galaxies with cusped luminosity
profiles and a massive central black hole, they do catch much of the
orbital structure at larger radii, and in some cases even provide a
good approximation of the galaxy potential. 
The solutions for the mean streaming motions, i.e., the first velocity
moments of the distribution function, are quite helpful in
understanding the variety of observed velocity fields in giant
elliptical galaxies and constraining their intrinsic shapes 
(e.g., Statler 1991\nocite{1991AJ....102..882S},
1994b\nocite{1994ApJ...425..500S}; 
Arnold et al.1994\nocite{1994MNRAS.271..924A};
Statler, DeJonghe \& Smecker-Hane 1999\nocite{1999AJ....117..126S}; 
Statler 2001\nocite{2001AJ....121..244S}). 
We expect that the projected velocity dispersion fields that can be
derived from our Jeans solutions will be similarly useful, and, in
particular, that they can be used to establish which combinations of
viewing directions and intrinsic axis ratios are firmly ruled out by
the observations. 
As some of the projected properties of the St\"ackel models can be
evaluated by analytic means (Franx 1988\nocite{1988MNRAS.231..285F}),
it is possible that this holds even for the intrinsic moments
considered here. 
Work along these lines is in progress.

The solutions presented here constitute a significant step towards
completing the analytic description of the properties of the separable
triaxial models, whose history by now spans more than a century. It is
remarkable that the entire Jeans solution can be written down by means
of classical methods. This suggests that similar solutions can be
found for the higher dimensional analogues of
\eqref{eq:jeanstriaxial}, 
most likely involving hyperelliptic integrals of higher order. 
It is also likely that the higher-order velocity moments for the
separable triaxial models can be found by similar analytic means, but
the effort required may become prohibitive. 

% ===== ACKNOWLEDGMENTS ===== %

\section*{acknowledgments}
\label{sec:acknowledgments}

This paper owes much to Donald Lynden--Bell's enthusiasm and
inspiration. 
This work begun during a sabbatical visit by CH to Leiden
Observatory in 1992, supported in part by a Bezoekersbeurs from NWO,
and also by NSF through grant DMS 9001404. 
CH is currently supported by NSF through grant DMS 0104751. 
This research was supported in part by the Netherlands Research School
for Astronomy NOVA. 
The authors gratefully acknowledge stimulating discussions with Wyn
Evans during the initial phases of this work. %\looseness=-2

% ===== REFERENCES ===== %

\bibliographystyle{mn2e}

% ===== APPENDICES ===== %

\appendix

% ===== APPENDIX A ===== %

\section{Solving for the difference in stress}
\label{sec:solving4diff}

We compare our solution for the stress components $T_{\lambda\lambda}$
and $T_{\mu\mu}$ with the result derived by EL89. They combine the two
Jeans equations \eqref{eq:jeanselliptic} into the single equation
\begin{equation}
  \label{eq:PDEdiff}
  \frac{\partial^2 \Delta}{\partial \lambda \partial \mu}
  + \biggl( \frac{\partial}{\partial \mu}
  \!-\! \frac{\partial}{\partial \lambda} \biggl)
  \frac{ \Delta }{ 2 (\lambda\!-\!\mu) } =
  \frac{\partial \rho}{\partial \lambda}
  \frac{\partial V_S }{\partial \mu}
  \!-\!
  \frac{\partial \rho}{\partial \mu}
  \frac{\partial V_S }{\partial \lambda},
\end{equation}
for the difference $\Delta \equiv T_{\lambda\lambda} - T_{\mu\mu}$ of
the two stress components.  Eq.~\eqref{eq:PDEdiff} is of the form
\begin{equation}
  \label{eq:PDEdifftwo}
  \mathcal{L}^\star \Delta = 
  \frac{\partial \rho}{\partial \lambda}
  \frac{\partial V_S }{\partial \mu}
  \!-\!
  \frac{\partial \rho}{\partial \mu}
  \frac{\partial V_S }{\partial \lambda},
\end{equation}
where $\mathcal{L}^\star$ is the adjoint operator defined in
eq.~\eqref{eq:adjointoperatorL*}.  As in \S\ref{sec:riemannsmethod},
eq.~\eqref{eq:PDEdiff} can be solved via a Riemann--Green function.

\subsection{The Green's function}
\label{sec:greensfunction}

In order to obtain the Riemann--Green function $\mathcal{G}^\star$ for
the adjoint operator $\mathcal{L}^\star$, we use the reciprocity
relation (Copson 1975\nocite{C75}, \S5.2) to relate it to the Riemann--Green function
$\mathcal{G}$, derived in \S\ref{sec:findingtheriemanngreenfunction}
for $\mathcal{L}$. With $c_1=c_2=-\frac12$ in this case, we get
\looseness=-2
\begin{eqnarray}
  \label{eq:adjointRgreenfunc}
  \mathcal{G}^\star(\lambda,\mu;\lambda_0,\mu_0) 
  \hspace{-7pt} & = & \hspace{-7pt}
  \mathcal{G}(\lambda_0,\mu_0;\lambda,\mu) \nonumber \\
  \hspace{-7pt} & = & \hspace{-7pt}
  \left( 
    \frac{\lambda_0\!-\!\mu_0}{\lambda\!-\!\mu}
  \right)^\frac{1}{2} 
  {}_2F_1(-\scriptstyle\frac{1}{2}\displaystyle,
  \scriptstyle\frac{3}{2} \displaystyle;1;w),
\end{eqnarray}
where $w$ as defined in \eqref{eq:definitionofw}.  EL89 seek to solve
eq.~\eqref{eq:PDEdifftwo} using a Green's function $G$ which satisfies
the equation
\begin{equation}
  \label{eq:ELPDEdiff}
  \mathcal{L}^\star G
  =\delta(\lambda_0\!-\!\lambda)\delta(\mu_0\!-\!\mu). 
\end{equation}
That they impose
the same boundary conditions that we do is evident from their remark
that, if $\mathcal{L}^\star$ were the simpler operator
$\partial^2/\partial\lambda\partial\mu$, $G$ would be
$\mathcal{H}(\lambda_0\!-\!\lambda) \mathcal{H}(\mu_0\!-\!\mu)$. This
is the same result as would be obtained by the singular solution
method of \S\ref{sec:singularsolution2D}, which, as we showed there,
is equivalent to the Riemann--Green analysis. Hence their $G$ should
match the $\mathcal{G}^\star$ of eq.~\eqref{eq:adjointRgreenfunc}.  We
show in \S\ref{sec:compwithEL89} that it does not.

%%%FIG
\begin{figure}
  \begin{center} 
  \includegraphics[draft=false,scale=0.5,trim=0.75cm 1cm
    0cm 2cm]{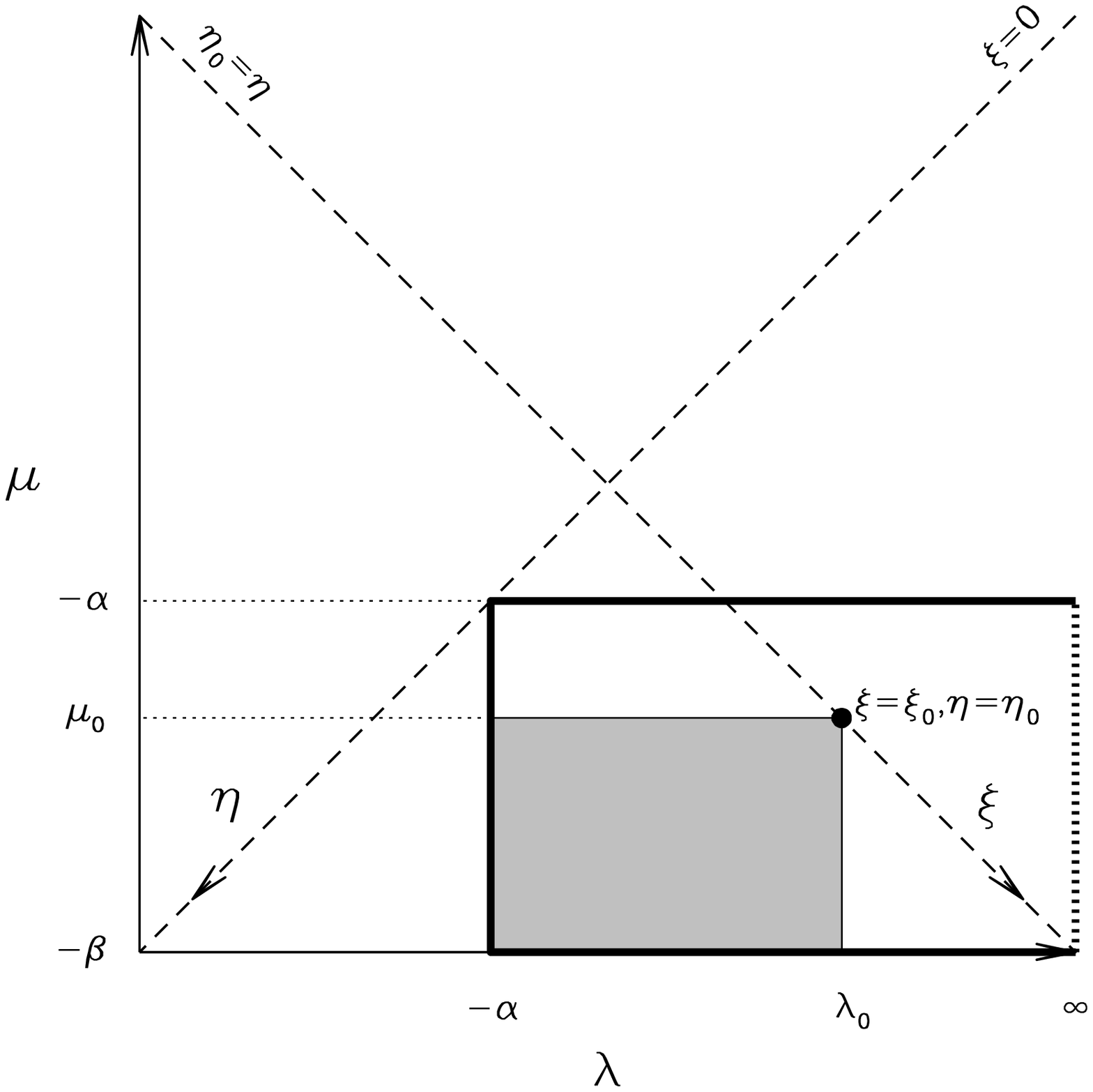} 
  \end{center}
    \caption[]{\slshape The physically relevant region of the
    $(\lambda,\mu)$-plane for the determination of the Riemann--Green
    function $G$, overlayed with the new coordinates $\xi$ and $\eta$ 
    \eqref{eq:newcoordsxieta}. The dot marks the source point of the
    Riemann--Green function $G$ at $(\lambda_0,\mu_0)$. This function
    is non-zero only in the shaded region, which denotes the domain of 
    influence in the $(\lambda,\mu)$-plane 
    of that source point. Fig.~\ref{fig:pointmasscontribution} on the
    other hand shows the $(\lambda_0,\mu_0)$-plane. It is relevant to
    the solution for the stress at a single field point $(\lambda,\mu)$. 
    The hatched region $D$ of Fig.~\ref{fig:pointmasscontribution} shows
    the domain of dependence of the field point, that is the portion of the
    source plane on which the solution at the field point depends.}
    \label{fig:intgregionxiandetaplane}
\end{figure}
%%%FIG 

\subsection{Laplace transform}
\label{sec:laplacetransform}

We use a Laplace transform to solve \eqref{eq:ELPDEdiff} because the
required solution is that to an initial value problem to which Laplace
transforms are naturally suited.  The PDE is hyperbolic with the lines
$\lambda=\mathrm{const}$ and $\mu=\mathrm{const}$ as
characteristics, and its solution is non-zero only in the rectangle
bounded by the characteristics $\lambda=\lambda_0$ and $\mu=\mu_0$,
and the physical boundaries $\lambda=-\alpha$ and $\mu=-\beta$
(Fig.~\ref{fig:intgregionxiandetaplane}).  We introduce new
coordinates
\begin{equation}
  \label{eq:newcoordsxieta}
  \xi = (\lambda\!-\!\mu)/\sqrt{2}, \qquad 
  \eta = -(\lambda\!+\!\mu)/\sqrt{2},
\end{equation}
so that eq.~\eqref{eq:ELPDEdiff} simplifies to
\begin{equation}
  \label{eq:ELPDEdifftwo}
  \mathcal{L}^\star G \equiv
  \frac{\partial^2 G}{\partial \eta^2}
  - \frac{\partial^2 G}{\partial \xi^2}
  - \frac{\partial}{\partial \xi} \left( \frac{G}{\xi} \right) =
  2 \delta (\xi-\xi_0)\delta (\eta-\eta_0),
\end{equation}
where $\xi_0 = (\lambda_0\!-\!\mu_0)/\sqrt{2}$ and $\eta_0 =
-(\lambda_0\!+\!\mu_0)/\sqrt{2}$ are the coordinates of the source point.
The factor of 2 arises from the transformation of the derivatives; the
product of the delta functions in \eqref{eq:ELPDEdiff} transforms into
that of \eqref{eq:ELPDEdifftwo} because the Jacobian of the
transformation \eqref{eq:newcoordsxieta} is unity. The reason for our
choice of $\eta$ is that $G \equiv 0$ for $\eta<\eta_0$, that is
$\lambda+\mu>\lambda_0+\mu_0$.  Hence $\eta$ is a time-like variable
which increases in the direction in which the non-zero part of the
solution propagates.  We take a Laplace transform in $\tilde{\eta}
=\eta-\eta_0$, and transform $G(\xi,\eta)$ to
\begin{equation}
  \label{eq:laptrafoG}
  \hat{G}(\xi,p) = \int\limits_0^{\infty} e^{-p\tilde{\eta}}
  G(\xi,\tilde{\eta}) \d\tilde{\eta}.
\end{equation}
There are two equally valid ways of taking proper account of the
$\delta(\eta-\eta_0)$ in taking the Laplace transform of equation
\eqref{eq:ELPDEdifftwo}. One can either treat it as
$\delta(\tilde{\eta}-0+)$, in which case it has a Laplace transform of
1, or one can treat it as $\delta(\tilde{\eta}-0-)$, in which case it
contributes a unit initial value to $\partial G/\partial \eta$ which
must be included in the Laplace transform of $\partial^2 G/\partial
\eta^2$ (Strauss 1992\nocite{S92}).  Either way leads to a transformed
equation for $\hat{G}(\xi,p)$ of
\begin{equation}
  \label{eq:laptrafoG_PDE}
  p^2 \hat{G} - \frac{\d^2 \hat{G} }{\d \xi^2} - \frac{\d}{\d \xi}   
  \left( \frac{\hat{G}}{\xi} \right) = 2\delta(\xi-\xi_0).
\end{equation}
The homogeneous part of eq.~\eqref{eq:laptrafoG_PDE} is the modified
Bessel equation of order one in the variable $p\xi$.  Two
independent solutions are the modified Bessel functions $I_1$ and
$K_1$. The former vanishes at $\xi=0$ and the latter decays
exponentially as $\xi \to \infty$.  We need $\hat{G}$ to decay
exponentially as $\xi \to \infty$ because $G(\xi,\eta)$ vanishes
for $\tilde{\eta}<\xi-\xi_0$, and hence its Laplace transform
$\hat{G}$ is exponentially small for large $\xi$. We also need $\hat
G$ to vanish at $\xi=0$ where $\lambda=\mu$. The focus at which
$\lambda=\mu=-\alpha$ is the only physically relevant point at
which $\xi=0$. It lies on a boundary of the solution region in the
$\lambda_0 \to-\alpha$ limit (Fig.~\ref{fig:intgregionxiandetaplane}).
The focus is a point at which the difference $\Delta$ between the
stresses vanishes, and hence $G$ and $\hat{G}$ should vanish there.
The delta function in eq.~\eqref{eq:laptrafoG_PDE} requires that
$\hat{G}$ be continuous at $\xi=\xi_0$ and that $\d\hat{G}/\d\xi$
decrease discontinuously by 2 as $\xi$ increases through
$\xi=\xi_0$. Combining all these requirements, we obtain the result
\begin{equation}
  \label{eq:sollaptrafoG}
  \hat{G}(\xi,p) =
  \begin{cases}
    2\xi_0\,K_1(p\xi)\,I_1(p\xi_0), 
                              &\text{$\xi_0 \le \xi < \infty$}, \\
    2\xi_0\,K_1(p\xi_0)\,I_1(p\xi), 
                              & \text{$0 \le \xi \le \xi_0$}.
  \end{cases}
\end{equation}
We use the Wronskian relation $I_1(x)K_1'(x) - I_1'(x)K_1(x) = -1/x$
(eq.\ [9.6.15] of Abramowitz \& Stegun 1965\nocite{AS65}) 
in calculating the prefactor of the products
of modified Bessel functions.  The inversion of this transform is
obtained from formula (13.39) of Oberhettinger \& Badii
(1973\nocite{OB1973...laplacetrafo}) which gives
\begin{equation}
\label{eq:sollapinvtrafoG}
  G(\xi,\!\tilde{\eta}) \! = \!\!
  \begin{cases}
    \! \sqrt{\! \frac{\xi_0}{\xi}} \,
    {}_2F_1( -\scriptstyle \frac{1}{2} \displaystyle, \scriptstyle
    \frac{3}{2} \displaystyle ;\!1 ; w), \hspace{-3pt} &   
    \text{$|\xi_0\!\!-\!\xi| \!\le\! \tilde{\eta} 
                                   \!\le\! \xi_0\!\!+\!\xi$}, \\  
    0, & \text{$-\infty \!<\! \tilde{\eta} \!<\! |\xi_0\!\!-\!\xi|$}, 
  \end{cases}
\end{equation}
we have (cf.\ eq.~[\ref{eq:definitionofw}])
\begin{equation}
  \label{eq:defw_xieta}
  w \equiv \frac{\tilde{\eta}^2-(\xi_0\!-\!\xi)^2}{ 4\xi_0\xi}  
  = \frac{(\lambda_0\!-\!\lambda) (\mu_0\!-\!\mu) }{
  (\lambda_0\!-\!\mu_0) (\lambda\!-\!\mu) }.
\end{equation}
The second case of eq.~\eqref{eq:sollapinvtrafoG} shows that $G$ does
indeed va\-nish outside the shaded sector $\lambda<\lambda_0$,
$\mu<\mu_0$. The first case shows that it agrees with the adjoint
Riemann--Green function $\mathcal{G}^\star$ of
\eqref{eq:adjointRgreenfunc} which was derived from the analysis of
\S\ref{sec:riemannsmethod}. \looseness=-2

\subsection{Comparison with EL89}
\label{sec:compwithEL89}

EL89 use variables $s=-\eta$ and $t=\xi$, whereas we avoided
using $t$ for the non-time-like variable. They consider the Fourier
transform
\begin{equation}
  \label{eq:fouriertrafoG}
    \bar{G}(\xi,k) = \int\limits_{-\infty}^{\infty} e^{-ik\tilde{\eta}} 
      {G}(\xi,\tilde{\eta}) \d\tilde{\eta}.
\end{equation}
Because $G \equiv 0$ for $\tilde{\eta} \le 0$, we can rewrite our
Laplace transform as their Fourier transform. Setting $p=-ik$ gives
$\bar{G}(\xi,k) = i \hat{G}(\xi,-ik)$, and using the formulas
$I_1(x)=-J_1(ix)$ and $K_1(x)=\frac{1}{2}\pi i H_1^{(1)}(ix)$,
eq.~\eqref{eq:sollaptrafoG} yields
\begin{equation}
  \label{eq:solfouriertrafoGour}
  \bar{G}(\xi,k) =
  \begin{cases}
    \pi i \xi_0\,H_1^{(1)}(k\xi)\,J_1(k\xi_0),  
    & \text{$\xi_0 \le \xi < \infty$}, \\ 
    \pi i \xi_0\,H_1^{(1)}(k\xi_0)\,J_1(k\xi), 
    & \text{$0   \le \xi \le \xi_0$}.
  \end{cases}
\end{equation}
This formula differs from the solution for the Fourier transform given
in eq.~(70) of EL89. The major difference is that their solution has
Hankel functions of the second kind $H_1^{(2)}(kt)=H_1^{(2)}(k\xi)$
where ours has $J_1$ Bessel functions.  Consequently their solution
has an unphysical singularity at $t=\xi=0$, and so, in our opinion, is
incorrect. Our solution, which was devised to avoid that singularity,
gives a result which matches that derived by Riemann's method in
\S\ref{sec:riemannsmethod}. \looseness=-2

\subsection{The solution for $\Delta$}
\label{sec:soldiffandcompriemann}

The solution for $\Delta$ using the adjoint Riemann--Green function is
given by eq.~\eqref{eq:solutionstressT} with $\mathcal{G}$ replaced by
$\mathcal{G}^\star$ and the sign of $c_2$ changed for the adjoint case
(Copson 1975\nocite{C75}). 
The hypergeometric function of eq.~\eqref{eq:adjointRgreenfunc} for
$\mathcal{G}^\star$ is expressible in terms of complete elliptical
integrals as
\begin{equation}
{}_2F_1(-\scriptstyle\frac{1}{2}\displaystyle,
    \scriptstyle\frac{3}{2} \displaystyle;1;w)
    =\frac{2}{\pi}[E(w)+2wE'(w)].
\end{equation}
Hence, the solution for the difference $\Delta$ between the two
principal stresses is given by 
\begin{multline}
  \label{eq:solutiondiff}
     \Delta(\lambda,\mu) = 
     \frac{2}{\pi(\lambda\!-\!\mu)^\frac{1}{2}} \Biggl\{ \\
     \int\limits_\lambda^\infty \hspace{-4pt} \d\lambda_0 
     \hspace{-5pt} \int\limits_\mu^{-\alpha} \hspace{-4pt} \d\mu_0
          \!\Bigl[\! E(w) + 2wE'(w) \!\Bigr]\!
     (\lambda_0\!-\!\mu_0)^\frac{1}{2} \!
     \biggl( \!
     \frac{\partial \rho}{\partial \lambda_0} \!
     \frac{\partial V_S }{\partial \mu_0}
     \!-\! 
     \frac{\partial \rho}{\partial \mu_0} \!
     \frac{\partial V_S }{\partial \lambda_0}
     \! \biggr) \\
     \hspace{-5pt}
    - \!\! \int\limits_\lambda^\infty \hspace{-4pt} \d\lambda_0 
          \!\biggl[\! E(w) + 2wE'(w)
     \hspace{-12pt} \underset{\mu_0=-\alpha}{\biggr]} \hspace{-12pt}
     \frac{\d}{\d\lambda_0} \!\Bigl[\!
     (\lambda_0\!+\!\alpha)^\frac{1}{2} \Delta(\lambda_0,-\alpha)
     \!\Bigr]\! 
     \!\Biggr\}.
     \hspace{-4pt}
\end{multline}
The determined reader can verify, after some manipulation, that this
expression is equivalent to the difference between the separate
solutions \eqref{eq:solutionTlambdalambda} and
\eqref{eq:solutionTmumu}, derived in \S\ref{sec:riemannsmethod}.

\subsection*{Note added in manuscript}
\label{sec:addednote}

We agree with the amendment to our method of solution for $\Delta$
given in Appendix \ref{sec:soldiffandcompriemann}. 
Our Green's function, while solving the differential equation, had the
wrong boundary conditions.\\
{\small N.W. Evans \& D. Lynden-Bell}

% ===== END MATTER ===== %

\bsp % ``This paper has been produced using the ...''

\label{lastpage}

\end{document}